\documentclass[aps,floatfix,superscriptaddress,twocolumn]{revtex4}
\pdfoutput=1 
\usepackage{amsmath,amssymb,eucal,graphicx}
\usepackage{float}
\usepackage{curves}
\usepackage{epstopdf}

\usepackage{multirow}
\usepackage{rotating}
\usepackage{color}

\def\l{{\mathcal L}}

\def\ve{\varepsilon}

\def\Rve{{\mathcal R}_\ve}
\def\r{{\bf r}}
\def\f{{\bf f}}
\def\e{{\bf e}}
\def\Li{{\rm Li}}
\def\etam{\omega}  
\def\etamm{\omega}   

\begin{document}

\title{Diffusive escape through a narrow opening: new insights into a classic problem}

\author{Denis S Grebenkov}
\affiliation{Laboratoire de Physique de la Mati\`{e}re Condens\'{e}e, CNRS, Ecole Polytechnique, Universit\'e Paris Saclay, F-91128 Palaiseau Cedex, France}
\email{denis.grebenkov@polytechnique.edu}
\author{Gleb Oshanin}
\affiliation{Laboratoire de Physique Th\'{e}orique de la Mati\`{e}re Condens\'{e}e, UPMC,
CNRS UMR 7600, Sorbonne Universit\'{e}s, 4 Place Jussieu, 75252 Paris Cedex 05, France}
\email{oshanin@lptmc.jussieu.fr}

\begin{abstract}

We study the mean first exit time $T_{\ve}$ of a particle diffusing in
a circular or a spherical micro-domain with an impenetrable confining
boundary containing a small escape window (EW) of an angular size
$\ve$.  Focusing on the effects of an energy/entropy barrier at the
EW, and of the long-range interactions (LRI) with the boundary on the diffusive search for the EW, we
develop a self-consistent approximation to derive for $T_{\ve}$ a
general expression, akin to the celebrated Collins-Kimball relation in
chemical kinetics and accounting for both rate-controlling factors in
an explicit way.  Our analysis reveals that the barrier-induced
contribution to $T_{\ve}$ is the dominant one in the limit $\ve \to
0$, implying that the narrow escape problem is not
``diffusion-limited'' but rather ``barrier-limited''.  We present the
small-$\ve$ expansion for $T_{\ve}$, in which the coefficients in
front of the leading terms are expressed via some integrals and
derivatives of the LRI potential.  On example of a triangular-well
potential, we show that $T_{\ve}$ is non-monotonic with respect to the
extent of the attractive LRI, being minimal for the ones having an
intermediate extent, neither too concentrated on the boundary nor
penetrating too deeply into the bulk.  Our analytical predictions are
in a good agreement with the numerical simulations.  
\end{abstract}

\maketitle

\section{Introduction}

The narrow escape problem (NEP) is ubiquitous in molecular and
cellular biology and concerns diverse situations when a particle,
diffusing within a bounded micro-domain, has to search for a small
specific target on the domain's boundary
\cite{lin,zhou,grig,yan2,schuss2,ben1,kol1,kol2}.  A particle can be
an ion, a chemically active molecule, a protein, a receptor, a ligand,
etc.  A confining domain can be a cell, a microvesicle, a compartment,
an endosome, a caviola, etc, while the target can be a binding or an
active site, a catalytic germ, or a narrow exit to an outer space,
from which case the name of the problem originates.  The outer space
can be an extracellular environment or, as considered in recent
analysis of diffusion and retention of $Ca^{2+}$-calmoduline-dependent
protein kinase II in dendritic spines \cite{byr,dag}, be the dendrite
itself while the narrow tunnel is a neck separating the spine and the
dendrite.  For all these stray examples, called in a unified way as
the NEP, one is generally interested to estimate the time needed for a
particle, starting from a prescribed or a random location within the
micro-domain, to arrive for the first time to the location of the
target.  The history, results, and important advances in understanding
of the NEP have been recently
reviewed \cite{schuss,new,holc,holc2,holc3}.

Early works on NEP were focused on situations when the confining
boundary is a hard wall, i.e., is perfectly reflecting everywhere,
except for the escape window (or other specific target), which is also
perfect in the sense that there is neither an energy nor even an
entropy barrier which the particle has to overpass in order to exit
the domain (or to bind to the specific site).  In this perfect,
idealised situation, the mean first exit time (MFET) through this
window is actually the mean first passage time (MFPT) to its location.
This MFPT was calculated by solving the diffusion equation with mixed
Dirichlet-Neumann boundary conditions \cite{ward}.

Capitalising on the idea that the particle diffusion along the
bounding surface can speed up the search process, set forward for
chemoreception by Adam and Delbr\"uck \cite{adam} and for protein
binding to specific sites on a DNA by Berg {\it et al.}  \cite{berg},
as well as on the idea of the so-called intermittent search
\cite{ben2,ben3,osh1,osh2,osh3,met2,met,alj}, more recent analysis of
the NEP dealt with intermittent, surface-mediated diffusive search for
an escape window, which also was considered as perfect, i.e. having no
barrier at its location.  The MFPTs were determined, using a mean-field
\cite{osh20} and more elaborated approaches
\cite{greb1,greb2,greb3,greb4,budde0,albe,budde,ben4,ber1,ber2}, as a
function of the adsorption/desorption rates, and of the surface
($D_{\rm surf}$) and the bulk ($D$) diffusion coefficients.  It was
found that, under certain conditions involving the ratio between $D$
and $D_{\rm surf}$, the MFPT can be a non-monotonic function of the
desorption rate, and can be minimised by an appropriate tuning of this
parameter
\cite{greb1,greb2,greb3,greb4,budde0,albe,budde,ben4,ber1,ber2}.

\begin{figure}
\begin{center}
\includegraphics[width=60mm]{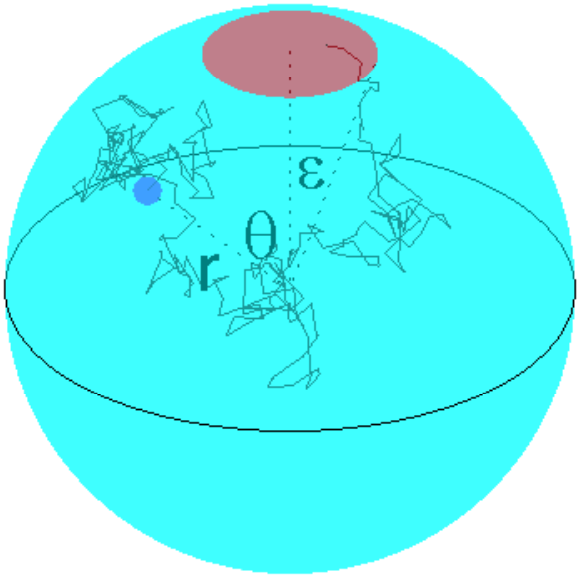}  
\end{center}
\caption{
Schematic picture of the narrow escape problem : An escape window (EW)
is a cap of polar angle $\ve$ and diameter $a = 2R\sin \ve$ located at
the North pole of a sphere of radius $R$. A diffusive particle starts
from a point $(r,\theta)$ (small filled circle) and eventually arrives
to the EW.}
\label{fig:scheme}
\end{figure}

In this paper we analyse the problem of a narrow escape of a Brownian
particle through a small window located on the boundary of a
three-dimensional spherical or a two-dimensional circular domain (see
Fig.~\ref{fig:scheme}) taking into account {\it explicitly} a finite
energy and/or an entropy barrier at the EW and the presence of
long-range (beyond the usual hard-core repulsion) particle-boundary
interactions characterised by a radially-symmetric potential $W(r)$.
We will use the term ``escape window'' (EW) in what follows, but
clearly our results will also apply to situations when this targeted
area is an extended binding region or an active site.  Our motivations
here are as follows:

i) The adsorption/desorption rates cannot be tuned independently as
they are linked on the microscopic level by the particle-boundary
interactions.  As a consequence, not all values of the latter
parameters are physically possible and it is unclear \textit{a priori}
if the non-monotonic behaviour of the MFPT observed in
earlier works can indeed take place in physical systems.

ii) If diffusion along the surface may indeed speed up the search for
the EW in case of contact\footnotemark\footnotetext{The term ``contact''  means
that a diffusing particle feels the boundary only appearing in its
immediate vicinity. The particle may then adsorb onto the boundary in
a non-localised way, perform surface diffusion and desorb back to the
bulk, re-appear in the vicinity of the boundary again, and etc.  This is
precisely the physical picture underlying the idea of the so-called
intermittent, surface-mediated search for the EW.}  attractive
interactions with the surface, it is natural to expect that, in the
presence of long-range attractive interactions with the surface, these
typical search times will be further reduced since the particle will
feel the surface on longer distances and will experience a drift
towards it.  Moreover, on intuitive grounds, one may expect the
existence of some optimal extent of the interaction potential since
for potentials with an infinite extent, the problem will reduce to the
one with purely hard-wall interactions.  Consequently, it may turn out
that the mean search time will be a non-monotonic function of the
extent of the interaction potential.

iii) In realistic systems not every passage to the EW results in the
escape from the domain, but this event happens only with some finite
probability, due to an energy or an entropy barrier, the latter being
dominated by the window size \cite{zhou,paolo1,paolo2}.  Similarly,
chemical reactions with a specific site on the boundary of the
micro-domain are also never perfect but happen with a finite
probability defining some rate constant $\kappa$ \cite{col}.  This rate
constant has already been incorporated into the theoretical analysis
of the NEP for systems with hard-wall or contact interactions with the
surface \cite{budde}, but to the best of our knowledge, its effect on
the MFET has never been appreciated in full detail\footnotemark\footnotetext{See also
\cite{holc3,1,2,3} for the analysis of the NEP with a
stochastically-gated EW, which situation is equivalent, in the limit
when the stochastic gating process has no memory, to the presence of a
partial reflection \cite{4}.}.  We will revisit this analysis for the
situations with (and without) the long-range particle-boundary
interaction.


These are the focal questions of our analysis.  We proceed to show
that the MFET naturally decomposes into two terms: the mean first
passage time to the EW and the time necessary to overcome the barrier
at the entrance to the EW.  We realise that the MFPT to the EW is
indeed an optimisable function of the interaction potential, when
these interactions are attractive, the effects being more pronounced
for spherical micro-domains than for circular micro-domains.  More
specifically, we find that while the MFPT appears to be a monotonic
function of the strength of the interaction potential at the boundary,
it exhibits a minimum with respect to the spatial extent of the
potential, which defines the force acting on the particle in the
vicinity of the boundary.  We show that the optimum indeed occurs for
interactions of an intermediate extent, such that they are neither
localised too close to the confining boundary, nor extend too deeply
into the bulk of the micro-domain.  In a way, the observation that the
MFPT exhibits a non-monotonic behaviour as the function of the extent
of the potential is consistent with the earlier predictions for the
optimum of the MFPT with respect to the desorption rate, based on the
models with intermittent motion.  As a matter of fact, the extent of
the potential defines the barrier against desorption from the
confining boundary and hence, controls the desorption rate.  There
are, however, some discrepancies between our predictions here and
those based on intermittent modelling on which we will comment in what
follows.

Further on, focusing on the effect of a barrier at the entrance to the
EW, which is always present in realistic situations, we demonstrate
that the contribution to the MFET stemming out from the passage
through the barrier is more singular in the limit $\ve \to 0$ than the
MFPT to the EW.  This means that, mathematically speaking, the former
provides the dominant contribution to the MFET implying that the
narrow escape problem is not ``diffusion-limited'' but rather
``barrier-limited''.  This observation has significant consequences
for chemical and biological applications.

Our analytical approach is based on the backward Fokker-Planck
equation with a long-range potential that governs the MFET to an
imperfect (partially-reflecting, $\kappa < \infty$) EW for a particle
starting from a given location within the micro-domain.  We obtain an
explicit approximate solution of the resulting mixed boundary value
problem by resorting to an approximation devised originally by Shoup,
Lipari and Szabo \cite{shoup} for the analysis of reaction rates
between particles with inhomogeneous reactivity \cite{solc}.  Within
this approximation, the exact boundary conditions are replaced by some
effective ones, reducing the problem to finding self-consistent
solutions.  The original self-consistent approximation was shown to be
in a good agreement with numerical solutions of the original problem
\cite{shoup,trey}.

We adapt this approximation to the NEP and also incorporate the
long-range potential interactions between the particle and the
boundary of the confining domain.  Our theoretical predictions, based
on this self-consistent approximation, will be checked against
available exact asymptotic results for the case when the boundary is
an impenetrable hard wall (long-range interactions are absent) and
when there is no barrier at the entrance to the EW.  For the general
case, we will verify our analytical predictions by extensive Monte
Carlo simulations and an accurate numerical solution of the original
mixed boundary problem by a finite elements method.

The paper is organised as follows: In Sec. \ref{mod} we describe our
model and introduce the main idea of the self-consistent approximation
(SCA).  In Sec. \ref{3Dgen} we obtain the solution of the NEP with the
modified boundary conditions, as prescribed by the SCA.  In
Sec. \ref{disc} we first present the general expressions for the MFET
for three-dimensional spherical and two-dimensional circular
micro-domains, for arbitrary particle-wall interactions, an arbitrary
$\kappa$ and the EW of an arbitrary angular size.  Discussing its
physical significance, we highlight the crucial role of the partial
reactivity and its effect on the MFET. Further on, we show that in the
narrow escape limit $\ve \to 0$ one can straightforwardly derive an
asymptotic small-$\ve$ expansion for the MFET in which the expansion
coefficients of the leading terms are explicitly defined via some
integrals and derivatives of a rather arbitrary interaction potential.
Lastly, on example of a representative triangular-well potential, we
discuss the role of repulsive and attractive particle-boundary
interactions and also demonstrate that the contribution to the MFET
due to the diffusive search for the location of the EW (i.e., the MFPT
to the EW), can be optimised by an appropriate tuning of the spatial
extent of the interaction potential.  We also analyse here some subtle
issues related to the applicability of the Adam-Delbr\"uck
dimensionality reduction scheme \cite{adam}.  Section \ref{conclusion}
concludes the paper with a brief recapitulation of our most
significant results.  Mathematical and technical details are presented
in Supplemental Materials (SM), where we describe the numerical
approaches used to verify our analytical predictions (\ref{SM1});
derive the asymptotic small-$\ve$ expansions and determine the
asymptotic behaviour for short-range potentials (\ref{SM2}); present
solutions in absence of particle-wall interactions (\ref{SM3}); and
discuss the particular case of a triangular-well interaction potential
for three-dimensional (\ref{SM4}) and two-dimensional (\ref{SM5})
micro-domains.

\section{Model and basic equations}
\label{mod}
 
Consider a point particle diffusing, with a diffusion coefficient $D$,
in a three-dimensional sphere (3D case) of radius $R$ or a
two-dimensional disk (2D case) of radius $R$, containing on the
confining boundary a small EW characterised by a polar angle $\ve$ and
having a diameter $a = 2 R \sin \ve$, see Fig. \ref{fig:scheme}.  We
stipulate that, in addition to the hard-core repulsion at the
boundary, the particle interacts with the confining wall via a
radially-symmetric potential\footnotemark\footnotetext{Note that $W(r)$ defines the
long-range interaction potential \textit{beyond} the hard-wall
repulsion. Therefore, the (repulsive) part of realistic interaction
potentials, say that of the Lennard-Jones $12-6$ potential, which
strongly diverges near the boundary can be thought off as already
included into the hard-core part of the interaction potential upon an
appropriate choice of some effective location of the boundary. In this
case $W(R)$ be will be a regular function and all its derivatives at
$r=R$ will exist.  We refer to \cite{pop} for a more detailed
discussion of this issue.} $W(r)$.  We hasten to remark that an
assumption that the interaction potential $W(r)$ depends only on the
radial distance from the origin is only plausible in the narrow escape
limit, i.e., when the polar angle is small ($\ve \ll \pi$) or,
equivalently, the linear extent $a$ of the EW is much smaller than the
radius $R$.  Otherwise, the interaction potential can acquire a
dependence on the angular coordinates.  Therefore, from a physical
point view, this model taking into account the long-range interactions
with the boundary in a spherically-symmetric form is representative
only when $\ve \ll \pi$.  On the other hand, for situations with $W(r)
\equiv 0$ the latter constraint can be relaxed and our analysis is
applicable for any value of $\ve$.

We are mainly interested in the MFET through the EW, $T_{\ve}$, from a
uniformly distributed random location, sometimes called the global
MFET \cite{bonito} and defined as
\begin{equation}
\label{definition_global}
T_{\ve} = \dfrac{1}{V_d} \int_{\Omega} dV_{d} \, t(r,\theta) \,,
\end{equation}
where $V_d$ is the volume of the micro-domain $\Omega$ ($V_{3} = 4
\pi R^3/3$ in 3D case and $V_2 = \pi R^2$ for 2D case) and
$t(r,\theta)$ is the mean time necessary for a particle, started from
some fixed location $(r,\theta)$ inside the domain, to exit through
the EW.  Due to the symmetry of the problem, $t(r,\theta)$ depends on
the radial distance $r$ to the origin ($0 \leq r \leq R$) and polar
angle $\theta$ ($0 \leq \theta \leq \pi$) but is independent of the
azimuthal angle $\phi$ for the 3D case.  In the 2D case, the
reflection symmetry of the circular micro-domain with respect to the
horizontal axis allows one to restrict the polar angle to $(0,\pi)$ as
well.  Although we focus in this paper exclusively on the global MFET
$T_\ve$, the SCA yields the explicit form of $t(r,\theta)$ too.

The function $t = t(r,\theta)$ satisfies the \textit{backward}
Fokker-Planck equation.  For the 3D case, the MFET $t$ obeys
\begin{equation}
\label{eq:Poisson}
t''  + \left(\frac{2}{r} - U' \right)  t'  
+ \frac{1}{r^2 \sin\theta} \frac{\partial}{\partial\theta} \left(\sin\theta \frac{\partial t}{\partial\theta}\right) = - \frac{1}{D} \,,
\end{equation}
where the prime here and henceforth denotes the derivative with
respect to the variable $r$, and $U(r)$ is a dimensionless, reduced
potential: $U(r) = \beta W(r)$, where $\beta$ is the reciprocal
temperature measured in the units of the Boltzmann constant.  For the
2D case, the MFET $t$ is governed by
\begin{equation}
\label{eq:Poisson_2d}
t''  + \left(\frac{1}{r} - U'(r) \right) t'  
+ \frac{1}{r^2} \frac{\partial^2 t}{\partial\theta^2} = - \frac{1}{D} .
\end{equation}
In both cases, the backward Fokker-Planck equation is to be solved
subject to the reflecting boundary condition which holds everywhere on
the wall ($r = R$), except for the location of the EW, on which a
partially-adsorbing boundary condition is imposed.  For the
\textit{backward} Fokker-Planck equation, these {\it mixed} boundary
conditions have the form\footnotemark\footnotetext{Note that the boundary condition (\ref{eq:BC_flux})
for the \textit{forward} Fokker-Planck equation will have a different
form \cite{Gardiner}:
\begin{equation}
- D \, (t' + t \, U')_{r=R} = \begin{cases} \kappa \, t(R,\theta), \, \qquad  0 \leq \theta \leq \ve ,\cr
0, \hskip 16mm  \ve < \theta \leq \pi , \end{cases} \nonumber\\
\end{equation}
i.e., it will include the derivative of the potential.}:
\begin{equation}
\label{eq:BC_flux}
- \left. D \, t'\right|_{r=R} = \begin{cases}  \kappa \, t(R,\theta) , \qquad  0 \leq \theta \leq \ve ,\cr
0 ,\hskip 18mm  \ve < \theta \leq \pi, \end{cases}
\end{equation}
where $\kappa$ is the above mentioned proportionality factor (in units
length/time), which accounts for an energy or an entropy barrier at
the entrance to the EW.  If such a barrier is absent, $\kappa =
\infty$ and one has a perfectly adsorbing boundary condition at the
location of the EW, in which case the first line of (\ref{eq:BC_flux})
becomes $t(R,\theta) = 0$ for $ 0 \leq \theta \leq \ve$.  Evidently,
in this case $T_{\ve}$ is entirely determined by the first passage to
the EW.

Within the context of chemical reactions with an active site located
on the inner surface of the micro-domain, the condition in the first
line of (\ref{eq:BC_flux}) can be considered as the
partially-reflecting reactive boundary condition put forward in the
seminal paper by Collins and Kimball
\cite{col}, and then extensively studied in the context of chemical
reactions \cite{Sano79,Sapoval94,Grebenkov06,Singer08,Bressloff08} and
search processes \cite{Grebenkov10a,Grebenkov10b,budde}.  In this
framework, $\kappa$ can be written down as $\kappa = K/(4 \pi R^2
s_g)$, where $K$ is the usual elementary reaction act constant (in
units of volume times the number of acts per unit of time within this
volume) and $s_g =\sin^2(\ve)$ is the geometric steric factor
characterising the fraction of the boundary area ``covered'' by the
active site, so that $4\pi R^2 s_g$ is simply the area of the active
site.  In turn, $K$ can be written\cite{berd} as $K = \textit{f} \,
V_s$ (see also the discussion in \cite{pop}), where $\textit{f}$  is
the rate describing the number of reaction acts per unit of time
within the volume $V_s$ of the reaction zone around an active site.
If the reaction takes place within a segment of a spherical shell,
defined by $R - \rho \leq r \leq R$ and $\theta \in (0,\ve)$, where
$\rho$ is the capture radius, one has $V_s = 4 \pi R^2 \rho s_g$, so
that $\kappa = \textit{f} \, \rho$.  Clearly, a similar argument holds
for the case of the EW with an energy barrier; in this case
$\textit{f}$ can be interpreted as the rate of successful passages
through the EW and is dependent on the energy barrier via the standard
Arrhenius equation.

Last but not least, even when the confining boundary is a
structureless hard wall so that an energy barrier is absent, a
particle, penetrating from a bigger volume (micro-domain) to a
narrower region of space, will encounter an entropy barrier $\Delta S$
at the entrance to the EW; in this case $\beta
\Delta S \sim \ln(a/R)$ \cite{paolo1,paolo2}, $a$ being the lateral size of the escape
window.  This suggests, in turn, that $\kappa$, associated with an
entropy barrier, depends linearly on $a$ and hence, linearly on $\ve$.
Therefore, for realistic situations one may expect that $\kappa$
associated with an entropy barrier is rather small, and gets
progressively smaller when $\ve \to 0$, becoming a rate-controlling
factor.  However, the absolute majority of earlier works on the NEP
has been devoted so far to the limit $\kappa = \infty$, which
corresponds to an idealised situation when there is neither energy,
nor even entropy barrier at the entrance to the EW so that the
particle escapes from the micro-domain upon the first arrival to the
location of the EW.  In this case the MFET is just the MFPT to the EW.
As we will show, even if $\kappa$ is independent of $\ve$, the passage
through the barrier provides the dominant contribution to the MFET in
the narrow escape limit $\ve \to 0$.

Our approach to the solution of the mixed boundary-value problem in
(\ref{eq:Poisson}, \ref{eq:BC_flux}) for the 3D case or in
(\ref{eq:Poisson_2d}, \ref{eq:BC_flux}) for the 2D case in the
presence of a general form of the reduced potential $U(r)$ hinges on
the SCA developed by Shoup, Lipari and Szabo \cite{shoup}, who studied
rates of an association of particles (e.g., ligands), diffusing in the
bath outside of an impenetrable sphere without LRI potential (i.e.,
$U(r) \equiv 0$), to some immobile specific site covering only a
portion of the outer surface of this sphere.  Within this
approximation one replaces the actual, mixed, partially adsorbing
boundary condition in the first line of (\ref{eq:BC_flux}) by an
\textit{effective} one -- a condition of a constant flux through the
boundary at the location of the EW.  More specifically, the mixed
boundary condition in (\ref{eq:BC_flux}) is replaced by an
inhomogeneous Neumann boundary condition:
\begin{equation}
\label{eq:BC_approx}
\left. D t'\right|_{r=R} = Q \, \Theta(\ve - \theta) \,,
\end{equation}
where $\Theta(\theta)$ is the Heaviside function and $Q$ is the
unknown flux which is to be determined self-consistently using an
appropriate closure relation.

It is important to emphasise two following points: (i) the solution of
the modified problem (\ref{eq:Poisson}, \ref{eq:BC_approx}) is defined
up to a constant while the solution of the original problem
(\ref{eq:Poisson}, \ref{eq:BC_flux}) is unique.  However, in the
narrow escape limit $\ve\to 0$, the leading terms of the MFET diverge
so that a missing constant would give a marginal contribution.  (ii)
the replacement of the boundary condition (\ref{eq:BC_flux}) by an
effective one (\ref{eq:BC_approx}) does not guaranty, in principle,
that the solution $t$ will be positive in the vicinity of the EW,
since the effective boundary condition requires that
(\ref{eq:BC_flux}) holds only on average (see also \ref{SM3}).  We
will show that this approximation provides nonetheless accurate
results for the global MFET with zero and non-zero $U(r)$. 

\section{Self-consistent approximation}
\label{3Dgen}

In this section we adapt the SCA for the NEP and also incorporate
within this approach the long-range potential interactions with the
confining boundary
\footnotemark\footnotetext{
See also recent \cite{pop} in which an analogous approach
combined with a hydrodynamic analysis has been developed to calculate
the self-propulsion velocity of catalytically-decorated colloids.}.
Within this extended approach, we derive general expressions for
$T_{\ve}$ for arbitrary potentials in both 2D and 3D cases, as
functions of the radius $R$ of the micro-domain, the angular size
$\ve$ of the EW, the constant $\kappa$, and the bulk diffusion
coefficient $D$.

\subsection{3D Case} 

The general solution of (\ref{eq:Poisson}) can be written as the
following expansion
\begin{equation}
\label{eq:u_ansatz}
t(r,\theta) = t_0(r) + \sum\limits_{n=0}^\infty a_n g_n(r) P_n(\cos(\theta)) \,,
\end{equation}
where $P_n(\cos(\theta))$ are the Legendre polynomials, $a_n$ are the
expansion coefficients which will be chosen afterwards to fulfil the
boundary conditions, $g_n(r)$ are the radial functions obeying
\begin{equation}
\label{eq:gn}
g''_n(r) + \left(\frac{2}{r} - U'(r)\right) g'_n(r) - \frac{n(n+1)}{r^2} g_n(r) = 0 \,,
\end{equation}
and lastly, $t_0(r)$ is the solution of the inhomogeneous problem that
can be written down explicitly as
\begin{equation}
t_0(r) = \frac{1}{D} \int\limits_{r}^{c_1} dx~ \frac{e^{U(x)}}{x^2} \int\limits_{c_2}^x dy~ y^2 e^{-U(y)} ,
\end{equation}
where $c_1$ and $c_2$ are two adjustable constants.  We set $c_2 = 0$
to ensure the regularity of solution at $r = 0$.  In order to fix the
constant $c_1$, we impose the Dirichlet boundary condition at $r = R$
to get
\begin{equation}
\label{eq:u0}
t_0(r) = \frac{1}{D} \int\limits_{r}^{R} dx~ \frac{e^{U(x)}}{x^2} \int\limits_{0}^x dy~ y^2 e^{-U(y)} .
\end{equation}
We emphasise that the Dirichlet boundary condition for $t_0(r)$ is
chosen here for convenience only.  As mentioned earlier, the solution
of the modified problem (\ref{eq:Poisson}, \ref{eq:BC_approx}) is
defined up to a constant which can be related to the constant $c_1$
here.  An evident advantage of such a choice is that $t_0(r)$ in
(\ref{eq:u0}) is the {\it exact} solution of the {\it original}
problem in case when $\ve = \pi$ (i.e., the EW is the entire boundary
of the sphere).

We turn next to the radial functions $g_n(r)$ defined in
(\ref{eq:gn}).  For $n=0$ the radial function can be defined
explicitly for an arbitrary potential $U(r)$ to give
\begin{equation}
g_0(r) = c_1 + c_2 \int\limits^r dx \frac{e^{U(x)}}{x^2}.
\end{equation}
To ensure that this solution is regular at $0$, we again set $c_2 =
0$, so that $g_0(r) \equiv 1$ (we set $c_1 = 1$ for convenience).  For
$n > 0$, explicit solutions of (\ref{eq:gn}) can be calculated only
when one makes a specific choice of the interaction potential.  In the
next section and in the SM, we will discuss the forms of $g_n(r)$ for
a triangular-well potential $U(r)$.  In general, we note that $g_n(r)$
are also defined up to two constants.  One constant is fixed to ensure
the regularity of the solution at $r = 0$.  The second constant can be
fixed by the choice of their value at $r = R$.  Without any lack of
generality, we set $g_n(R)=1$.  As a matter of fact, the final results
will include only the ratio of $g_n(r)$ and of its first derivative,
and hence, will not depend on the particular choice of the
normalisation.

Next, substituting (\ref{eq:u_ansatz}) into (\ref{eq:BC_approx}) we get
\begin{equation}
\label{eq:condition}
\sum\limits_{n=1}^\infty a_n P_n(\cos(\theta)) g'_n(R) = \frac{Q}{D} \Theta(\ve - \theta) - t_0'(R),
\end{equation}
where we have used $g_0(r) \equiv 1$.  Multiplying the latter equation
by $\sin(\theta)$ and integrating over $\theta$ from $0$ to $\pi$, we
find the following expression for the flux $Q$:
\begin{equation}
\label{eq:Q}
Q = \frac{2D t'_0(R)}{1-\cos(\ve)} \,,
\end{equation}
where, in virtue of  (\ref{eq:u0}),
\begin{equation}
\label{eq:t0prime_3d}
D \,  t'_0(R) = - \frac{e^{U(R)}}{R^2} \int\limits_0^R dr~ r^2 e^{-U(r)} .
\end{equation}
Note that for $U(r) \equiv 0$, the expression in (\ref{eq:Q})
coincides with the standard compatibility condition for the interior
Neumann problem.  Note also that $Q$ depends only on the form of the
interaction potential, $R$ and the angular size $\ve$ of the EW but it
is independent of the kinetic parameters $\kappa$ and $D$.

Further on, multiplying both sides of (\ref{eq:condition}) by
$P_m(\cos(\theta)) \sin(\theta)$ and integrating the resulting
equation over $\theta$ from $0$ to $\pi$, we get, taking advantage of
the orthogonality of the Legendre polynomials, the following
representation of the expansion coefficients $a_n$:
\begin{equation}
a_n = \frac{t'_0(R) \phi_n(\ve)}{g'_n(R)} \,,  \quad n > 0 \,,
\end{equation}
where we used (\ref{eq:Q}) and defined
\begin{equation}
\label{eq:phin3d}
\phi_n(\ve) = \frac{P_{n-1}(\cos(\ve)) - P_{n+1}(\cos(\ve))}{1-\cos(\ve)} \,, \quad n > 0.
\end{equation}
Gathering the above expressions, we rewrite (\ref{eq:u_ansatz}) as
\begin{equation}
\label{eq:urtheta}
\begin{split}
t(r,\theta) & = t_0(r) + a_0 + t'_0(R) \sum\limits_{n=1}^\infty \frac{g_n(r)}{g'_n(R)} \phi_n(\ve) P_n(\cos(\theta)) \,,  \\
\end{split}
\end{equation}
in which only $a_0$ remains undefined.  Since the solution of the
modified problem is defined up to a constant, one could stop here,
leaving $a_0$ as a free constant.  To be closer to the original
problem, we fix $a_0$ through the self-consistency condition by
plugging (\ref{eq:urtheta}) into (\ref{eq:BC_flux},
\ref{eq:BC_approx}), multiplying the result by $\sin\theta$ and
integrating over $\theta$ from $0$ to $\ve$, to get the following
closure relation:
\begin{eqnarray}
\label{eq:a0_selfconsistent}
- \kappa \hspace*{-1mm} \int\limits_0^{\ve} d\theta \sin(\theta) \, t(R,\theta) &=& D \hspace*{-1mm}
\int\limits_0^{\ve} d\theta \sin(\theta)  \left.t'\right|_{r=R} \nonumber\\
&=& Q \hspace*{-1mm} \int\limits_0^{\ve} d\theta \sin(\theta) \,.
\end{eqnarray}
Using the latter relation, noticing that $t_0(R) = 0$ by construction (see
(\ref{eq:u0})) and excluding $Q$ via (\ref{eq:Q}), we arrive at
\begin{equation}
\label{eq:a0}
a_0 = - R \, t'_0(R) \biggl[\frac{2D}{R\kappa(1-\cos(\ve))} + \Rve^{(3)} \biggr] \,, 
\end{equation}
where
\begin{equation}
\label{eq:Rd_3d}
\Rve^{(3)}  = \sum\limits_{n=1}^\infty \frac{g_n(R)}{Rg'_n(R)} \frac{\phi_n^2(\ve)}{(2n+1)} .
\end{equation}
Equations (\ref{eq:u0}, \ref{eq:phin3d}, \ref{eq:a0}, \ref{eq:Rd_3d})
provide an exact, closed-form solution (\ref{eq:urtheta}) of the
modified problem in (\ref{eq:Poisson}, \ref{eq:BC_approx}) for the 3D
case.  This solution is valid for an arbitrary initial location of the
particle, an arbitrary reaction rate $\kappa$ and an arbitrary form of
the interaction potential $U(r)$.  We also note that $\Rve^{(3)} $ in
(\ref{eq:Rd_3d}) is a non-trivial function which encodes all the
relevant information about the fine structure of the interaction
potential.

\subsection{2D case} 

We follow essentially the same line of thought like in the previous
subsection.  In two dimensions the equation (\ref{eq:gn}) for the
radial functions becomes
\begin{equation}
\label{eq:gn2d}
g''_n + \left(\frac{1}{r} - U'\right) g'_n - \frac{n^2}{r^2} g_n = 0 \,.
\end{equation}
The general solution for $t(r,\theta)$ then reads
\begin{equation}
\label{2d}
t(r,\theta) = t_0(r) + a_0 + 2 t'_0(R) \sum\limits_{n=1}^\infty  \frac{g_n(r)}{g'_n(R)} ~ \frac{\sin(n\ve)}{n \ve}  \cos(n\theta) ,
\end{equation}
where $\cos(n\theta)$ replace the Legendre polynomials from the 3D
case, and the solution of the inhomogeneous problem has the form:
\begin{equation}
\label{eq:u0_2d}
t_0(r)  = \frac{1}{D} \int\limits_{r}^{R} dx~ \frac{e^{U(x)}}{x} \int\limits_{0}^x dy~ y ~e^{-U(y)} .
\end{equation}
The coefficient $a_0$ in (\ref{2d}) is given explicitly by
\begin{equation}
\label{a_02d}
a_0  = - Rt'_0(R) \biggl(\frac{\pi D}{R\kappa \, \ve} + \Rve^{(2)}  \biggr),  
\end{equation}
with
\begin{equation}
\label{rv2d}
\Rve^{(2)}  = 2\sum\limits_{n=1}^\infty  \frac{g_n(R)}{Rg'_n(R)} ~ \left(\frac{\sin(n\ve)}{n \ve}\right)^2  \,. 
\end{equation}
Equations (\ref{2d}) to (\ref{rv2d}) determine an exact, closed-form
solution $t(r,\theta)$ in the modified problem for the 2D case.  Like
in the 3D case, $\Rve^{(2)}$ in (\ref{rv2d}) contains all the relevant
information about the long-range interaction potential.

\section{Results and discussion}
\label{disc}

Capitalising on the results of the previous section, we find the
following general expressions for the global MFET defined in
(\ref{definition_global}): 
\begin{eqnarray}
\label{ckI}
T^{(3)}_{\ve} &=& \underbrace{\dfrac{R^2 }{3 D} \Rve^{(3)} \l^{(3)}_U(R)
+ \dfrac{R^2}{3 D} \int^1_0 dx \, x^4  \, \l^{(3)}_U(xR)}_{\textrm{diffusion to the EW}}  \nonumber \\
&+& \underbrace{\dfrac{2R \l_U^{(3)}(R)}{3 \, \kappa \, (1-\cos(\ve))} }_{\textrm{barrier at the EW}}  \, , 
\end{eqnarray}
where $\l^{(3)}_U(r)$ is the functional of the potential $U(r)$,
given explicitly by
\begin{equation}
\l^{(3)}_U(r) = 3 \dfrac{e^{U(r)}}{r^3} \int^r_0 d\rho \, \rho^2 \, e^{- U(\rho)} \,,
\end{equation}
and
\begin{eqnarray}
\label{ckII}
T^{(2)}_{\ve} &=& \underbrace{\dfrac{R^2}{2 D} \Rve^{(2)} \l^{(2)}_U(R)
 + \dfrac{R^2}{2 D} \int^1_0 dx \,  x^3 \, \l^{(2)}_U(xR)}_{\textrm{diffusion to the EW}}  \nonumber \\
&+& \underbrace{\dfrac{\pi  R \, \l^{(2)}_U(R)}{2 \, \kappa \, \sin(\ve)} }_{\textrm{barrier at the EW}}  \, , 
\end{eqnarray}
with
\begin{equation}
\l^{(2)}_U(r) = 2 \dfrac{e^{U(r)}}{r^2} \int^r_0 d\rho \, \rho \, e^{-U(\rho)} \,.
\end{equation}
For $U(r) \equiv 0$, both $\l^{(2)}_U(r)$ and $\l^{(3)}_U(r)$ are
simply equal to $1$.

We note next that $\Rve^{(3)}$ and $\Rve^{(2)}$ vanish when $\ve =
\pi$, so that the second terms in the first line in (\ref{ckI}) and
(\ref{ckII}), i.e.
\begin{equation}
\label{boundary3}
T^{(3)}_{\pi}(\kappa=\infty) =\dfrac{R^2}{3 D} \int^1_0 dx \, x^4  \, \l^{(3)}_U(xR)
\end{equation}
and
\begin{equation}
\label{boundary2}
T^{(2)}_{\pi}(\kappa=\infty) =\dfrac{R^2}{2 D} \int^1_0 dx \,  x^3 \, \l^{(2)}_U(xR) \,,
\end{equation}
can be identified as the MFPTs from a random location to any point on
the boundary of the micro-domain, in presence of the
radially-symmetric interaction potential $U(r)$.  In what follows we
will show that these (rather simple) $\ve$-independent contributions
to the overall MFET exhibit quite a non-trivial behaviour as functions
of the parameters of the interaction potential $U(r)$.

Equations (\ref{ckI}) and (\ref{ckII}) are the main general result of
our analysis.  We emphasise that these expressions have the same
physical meaning as the celebrated relation for the apparent rate
constant due to Collins and Kimball \cite{col} and define the global
MFET as the sum of two contributions: the first one is the time
necessary for a diffusing particle (starting from a random location
within the micro-domain) to find the EW (i.e., the MFPT to the EW),
while the second one describes the time necessary to overcome a finite
barrier at the entrance to the EW, once the particle appears in its
vicinity.  Clearly, the last contribution vanishes when $\kappa \to
\infty$, i.e., in the perfect EW (reaction) case, while the first one
vanishes for an infinitely fast diffusive search, i.e., when $D \to
\infty$.  The additivity of the two controlling factors will permit us
to study separately the effects due to a finite $\kappa$, and the
effects associated with the diffusive search for the EW, biased by the
long-range potential $U(r)$.  We proceed to show that, interestingly
enough, the last term in (\ref{ckI}) and (\ref{ckII}) is always
dominant in the limit $\ve \to 0$, i.e., the rate-controlling factor
for the NEP is the barrier at the entrance, not the diffusive search
process.  To the best of our knowledge, this important conclusion has
not been ever spelled out explicitly, but may definitely have
important conceptual consequences for biological and chemical
applications.

One notices next that the first terms in (\ref{ckI}) and (\ref{ckII})
contain infinite series $\Rve^{(3)}$ and $\Rve^{(2)}$ implying that
these terms can be explicitly determined only when one (i) specifies
the interaction potential, (ii) manages to solve exactly the
differential equations (\ref{eq:gn}) or (\ref{eq:gn2d}) for the radial
functions $g_n(r)$ corresponding to the chosen $U(r)$ and, (iii) is
able to sum the infinite series.  At the first glance, this seems to
be a severe limitation of the approach because equations (\ref{eq:gn})
or (\ref{eq:gn2d}) can be solved exactly only for a few basic
potentials.  Quite remarkably, however, we managed to bypass all these
difficulties and to determine the asymptotic behaviour of the MFET in
the narrow escape limit $\ve \to 0$ for a rather general class of the
interaction potentials without solving the differential equations
(\ref{eq:gn}) or (\ref{eq:gn2d}).  The circumstance, which allows us
to circumvent solving these equations, is that the small-$\ve$
behaviour of the infinite series $\Rve^{(3)}$ and $\Rve^{(2)}$ is
dominated by the terms $g_n(R)/g'_n(R)$ with $n\to \infty$, whose
asymptotic behaviour can be directly derived from (\ref{eq:gn},
\ref{eq:gn2d}) for potentials $U(r)$ which have a bounded first
derivative by constructing an appropriate perturbation theory
expansion.  For such potentials, we obtain (see
\ref{SM2} for more details):
\begin{eqnarray}
\label{asympR3D}
&& \Rve^{(3)} = \dfrac{32}{3 \pi} \ve^{-1} + \bigl(1 - R \, U'(R)\bigr) \ln(1/\ve)  \nonumber\\
&+& \ln 2 - \frac{7}{4} - \left(\frac{1}{4} + \frac{\pi^2}{12} + \ln 2\right) R \, U'(R) \\
&+& \sum_{n=1}^{\infty} (2 n+1) \left(\dfrac{g_n(R)}{R g'_n(R)} - \dfrac{1}{n} + \dfrac{R \, U'(R)}{2 n^2}\right) + O(\ve) \,,   \nonumber
\end{eqnarray}
where $U'(R)$ denotes the derivative of the interaction potential
right at the boundary, and the symbol $O(\ve)$ signifies that the
omitted terms, in the leading order, vanish linearly with $\ve$ when
$\ve \to 0$.  Note that one needs the precise knowledge of the radial
functions $g_n(r)$ only for the calculation of the subdominant,
$\ve$-independent term in the third line in (\ref{asympR3D}).

Analogous calculations for the 2D case (see the \ref{SM2}) entail the
following small-$\ve$ expansion:
\begin{eqnarray}
\label{asympR2D}
\Rve^{(2)} &=& 2 \ln(1/\ve) + 3 - 2 \ln 2  \nonumber\\
&+& 2 \sum_{n=1}^{\infty} \left(\dfrac{g_n(R)}{R g'_n(R)}-\dfrac{1}{n}\right) + O(\ve) \,.
\end{eqnarray}
Again, the precise knowledge of the radial functions $g_n(r)$ is only
needed for the calculation of the $\ve$-independent term, which
embodies all the dependence on the interaction potential; the leading
term in this small-$\ve$ expansion appears to be completely
independent of $U(r)$.

Combining Eqs. (\ref{ckI}) and (\ref{asympR3D}) we find that for an
arbitrary potential $U(r)$ possessing a bounded first derivative
within the domain, the MFET for the 3D case admits the following
small-$\ve$ asymptotic form
\begin{align}
\label{gen3}
&T^{(3)}_{\ve} = \dfrac{4 R \l_U^{(3)}(R)}{3  \, \kappa} \ve^{-2} + \dfrac{32 R^2 \l_U^{(3)}(R)}{9 \pi D} \ve^{-1}  \nonumber\\
&+ \dfrac{R^2 \l_U^{(3)}(R)}{3 D} \bigl(1 - R U'(R)\bigr) \ln(1/\ve) + \xi_U^{(3)} + O(\ve) \,,
\end{align}
where the $\ve$-independent, sub-leading term $\xi_U$ is given
explicitly by
\begin{align}
\label{gen3as}
&\xi_U^{(3)} =  T^{(3)}_{\pi}(\kappa=\infty) + \frac{R \l_U^{(3)}(R)}{9\kappa}  \nonumber\\
& + \dfrac{R^2 \l_U^{(3)}(R)}{3 D} \Bigg[\ln 2 - \frac{7}{4} - \left(\ln 2 + \frac{1}{4} + \frac{\pi^2}{12} \right) R U'(R)  \nonumber\\
&+ \sum_{n=1}^{\infty} (2 n+1) \left(\dfrac{g_n(R)}{R g'_n(R)} - \dfrac{1}{n} \left(1- \dfrac{R U'(R)}{2 n}\right)\right)\Bigg] \,,
\end{align}
with the MFPT $T^{(3)}_{\pi}(\kappa=\infty)$ to any point on the
boundary being defined in (\ref{boundary3}).

For the 2D case an analogous small-$\ve$ expansion reads
\begin{eqnarray}
\label{gen2}
T^{(2)}_{\ve} &=& \dfrac{\pi R \l_U^{(2)}(R)}{2\, \kappa}  \ve^{-1} + \dfrac{R^2 \l_U^{(2)}(R)}{D} \ln(1/\ve) + \nonumber\\
&+& \xi_U^{(2)} + O(\ve) \,,
\end{eqnarray} 
where the $\ve$-independent term $\xi_U^{(2)}$ is given explicitly by
\begin{align}
\label{gen2as}
\xi_U^{(2)} &= T^{(2)}_{\pi}(\kappa=\infty)  \nonumber\\
& +  \dfrac{\l_U^{(2)}(R) R^2}{D} \left(\frac{3}{2} -  \ln 2 + \sum_{n=1}^{\infty} \left(\dfrac{g_n(R)}{R g'_n(R)}-\dfrac{1}{n}\right)\right) \,,
\end{align}
with $T^{(2)}_{\pi}(\kappa=\infty)$ being defined in (\ref{boundary2}).

The expressions in (\ref{gen3}) and (\ref{gen2}) constitute our second
general result in the narrow escape limit.  This result has several
interesting features which have to be emphasised.

i) For quite a general class of the interaction potentials $U(r)$,
these expressions make explicit our claim that in the narrow escape
limit $\ve \to 0$ the dominant contribution to the MFET comes from the
passage through the barrier at the EW, rather than from the diffusive
search for the location of the EW.  Indeed, we observe that the terms
originating from the presence of a barrier have at least one (or two
if $\kappa \sim \ve$) more inverse power of $\ve$, as compared to the
terms defining the MFPT to the EW.  Clearly, in the narrow escape
limit $\ve \to 0$, a passage through the EW is the dominant
rate-controlling factor.

ii) Remarkably, it appears that in order to determine the coefficients
in front of the leading terms (diverging in the limit $\ve \to 0$), we
do not have to solve the differential equations for the radial
functions $g_n(r)$ but merely to integrate and to differentiate the
interaction potential.  The resulting expressions for $T_{\ve}^{(d)}$
have quite a transparent structure and all the terms entering
(\ref{gen3}) and (\ref{gen2}) have a clear physical meaning.

For both 3D and 2D cases, the coefficients in front of the leading
term associated with the barrier are entirely defined by
$\l_U^{(3)}(R)$ and $\l_U^{(2)}(R)$.  For the 3D case, the coefficient
in front of the leading term in the MFPT to the EW, which diverges as
$1/\ve$, is also entirely defined by the integrated particle-wall
potential $U(r)$ via $\l_U^{(3)}(R)$, while the sub-leading diverging
term ($\sim \ln(1/\ve)$) depends also on the force, $-U'(R)$, acting
on the particle at the boundary of the micro-domain.  In the 2D case,
the coefficient in front of the leading term in the MFPT to the EW,
diverging as $\ln(1/\ve)$, again is defined by the integrated
potential via $\l_U^{(2)}(R)$.

Therefore, details of the ``fine structure'' of the interaction
potential $U(r)$ (i.e., possible maxima or minima for $r$ away from
the boundary) which are embodied in the radial functions, have a minor
effect on $T^{(3)}_{\ve}$ and $T^{(2)}_{\ve}$ appearing only in the
subdominant terms, which are independent of $\ve$ in the narrow escape
limit.

iii) The $1/\ve$ singularity of the MFPT to the EW is a specific
feature of the 3D case and stems from such diffusive paths, starting
at a random location and ending at the EW, which spend most of the
time in the bulk far from the boundary.  Our analysis shows how the
presence of the long-range interaction potential modifies the
amplitude of the corresponding contribution to the MFET.  For the 3D
case, the sub-leading, logarithmically diverging term accounts for the
contribution of the paths which are most of the time localised near
the confining boundary.  In what follows, we will discuss the relative
weights of these contributions considering the triangular-well
interaction potential as a particular example.

iv) When the potential $U(r)$ is a monotonic function of $r$, the
coefficients $\l^{(2)}_U(r)$ and $\l^{(3)}_U(r)$ are monotonic
functions of the amplitude $U_0$ of the potential $U(r)$.  In fact,
setting $U(r) = U_0 {\mathcal U}(r)$, one gets
\begin{equation}
U_0 \frac{\partial \l^{(d)}_U(r)}{\partial U_0} = \frac{d}{r^d} e^{U(r)} \int\limits_0^r d\rho~ \rho^{d-1} ~ e^{-U(\rho)} ~ \bigl[U(r) - U(\rho)\bigr] 
\end{equation}
so that the derivative in the left hand side does not change the sign.
For instance, if ${\mathcal U}(r)$ is an increasing function, then
$\l^{(d)}_U(r)$ is also increasing for any $U_0$ (even negative).  As
a consequence, the coefficients in front of each term in (\ref{gen3})
and (\ref{gen2}) are monotonic functions of the amplitude $U_0$ of the
potential so that the global MFET is expected to be a monotonic
function of $U_0$, at least in the narrow escape limit.  In
Sec. \ref{sec:triangular}, on example of a triangular-well potential,
we will quest the possibility of having a minimum of the MFET with
respect to the extent of the interaction potential $U(r)$.

\subsection{Global MFET for systems without long-range potentials}
\label{sec:nopotential}

To set up the scene for our further analysis, we consider our
expressions in (\ref{ckI}) and (\ref{ckII}) in case when the confining
boundary is a hard wall (i.e. $U(r) \equiv 0$), concentrating on the
dependence of the global MFET on $\ve$, $\kappa$, $D$ and $R$.  This
will permit us to check how accurate our approach is, by comparing our
predictions against few available exact results, and also to highlight
in what follows the role of the long-range interactions with the
confining boundary.

In absence of the interaction potential, equations (\ref{ckI}) and
(\ref{ckII}) considerably simplify:
\begin{eqnarray}
\label{ckI2}
T^{(3)}_{\ve} &=& \dfrac{2 R}{3 \, \kappa \, (1-\cos(\ve))}  + \dfrac{R^2}{3 D}\left(\Rve^{(3)} + \dfrac{1}{5} \right) \,, \\
\label{ckII2}
T^{(2)}_{\ve} &=& \dfrac{\pi \, R}{2 \, \kappa \, \sin(\ve)} + \dfrac{R^2}{2 D} \left(\Rve^{(2)}  + \dfrac{1}{4} \right) \,,
\end{eqnarray}
where $\Rve^{(3)} $ and $\Rve^{(2)}$ become
\begin{eqnarray}
\label{eq:Rve_3d_U0}
\Rve^{(3)} &=& \sum\limits_{n=1}^\infty \frac{\phi_n^2(\ve)}{n (2n+1)} , \\
\label{r2}
\Rve^{(2)} &=& 2 \sum_{n=1}^{\infty} \dfrac{1}{n} \left(\dfrac{\sin n \ve}{n \ve}\right)^2 .
\end{eqnarray}
The asymptotic small-$\ve$ behaviour of these series is discussed in
detail in
\ref{SM2}.

We focus first on the asymptotic behaviour of $T_{\ve}$ in the limit
$\ve \to 0$.  Using the asymptotic small-$\ve$ expansion, presented in
(\ref{sigma1}, \ref{sigma2}), we find that in the 3D case
\begin{align}
\label{asymp}
&T^{(3)}_{\ve} = \dfrac{4 R }{3 \, \kappa} ~\ve^{-2} + \dfrac{32 R^2}{9 \pi D} \ve^{-1} +
\dfrac{R^2}{3 D} \ln\left(1/\ve\right)   \nonumber\\
&+ \frac{R^2}{3 D} \left(\ln 2 - \frac{31}{20}\right) + \dfrac{R}{9 \, \kappa} +  O\left(\ve\right) \,.
\end{align}
The first term in (\ref{asymp}) is the contribution due to a finite
barrier, while the second and the third terms define the contribution
due to the diffusive search for the EW, stemming out of the
non-trivial term in (\ref{eq:Rve_3d_U0}) proportional to $\Rve^{(3)}$.

For $\kappa \equiv \infty$ (i.e., in an idealised situation when there
is neither an energy nor even an entropy barrier at the entrance to
the EW), the first term in (\ref{asymp}), proportional to $1/\kappa$
and exhibiting the strongest singularity in the limit $\ve \to 0$, is
forced to vanish, so that the leading $\ve$-dependence of $T_{\ve}$
becomes determined by the second term, diverging as $1/\ve$. The first
correction to this dominant behavior is then provided by the
logarithmically diverging term.

Such an asymptotic behaviour qualitatively agrees with the exact
asymptotics due to Singer {\it et al.} \cite{Singer06a}:
\begin{equation}
\label{eq:T3d_Singer1}
T_\ve^{(3)} \simeq \frac{\pi R^2}{3D} \biggl(\ve^{-1} + \ln(1/\ve) + O(1) \biggr)  \,,
\end{equation}
which contains both $1/\ve$- and logarithmically diverging terms.  We
notice however a small discrepancy in the numerical prefactors: the
coefficient $32/(9\pi) \approx 1.1318$ in (\ref{ckI2}) slightly
exceeds the coefficient $\pi/3 \approx 1.0472$ in
(\ref{eq:T3d_Singer1}), the relative error being as small as $8\%$.
In turn, the amplitude of the subdominant term, which is
logarithmically divergent as $\ve \to 0$, appears to be $\pi$ times
less, as compared to the coefficient in the logarithmically divergent
term in (\ref{eq:T3d_Singer1}).  Therefore, for $\kappa \equiv \infty$
and $\ve \to 0$, the SCA predicts correctly the dependence of the
leading term of $T_{\ve}$ on the pertinent parameters but slightly
overestimates its amplitude, and also underestimates the amplitude of
the sub-dominant term, associated with the contribution of the
diffusive paths localised near the confining boundary.

For the 2D case, the summation in (\ref{r2}) can be performed exactly
so that (\ref{ckII2}) can be written explicitly as
\begin{equation}
\label{teII}
T_{\ve}^{(2)} = \dfrac{\pi R}{2 \kappa \sin(\ve)} + \dfrac{R^2}{D} \dfrac{\xi(\ve)}{\ve^2} + \dfrac{R^2}{8 D} \,,
\end{equation}
where 
\begin{equation}
\label{eq:xi}
\xi(\ve) = \sum_{n=1}^{\infty} \dfrac{\sin^2 n \ve}{n^3} = \dfrac{1}{2} \zeta(3) 
- \dfrac{1}{4} \left({\rm Li}_3\left(e^{2 i \ve}\right) + {\rm Li}_3\left(e^{-2 i \ve}\right)\right) ,
\end{equation}
$\zeta$ being the Riemann zeta function, $\zeta(3) \approx 1.202$, and
${\rm Li}_3(y)$ being the trilogarithm: ${\rm Li}_3(y) =
\sum_{n=1}^\infty y^n/n^3$.  In the limit $\ve \to 0$, (\ref{teII})
admits the following asymptotic expansion
\begin{align}
\label{asymp2}
&T_{\ve}^{(2)} = \dfrac{\pi R}{2 \kappa} \ve^{-1} + \dfrac{R^2}{D} \ln\left(1/\ve\right) 
+ \dfrac{R^2}{D} \left(\dfrac{13}{8} - \ln2 \right)
+ O\left(\ve\right) \,.
\end{align}
Again, we notice that the first term, associated with the barrier at
the entrance to the EW, has a more pronounced singularity as $\ve \to
0$ than the second term stemming out of the diffusive search for the
EW.  Hence, similarly to the 3D case, for sufficiently small angular
sizes of the EW, the controlling factor is the passage via the EW, not
the diffusive search for the entrance to the latter.

\begin{figure}
\begin{center}
\includegraphics[width=85mm]{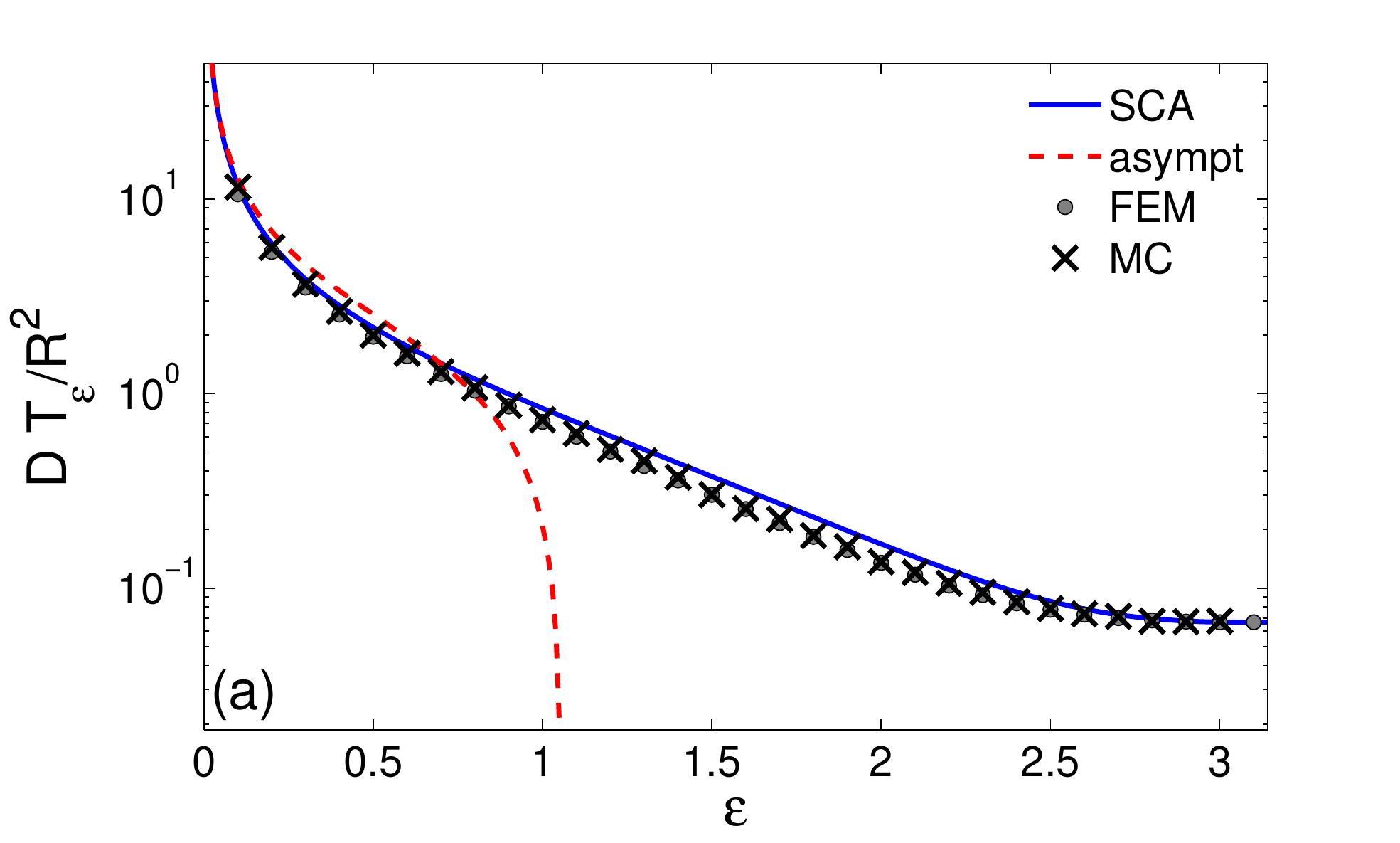}  
\includegraphics[width=85mm]{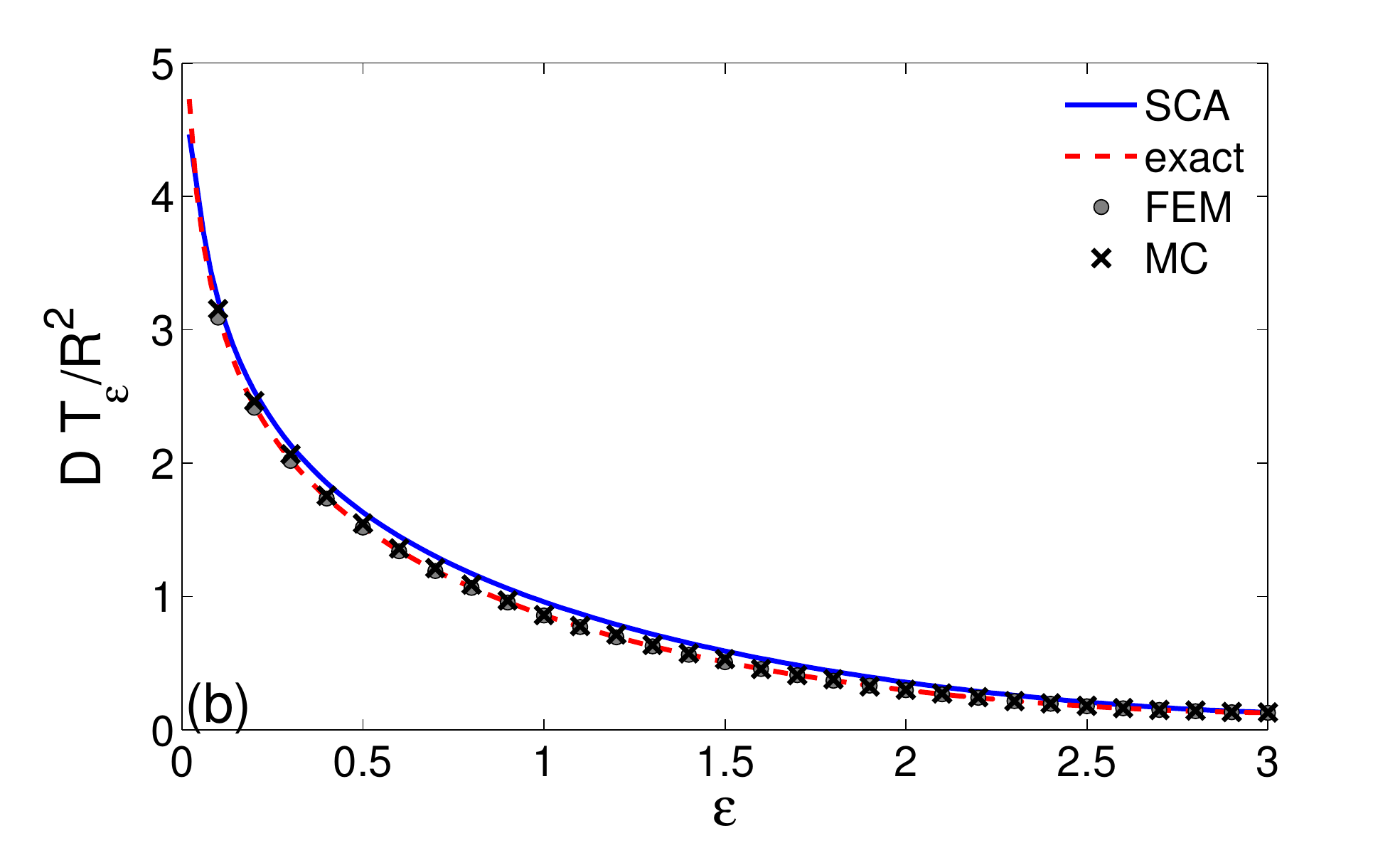}  
\end{center}
\caption{
Hard-wall interactions ($U(r) \equiv 0$) and no barrier at the EW
($\kappa = \infty$).  The dimensionless global MFET, $D
T_\ve^{(d)}/R^2$, as a function of the target angle $\ve$.  {\bf (a)}
3D Case.  Our SCA prediction in (\ref{ckI2}) and the asymptotic
relation in (\ref{eq:T3d_Singer1}) are shown by solid and dashed
lines, respectively.  The numerical solution by the FEM (with the mesh
size $0.01$) is shown by circles, while the Monte Carlo simulations
are shown by crosses.  {\bf (b)} 2D Case.  Our SCA prediction in
(\ref{teII}) (solid line) is compared against the exact result in
(\ref{eq:Tve_exact2}) (dashed line), numerical FEM solution (circles)
and MC simulations (crosses).}
\label{fig:t3d_eps}
\end{figure}

The first term in (\ref{asymp2}), which is proportional to $1/\kappa$
and diverges as $1/\ve$ in the limit $\ve \to 0$, is forced to vanish
in the idealised case $\kappa \equiv \infty$, and the leading
small-$\ve$ behaviour of $T_{\ve}^{(2)}$ in (\ref{asymp2}) becomes
determined by the second term, which exhibits only a slow, logarithmic
divergence with $\ve$.  At this point it is expedient to recall that
the original mixed boundary-value problem in (\ref{eq:Poisson_2d}) and
(\ref{eq:BC_flux}) for the 2D case with $\kappa \equiv \infty$ can be
solved exactly
\cite{Singer06b,Caginalp12,Rupprecht15} and here $T_\ve^{(2)}$ reads:
\begin{eqnarray}
\label{eq:Tve_exact2}
T_\ve^{(2)} &=& \frac{R^2}{D} \left(\frac{1}{8} - \ln(\sin(\ve/2)) \right) \nonumber\\ 
&=& \frac{R^2}{D} \ln\left(1/\ve\right) + \dfrac{R^2}{D} \left(\ln 2 + \dfrac{1}{8}\right) + O\left(\ve^2\right).
\end{eqnarray}
Comparing the solution of the modified problem in (\ref{asymp2}) with
$\kappa \equiv \infty$, and the exact small-$\ve$ expansion in the
second line in (\ref{eq:Tve_exact2}), we conclude that the leading
terms in both equations coincide.  This means that for the 2D case the
SCA developed in the present work determines exactly not only the
dependence of the leading term on the pertinent parameters, but also
the correct numerical factor.  The subdominant, $\ve$-independent term
in (\ref{asymp2}), appears to be slightly larger, by $3/2 - 2\ln 2
\approx 0.114$, than the analogous term in the exact result in
(\ref{eq:Tve_exact2}).

\begin{figure}
\begin{center}
\includegraphics[width=85mm]{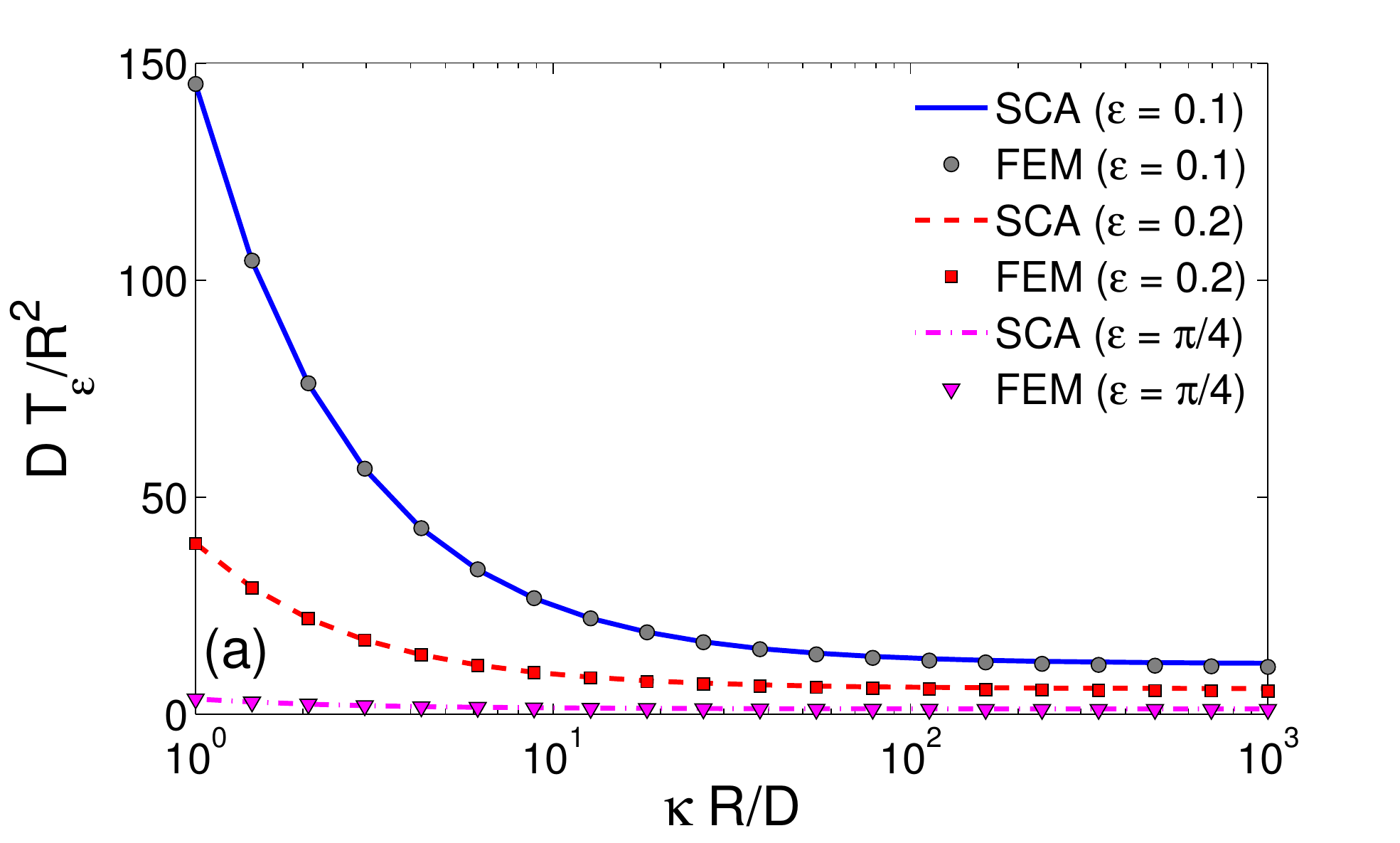}  
\includegraphics[width=85mm]{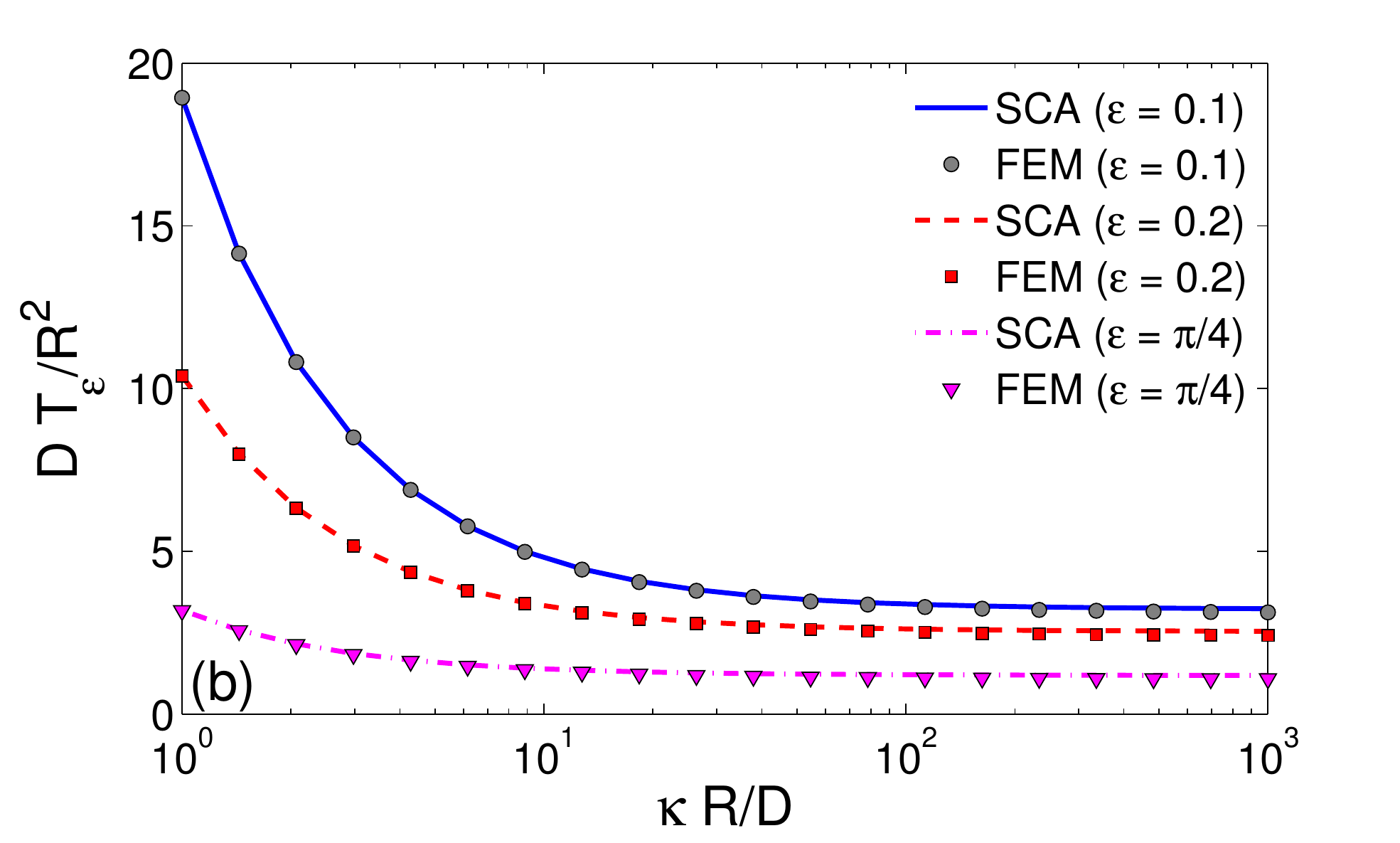}  
\end{center}
\caption{
Hard-wall interactions ($U(r) \equiv 0$) with a barrier at
the EW.  The dimensionless global MFET, $D T_{\ve}^{(d)}/R^2$, as a
function of a dimensionless parameter $\kappa R/D$ for $\ve = 0.1$
(solid line), $\ve = 0.2$ (dashed line) and $\ve = \pi/4$ (dotted
line): {\bf (a)} 3D case and {\bf (b)} 2D case.  The SCA (lines) is
compared to numerical solutions obtained by a FEM (with the mesh size
$0.01$), which are shown by symbols. }
\label{fig:t3d_mean_kappa}
\end{figure}

Since the SCA is applicable to any size of the EW, we will check its
accuracy for the EWs extended beyond the narrow escape limit.  As a
matter of fact, for chemical reactions involving specific sites
localised on the confining boundary, the angular size of the latter is
not necessarily small.  To this end, in the remaining part of this
subsection we will check the $\ve$-dependence of $T_{\ve}$ in
(\ref{ckI2}) and (\ref{teII}) for $\kappa \equiv \infty$, as well as
the $\kappa$-dependence of $T_{\ve}$ for fixed $\ve$, and will compare
our predictions against the results of numerical simulations (see
\ref{SM1}).


We first consider the well-studied case when there is no barrier at
the EW ($\kappa = \infty$), so that the first term in  (\ref{ckI2})
vanishes.  Figure \ref{fig:t3d_eps}a compares the SCA prediction in
(\ref{ckI2}), the asymptotic relation (\ref{eq:T3d_Singer1}) by Singer
{\it et al.} \cite{Singer06a}, and the numerical solution of the
original problem by a finite elements method (FEM) and by Monte Carlo
(MC) simulations, which are described in \ref{SM1}.  We observe a
fairly good agreement over the whole range of target angles, for $\ve$
varying from $0$ to $\pi$, which is far beyond the narrow escape
limit, and the dimensionless parameter $D T_{\ve}/R^2$ varying over
more than two decades.  We find that the SCA provides an accurate
solution even for rather large $\ve$, at which the asymptotic relation
(\ref{eq:T3d_Singer1}) completely fails.  A nearly perfect agreement
is also observed for the 2D case (see Fig. \ref{fig:t3d_eps}b).

Next, we study the $\kappa$-dependence of the global MFET.  As we have
already mentioned, the SCA predicts that the presence of a finite
barrier at the entrance to the EW (and hence, for a finite $\kappa$)
is fully captured by the first term in (\ref{ckI2}) and (\ref{ckII2}).
Figure \ref{fig:t3d_mean_kappa} illustrates how accurately the SCA
accounts for the effect of a partial reactivity $\kappa$ on the global
MFET $T_{\ve}$, even for not too small EW (e.g., for $\ve = \pi/4$)
and a broad range of reactivities $\kappa$.  For even larger EW, the
SCA is still accurate for small $\kappa$ but small deviations are
observed at larger $\kappa$ (not shown).  As expected, the global MFET
diverges as $\kappa \to 0$ because the EW becomes impenetrable, as the
remaining part of the boundary is.  Similar results are obtained for
the 2D case.  Overall, we observe a fairly good agreement between our
theoretical predictions and the numerical simulations, for both 3D and
2D cases.


\subsection{Global MFET for a system with a triangular-well potential.}
\label{sec:triangular}

We focus on a particular choice of the interaction potential $U(r)$ --
a triangular-well radial potential -- defined as (see also
Fig. \ref{Fig2} in \ref{SM4})
\begin{equation}
\label{triangular}
U(r) = \begin{cases}  0,  \hskip 21mm 0 \leq r \leq r_0, \cr
U_0 \dfrac{r-r_0}{R - r_0},  \qquad r_0 < r \leq R, \end{cases}
\end{equation}
where $0 \leq r_0 \leq R$, $R - r_0 \equiv l_{ext}$ is the spatial
extent of the potential (a characteristic scale) and $U_0$ is a
dimensionless strength of the potential at the boundary, $U(R) \equiv
U_0$.  Note that $- U_0/l_{ext}$ can be interpreted as a constant
force exerted on the particle when it appears within a spherical shell
of extent $l_{ext}$ near the boundary of the micro-domain.  The
strength of the potential can be negative (in case of attractive
interactions) or positive (in case of repulsive interactions).  For
this potential, we find explicit closed-form expressions for the
radial functions $g_n(r)$ (see \ref{SM4} and \ref{SM5}) and examine
the dependence of the global MFET on $U_0$, $r_0$ and other pertinent
parameters, such as, $\ve$, $D$, $\kappa$ and $R$.

\begin{figure}
\begin{center}
\includegraphics[width=85mm]{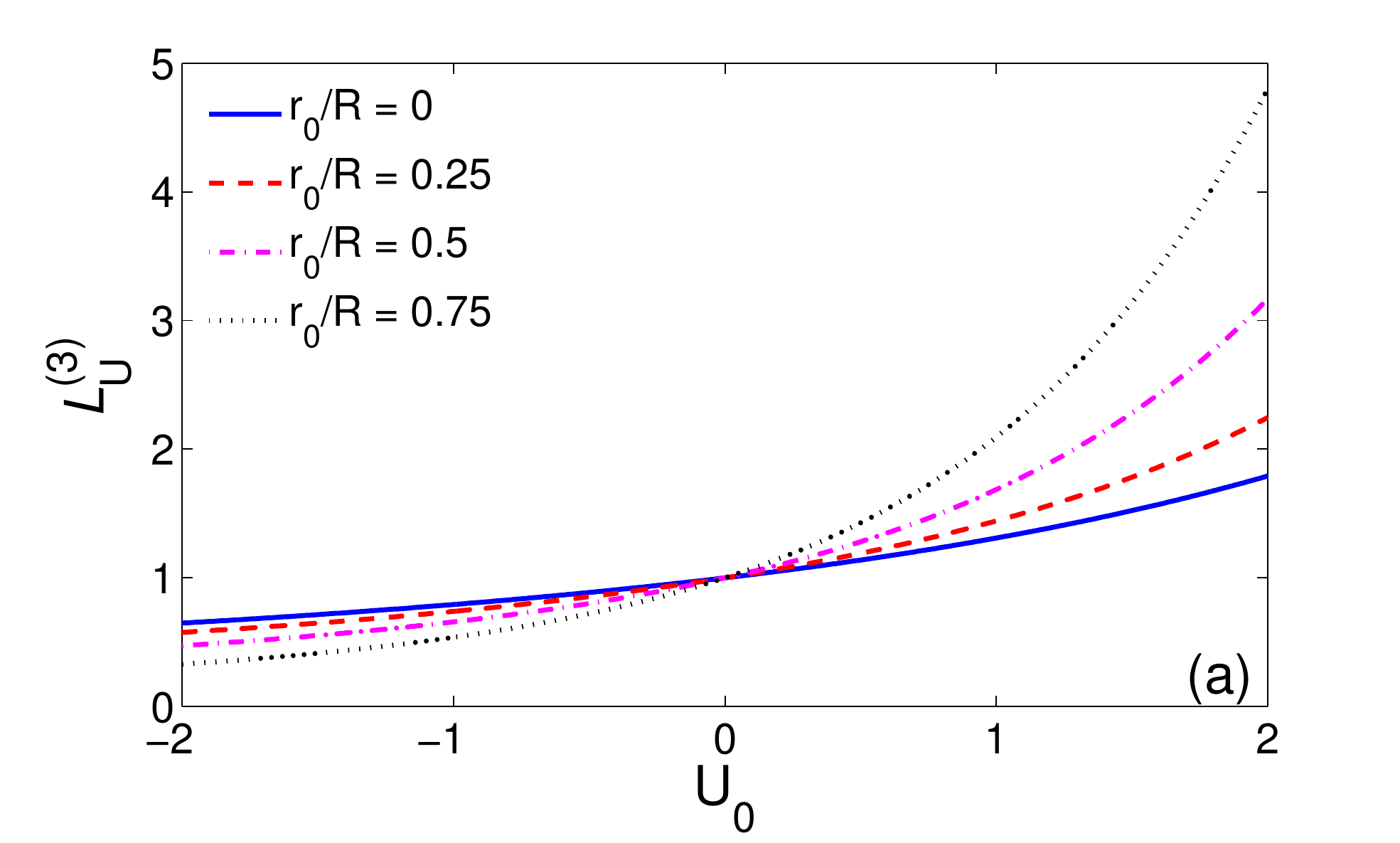} 
\includegraphics[width=85mm]{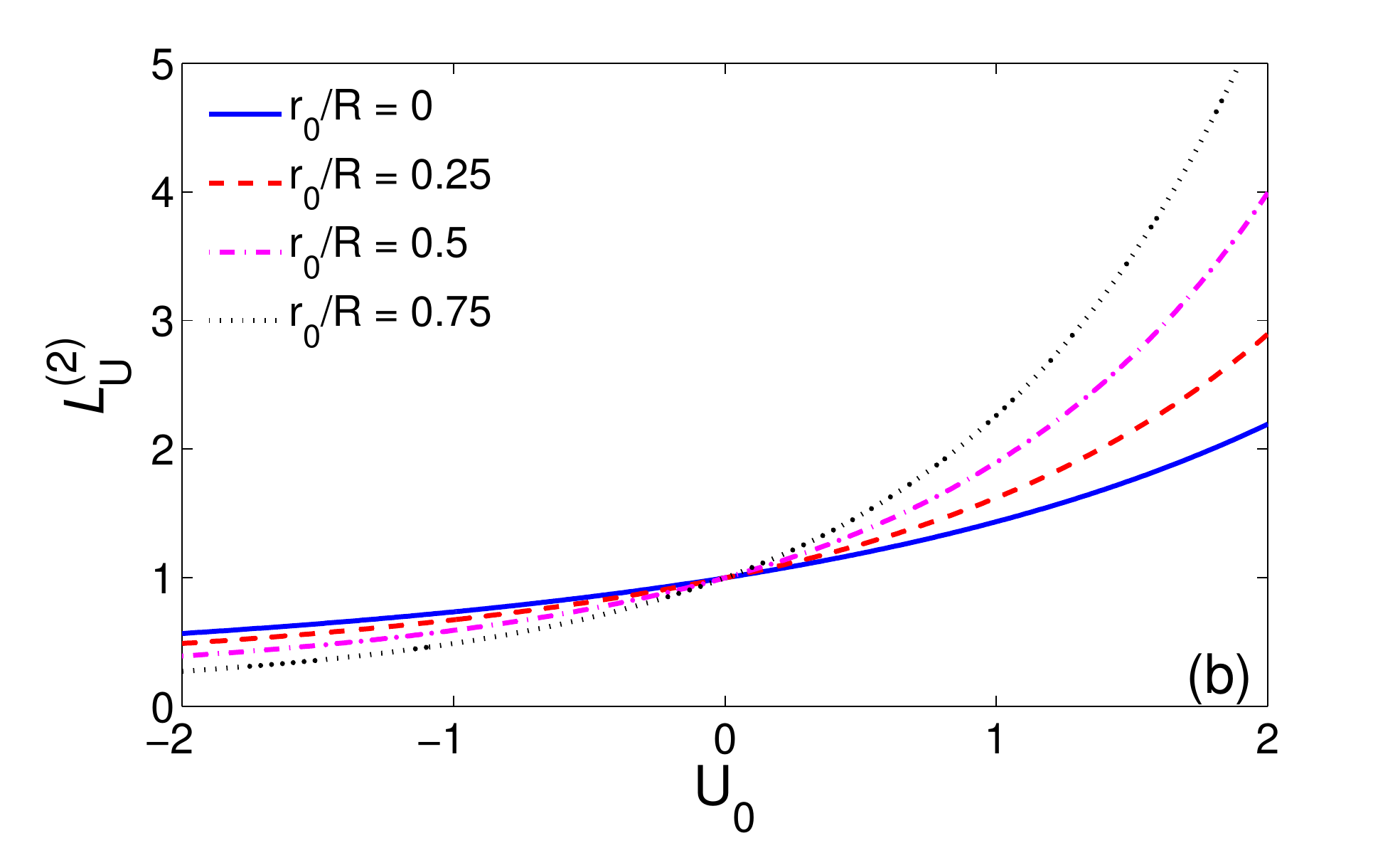} 
\end{center}
\caption{
The amplitudes $\l^{(3)}_U(R)$ {\bf (a)} and $\l^{(2)}_U(R)$ {\bf
(b)} in (\ref{3DtriangularL}) and (\ref{2DtriangularL}) vs $U_0$ for
$r_0 = 0$ (solid line), $r_0/R = 0.25$ (dashed line), $r_0/R = 0.5$
(dash-dotted line) and $r_0/R = 0.75$ (dotted line). }
\label{fig:denis2}
\end{figure}

We begin with the analysis of the functionals $\l^{(3)}_U(R)$ and
$\l^{(2)}_U(R)$, which define the amplitudes of all $\ve$-diverging
leading terms in (\ref{gen3}) and (\ref{gen2}).  For the
triangular-well potential in (\ref{triangular}), these functionals are
given explicitly by
\begin{equation}
\label{3DtriangularL}
\begin{array}{l}
\l^{(3)}_U(R) = \dfrac{3}{U_0^3} \biggl\{- 2 - 2 U_0 - U_0^2 + (6 + 4 U_0 + U_0^2) x_0  \\
- 2 (3 + U_0) x_0^2 + 2 x_0^3+ e^{U_0} \biggl[ 2 + 2 (U_0 - 3) x_0 \\
+ (6 - 4 U_0 + U_0^2) x_0^2 - \biggl(2 - 2 U_0 + U_0^2 - \dfrac{U_0^3}{3}\biggr) x_0^3 \biggr] \biggr\} \,,  \\ \end{array} 
\end{equation}
where $x_0 = r_0/R$, and
\begin{equation}
\label{2DtriangularL}
\begin{array}{l}
\l^{(2)}_U(R) = \dfrac{2}{U_0^2} \biggl\{ -1-U_0 +(U_0+2) x_0 - x_0^2  \\
+ e^{U_0} \biggl[1 + \left(U_0 - 2\right) x_0 +\biggl(1 - U_0 + \dfrac{U_0^2}{2}\biggr) x_0^2\biggr]\biggr\} \,, \\ \end{array}
\end{equation}
for the 3D and the 2D cases, respectively.  We plot expressions in
(\ref{3DtriangularL}) and (\ref{2DtriangularL}) in
Fig. \ref{fig:denis2} as functions of $U_0$ for different values of
$r_0$.  We observe that for any fixed $r_0$, $\l^{(3)}_U(R)$ and
$\l^{(2)}_U(R)$ are monotonically increasing functions of $U_0$, which
agrees with the general argument presented earlier in the text.  In
turn, for any fixed $U_0 >0 $ (repulsive interactions with the
boundary), they are increasing functions of $r_0$ which means that
they are decreasing functions of the extent of the potential,
$l_{ext}=R - r_0$.  Therefore, all the contributions to the global
MFET $T_{\ve}$ become larger for stronger repulsion (as they should)
and also get increased upon lowering the extent the potential (i.e.,
upon an \textit{increase} of the force $- U_0/l_{ext}$ pointing
\textit{towards} the bulk and hindering the passage to and through the
EW).  For sufficiently large positive $U_0$ and for any $r_0 > 0$, the
dominant contribution to $\l^{(d)}_U(R)$ is
\begin{equation}
\l^{(d)}_U(R) \sim (r_0/R)^d \exp(U_0) \,, \qquad d=2,3 \,,
\end{equation}
so that the coefficients in the small-$\ve$ expansions of $T_{\ve}$ in
(\ref{gen3}) and (\ref{gen2}) become exponentially large with $U_0$.

In turn, for negative values of $U_0$ (attractive interactions),
$\l^{(3)}_U(R)$ and $\l^{(2)}_U(R)$ decrease upon an increase of
$|U_0|$ and also decrease when $r_0$ approaches $R$, i.e., when the
interactions become short ranged.

\subsection{Adam-Delbr\"uck scenario: Limit $\etam = R \, U_0/(R - r_0) \to  - \infty$}

Before we proceed to the general case with arbitrary $U_0$ and $r_0$,
we discuss first the situations when the dimensionless parameter
\begin{equation}
\label{eta}
\etam = R\, U'(R) = \frac{R \, U_0}{R -r_0} ,
\end{equation}
has large negative values.  This can be realised for either big
negative $U_0$, which case is more of a conceptual interest but is
apparently not very realistic, or for short-range potentials with
fixed $U_0 < 0$ and $r_0$ close to $R$, the latter case being
physically quite meaningful.

For such values of $\etam$ the leading behaviour of the amplitudes
$\l^{(3)}_U(R)$ and $\l^{(2)}_U(R)$ is simply described by
\begin{equation}
\label{attr}
\l^{(d)}_U(R) \sim  \frac{d}{|\etam|} \,, \qquad d= 2, 3 \,,
\end{equation}
i.e., the amplitudes vanish as $\etam \to - \infty$ as a first inverse
power of $\etam$.  This means, in turn, that all the contributions to
$T_{\ve}^{(d)}$ which are multiplied by $\l^{(d)}_U(R)$ (i.e., both
the contribution due to a barrier at the EW and the MFPT to the EW)
decrease in presence of attractive interactions with the boundary of
the micro-domain.

We analyse next the behaviour of another key ingredient of
(\ref{gen3}, \ref{gen2}) -- the infinite series $\Rve^{(3)}$ and
$\Rve^{(2)}$.  We realise that there is a subtle point in the
behaviour of the latter which stems from the fact that the limit of
large negative $\etam$ and the narrow escape limit $\ve \to 0$ do not
commute.  Taking the limit $\ve \to 0$ first and then turning to the
limit $\etam \to - \infty$, we arrive at the expression which
describes correctly the behaviour of $\Rve^{(3)}$ and $\Rve^{(2)}$ for
small $\ve$ and moderately large $|\etam|$ such that $|\etam| \ll
1/\ve$.  If, on contrary, we first take the limit $\etam \to - \infty$
for a fixed $\ve$, and then take the limit $\ve \to 0$, we obtain the
correct large-$|\etam|$ behaviour, which yields physically meaningful
results for infinitely large negative $\etam$.  This point is
discussed in detail in \ref{SM2}.
  
Accordingly, for the 3D case in the narrow escape limit $\ve \to 0$
with large but finite negative $\etam$, such that $1 \ll |\etam| \ll
1/\ve$, we have
\begin{align}
\label{bigg}
T^{(3)}_\ve & \simeq \frac{R^2}{D} \biggl(\ln (1/\ve) + \ln |\etam| + (\gamma - 3/2)\biggr) + T_{\pi}^{(3)}(\kappa=\infty)  \nonumber\\
&+ \frac{1}{|\etam|}
\left(\frac{4R}{\kappa} \ve^{-2} + \frac{32R^2}{3\pi D} \ve^{-1} + O(1) \right) \,,
\end{align}
where $\gamma \approx 0.577$ is the Euler-Mascheroni constant and the
MFPT $T_{\pi}^{(3)}(\kappa=\infty)$ from a random location to any
point on the boundary is given explicitly for the triangular-well
potential by
\begin{equation}
\label{eq:Tpi_3d}
\begin{split}
& T_{\pi}^{(3)}(\kappa = \infty) = \frac{r_0^5}{15 D R^3} - \frac{R^2}{12 D \etam} \biggl(x_0^4 +3 \\
& + \frac{8}{\etam} + \frac{12}{\etam^2} - \frac{24}{\etam^4}  \biggr) + \dfrac{R^2 \, e^{U_0} }{3 \, D  \, \etam} \bigg(x_0^3 + 
\frac{x_0^2(3 - x_0)}{\etam}  \\
& + \frac{3x_0(2 - x_0)}{\etam^2} +\frac{6 (1 - x_0)}{\etam^3} - \frac{6}{\etam^4} \bigg) \,.
\end{split}
\end{equation}
Note that in the limit $\etam \to - \infty$, the MFPT
$T_{\pi}^{(3)}(\kappa = \infty)$ becomes
\begin{equation}
T_{\pi}^{(3)}(\kappa = \infty) = \frac{r_0^5}{15 D R^3}  + O\left(\frac{1}{|\etam|}\right) \,,
\end{equation}
where the leading in this limit term can be interpreted as the product
of the probability $(r_0/R)^3$ that the particle's starting point is
within the spherical region of radius $r_0$ (in which diffusion is not
influenced by the potential), and the MFPT, equal to $r_0^2/ (15 D)$,
from a random location within this spherical region to its boundary.
Clearly, when a particle reaches the extent of the infinitely strong
attractive potential (or if it was started in that region), it is
drifted immediately to the boundary, and this step does not increase
the MFPT.

On the other hand, for small $\ve$ and negative $\etam$ which can be
arbitrarily (even infinitely) large by an absolute value, we find (see
\ref{SM2})
\begin{align}
\label{big}
T^{(3)}_\ve & \simeq \frac{R^2}{D} \biggl(2 \, \ln (1/\ve)+ \ln 2 - 1/4\biggr) + T_{\pi}^{(3)}(\kappa = \infty)  \nonumber\\
&+ \frac{1}{|\etam|} \left(\frac{4R}{\kappa} \ve^{-2} +  O(1) \right) \,,
\end{align}
which differs from the expression in (\ref{bigg}) in two aspects: the
amplitude of the dominant term, which diverges logarithmically as $\ve
\to 0$, is twice larger, and the term $\ln |\omega|$, which diverges
logarithmically as $\etam \to - \infty$, is absent.

Therefore, we observe that attractive interactions with either large
negative $U_0$, or fixed negative $U_0$ and $r_0 \sim R$ (both giving
a large negative $\etam$ according to (\ref{eta})), effectively
suppress the contribution due to a finite barrier at the entrance to
the EW, and also the contribution due to the diffusive paths which
find the EW via 3D diffusion.  On contrary, such interactions do not
affect the term which is logarithmically diverging as $\ve \to 0$ and
is associated with the paths localised near the confining boundary.
In other words, in this limit the plausible scenario to find the EW is
that of the dimensionality reduction, suggested by Adam and Delbr\"uck
\cite{adam}: a particle diffuses in the bulk until it hits the
confining boundary at some random position (which takes time of order
of $r_0^5/(15 D R^3)$, see the first term in (\ref{eq:Tpi_3d})), and
then diffuses along the boundary, not being able to surmount the
barrier against desorption and to escape back to the bulk, until it
ultimately finds the EW (the term $2 R^2 \ln(1/\ve)/D$).

It is worthwhile to mention that our SCA approach reproduces the
leading in the limit $\ve \to 0$ behaviour of the MFPT due to the
surface diffusion exactly, i.e., not only the $\ve$-dependence but
also the numerical factor in the amplitude.  In fact, the mean time
$T_s$ needed for a particle, diffusing on a surface of a 3D sphere and
starting at a random point\footnotemark\footnotetext{    
For a starting point fixed by an angular coordinate $\theta$, the MFPT
for surface diffusion is known to be
$$t_s(\theta) = \frac{R^2}{D} \ln\left(\frac{1-\cos\theta}{1-\cos\ve}\right)$$
for $\ve \leq \theta \leq \pi$, and $0$ otherwise.  Averaging the latter expression
 over
the uniformly distributed starting point, i.e., performing the integral 
$$T_s = \frac{1}{4\pi} \int\limits_0^{2\pi}d\varphi \int\limits_0^{\pi} d\theta \sin\theta ~ t_s(\theta) ,$$
one obtains the expression in (\ref{eq:Tsurf_3d}). },
to arrive for the first time to a disc of an angular size $\ve$
located on the surface of this sphere has been calculated
exactly\cite{Sano79}:
\begin{eqnarray}
\label{eq:Tsurf_3d}
T_s & =& \frac{R^2}{D} \left(\ln\left(\frac{2}{1-\cos\ve}\right) - \frac{1+\cos\ve}{2}\right) \\
\label{eq:Tsurf_3d_asympt}
&\simeq& \frac{R^2}{D} \biggl(2\ln(1/\ve) + 2\ln 2 - 1 + O(\ve^2)\biggr).  
\end{eqnarray}
Comparing (\ref{big}) and (\ref{eq:Tsurf_3d_asympt}), we notice that
the leading terms in both expressions are identical, and that both
expressions differ slightly only by numerical constants in the
$\ve$-independent terms.

In the 2D case with large negative $\etam$, the dominant contribution
to the MFET comes not from the logarithmically diverging term in the
limit $\ve \to 0$, as one could expect, but rather from the infinite
sum in the second line in (\ref{gen2as}), which is independent of
$\ve$.  We discuss this observation in \ref{SM2} and show that this
sum diverges in proportion to $|\etam|$ when $\etam \to - \infty$.
Using the asymptotic relation (\ref{eq:Rve_asympA}), which presents
this contribution to $\Rve^{(2)}$ in an explicit form, we find
\begin{equation}
\label{gen22}
\begin{split}
T^{(2)}_{\ve} & =  T_{\pi}^{(2)}(\kappa=\infty) + \frac{\pi^2 R^2}{3D} + O(1) \\
& + \frac{1}{|\etam|} \left(\dfrac{\pi R}{\kappa}  \, \ve^{-1} + \dfrac{2 R^2}{D} \, \ln(1/\ve) \right) \,,\\
\end{split}
\end{equation} 
where $T_{\pi}^{(2)}(\kappa = \infty)$ is the MFPT from a random
location within the disc to any point on its boundary.  For the
triangular-well potential the latter is given explicitly by
\begin{align}
\label{eq:Tpi_2d}
&T_{\pi}^{(2)}(\kappa = \infty) = \frac{r_0^4}{8 D R^2} - \frac{R^2}{6 D \etam} \biggl( x_0^3 + 2
+ \frac{3}{\etam} - \frac{6}{\etam^3}\biggr) \nonumber\\
&+ \frac{R^2 \, e^{U_0}}{2 \, D \, \etam} \biggl(x_0^2 + \frac{x_0(2 - x_0)}{\etam} + \frac{2(1 - x_0)}{\etam^2} - \frac{2}{\etam^3} \biggr) \,.
\end{align}
For large negative $\etam$, the leading term in (\ref{eq:Tpi_2d}) is
$r_0^4/(8 D R^2)$.  This term can be simply interpreted as the mean
time, equal to $r_0^2/(8D)$, needed to appear for the first time
within the reach of the triangular-well potential for a particle whose
starting point is uniformly distributed within the subregion $r < r_0$
(in which diffusion is free), multiplied by the probability
$r_0^2/R^2$ that this starting point is within this subregion.

Hence, in the limit $\etam \to -\infty$, we obtain for the 2D case
\begin{equation}
\label{gen22a}
T^{(2)}_{\ve} \approx \left(\frac{r_0}{R}\right)^2 \,  \frac{r_0^2}{8 D} + \frac{\pi^2 R^2}{3D}  \,. 
\end{equation} 
Again, we observe that the contribution due to a barrier at the EW,
and the leading, logarithmically divergent contribution to the MFPT to
the EW due to a two-dimensional diffusive search within the disc, are
both suppressed by attractive interactions.  Indeed, in this case a
plausible Adam-Delbr\"uck-type argument \cite{adam} states that a
diffusive particle first finds the boundary of the disc (the first
term in (\ref{gen22a})), and then diffuses along the boundary until it
ultimately finds the EW.  For arbitrary $\ve$ the latter MFPT is
well-known and is given by
\begin{equation}
\label{eq:Tsurf_2d}
T_s = \frac{R^2}{D} ~ \frac{(\pi - \ve)^3}{3\pi} = \frac{\pi^2 R^2}{3D} + O(\ve) \,,
\end{equation}
where the leading term on the right-hand-side is exactly the same as
the second term in (\ref{gen22a}).  This implies that our SCA approach
predicts correctly the values of both MFPTs.

To summarise the results of this subsection, we note that the
behaviour observed in the limit $\etam \to - \infty$ follows precisely
the Adam-Delbr\"uck dimensionality reduction scenario for both 3D and
2D cases.  Our expressions (\ref{big}) and (\ref{gen22}) provide the
corresponding MFETs, and also define the correction terms due to
finite values of $\etam$.  We emphasise, however, that the limit
$\etam \to - \infty$ is an extreme case.  For modest negative $\etam$
the representative trajectories for both three-dimensional and
two-dimensions systems will consist of paths of alternating phases of
bulk diffusion and diffusion along the boundary.  We analyse this
situation below.

\subsection{Beyond the Adam-Delbr\"uck scenario: Arbitrary $U_0$ and $r_0$.}

Consider the behaviour of the global MFET defined in (\ref{gen3}) or
(\ref{gen2}) for arbitrary $U_0$ and $r_0$.  Taking into account the
explicit expressions for $\l^{(3)}_U(R)$ in (\ref{3DtriangularL}) and
for $\l^{(2)}_U(R)$ in (\ref{2DtriangularL}), we notice that the
coefficients in front of each term in (\ref{gen3}) and (\ref{gen2})
are monotonic functions of $U_0$
\footnotemark\footnotetext{
In principle, the coefficient in front of the logarithm, which
contains the factor $(1 - R \, U_0/(R-r_0))$, can become negative in
case of repulsive interactions but it appears that it has a little effect
on the whole expression and does not cause a non-monotonic behaviour.}.
This implies that in the narrow escape limit $\ve \to 0$, we do not
expect any non-monotonic behaviour of the global MFET with respect to
$U_0$.  A thorough analytical and numerical analysis of
$T_{\ve}^{(d)}$ shows that it is indeed the case.

\begin{figure}
\begin{center}
\includegraphics[width=85mm]{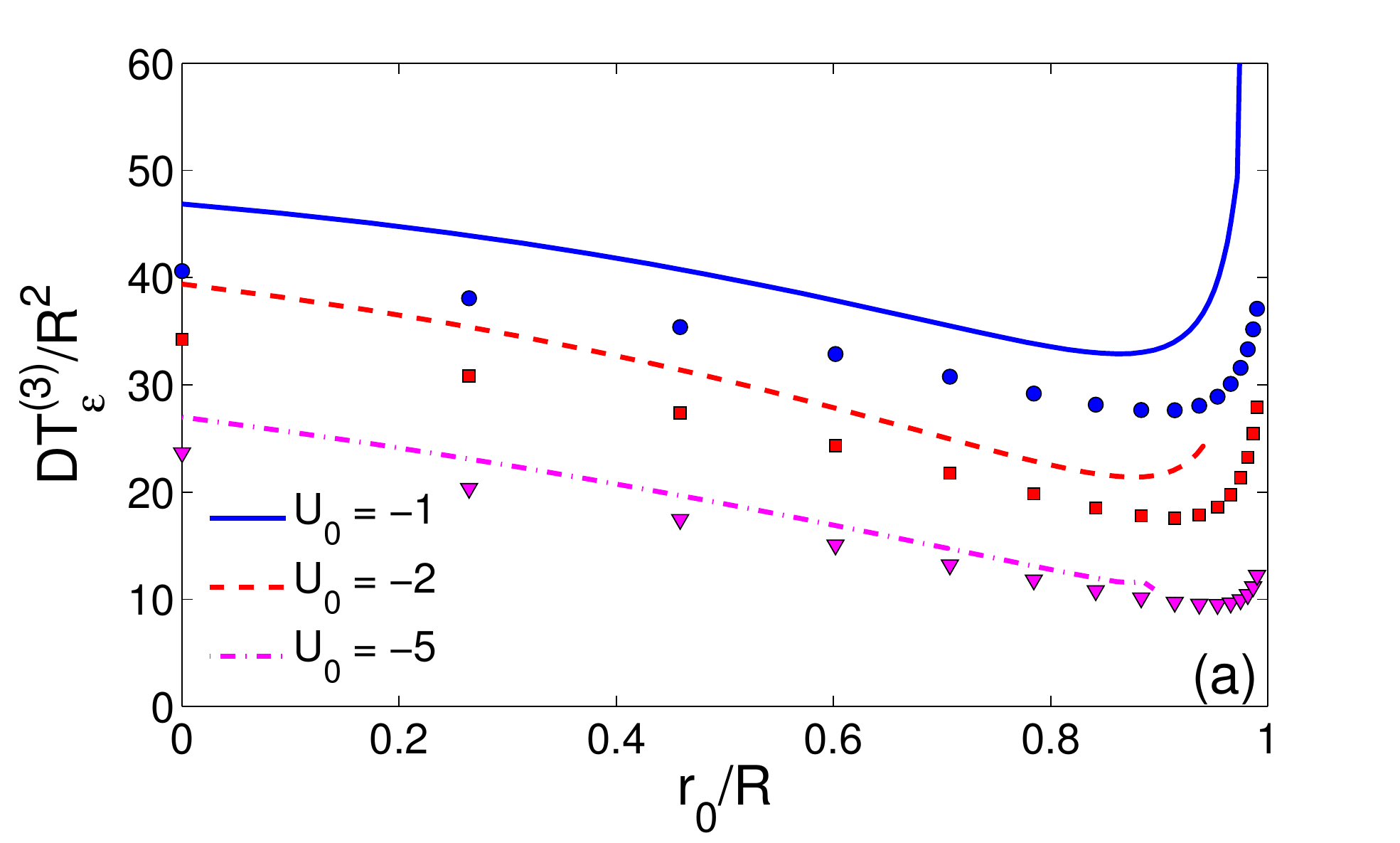} 
\includegraphics[width=85mm]{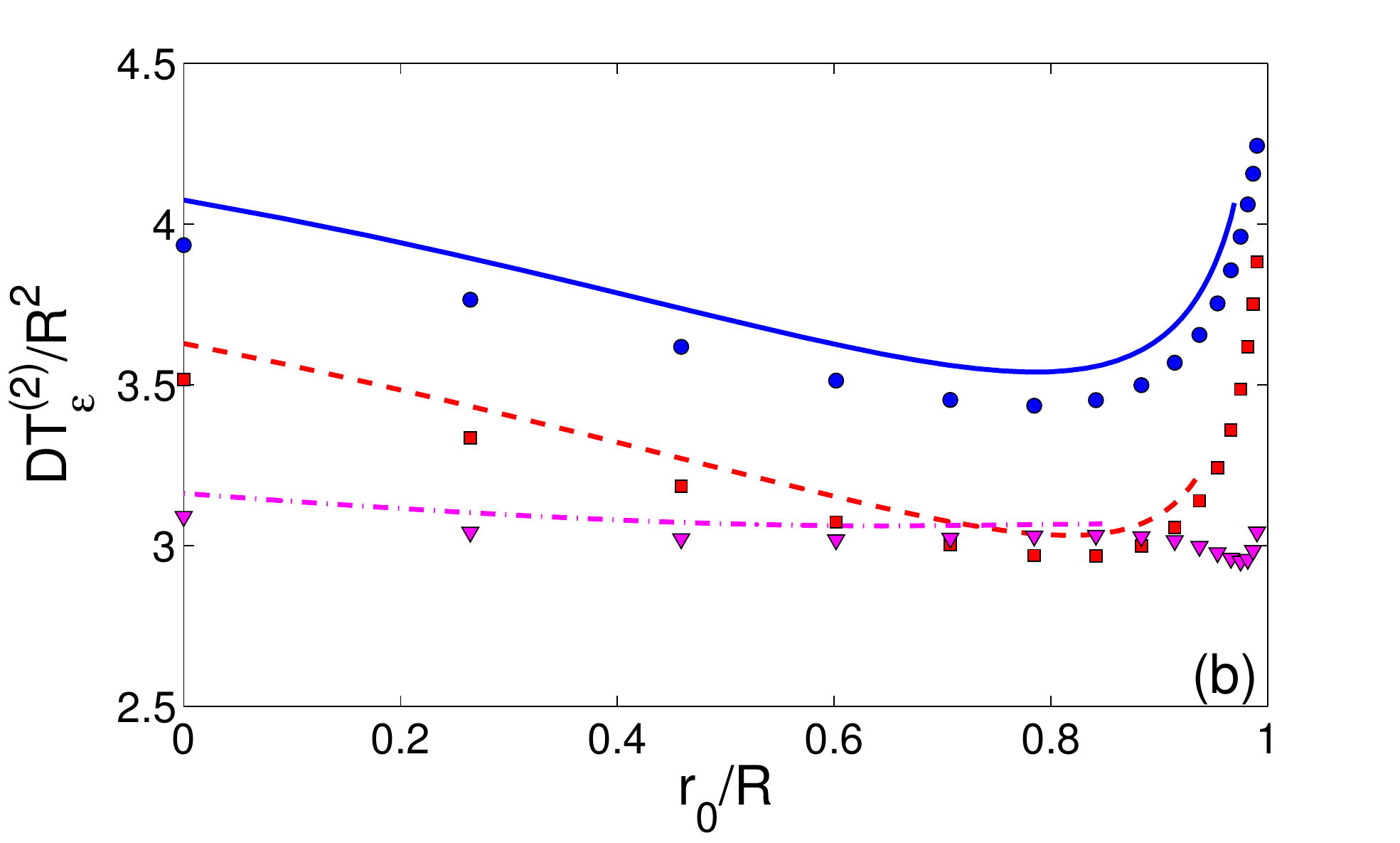} 
\end{center}
\caption{
The dimensionless global MFET, $D T^{(d)}_\ve/R^2$, as a function of
$x_0 = r_0/R$ for a triangular-well potential with no barrier at the
EW, $\kappa = \infty$.  Our analytical predictions are shown by curves
for three values of $U_0$ and $\ve = 0.02$.  Symbols present numerical
solutions obtained by a FEM (with the mesh size $0.005$).  {\bf (a)}
3D case, $T^{(3)}_\ve$ is given by (\ref{gen3}); {\bf (b)} 2D case,
$T^{(2)}_\ve$ is given by (\ref{gen2}).}
\label{fig:denis3}
\end{figure}

As we expected on intuitive grounds, for attractive particle-boundary
interactions the global MFET $T_{\ve}^{(d)}$ appears to be a
non-monotonic, and hence, an optimisable function of the extent of the
potential.  This is a new spectacular feature of the NEP unveiled by
our model involving spatially-extended interactions with the confining
boundary.  In Figs.~\ref{fig:denis3}a and \ref{fig:denis3}b we plot
$T_{\ve}^{(3)}$ and $T_{\ve}^{(2)}$, defined by (\ref{gen3}) and
(\ref{gen2}) with $\kappa = \infty$, as a function of $r_0$ for
several negative values of $U_0$ ($U_0 = -1$, $U_0 = - 2$ and $U_0 = -
5$) and $\ve = 0.02$.  We remark that the first terms in (\ref{gen3})
and (\ref{gen2}), associated with a finite barrier and dropped by
setting $\kappa = \infty$, are monotonic functions of $r_0$.


We observe that in the 3D case, the global MFET $T_{\ve}^{(3)}$, as a
function of $r_0$, exhibits a rather pronounced minimum for $r_0$ away
from $r_0 = 0$ and from $r_0 = R$ (contact interactions).  This means
that there exists an optimal extent of the potential: in order to
minimise the global MFET, the potential should neither extend too deep
into the bulk nor should be too localised on the boundary.  Further
on, we realise that the precise location of the minimum of
$T_{\ve}^{(3)}$ depends, in general, on $\ve$ and $U_0$.  Analysing
our result in (\ref{gen3}), we observe that when we gradually decrease
$\ve$, the minimum moves closer to the boundary (but never reaches it
and $T_{\ve}^{(3)}$ still exhibits a very abrupt growth when $r_0 $
becomes too close to $R$) and becomes more pronounced.  Conversely, if
we progressively increase $\ve$, the minimum moves away from the
boundary and becomes more shallow.  Increasing the strength of
attractive interactions (while keeping $\ve$ fixed) also pushes the
minimum closer to the boundary.

For the 2D case the global MFET $T_{\ve}^{(2)}$ shows a qualitatively
similar behaviour but the effects are much less pronounced.  In
principle, the optimum also exists in this case, but the minimum
appears to be much more shallow.  We also observe that for strong
attraction ($U_0 = - 5$), apart of some deep in the vicinity of $r_0 =
R$, $T_{\ve}^{(2)}$ appears to be almost independent of $r_0$.

As in the analysis of the NEP without the particle-boundary interactions
in Sec. \ref{sec:nopotential}, one can observe some small
discrepancies between the asymptotic relations (\ref{gen3},
\ref{gen2}) and numerical solutions of the original problem by a FEM.
As discussed earlier, they are related to the SCA approximation, in
which the mixed boundary condition is replaced by an effective
inhomogeneous Neumann condition.  On the other hand, the proposed
approximation provides explicit asymptotic formulae for arbitrary
potentials that captures qualitatively well all the studied features
of the NEP.

Two remarks are now in order:

a) Emergence of an optimum at an intermediate extent of the potential
implies that the Adam-Delbr\"uck dimensionality reduction scenario
(which corresponds to $\etam \to - \infty$ and hence, to the part of
the curves in Figs.~\ref{fig:denis3}a and \ref{fig:denis3}b close to
$r_0 = R$ where $T_{\ve}^{(3)}$ exhibits quite a steep growth
attaining large values) is not at all the optimal (i.e., less time
consuming) way of finding the EW.  In reality, the optimum corresponds
to situations when the barrier against desorption is not very large so
that the particle does not remain localised near the boundary upon
approaching it for the first time, but rather has a possibility to
overpass the barrier against desorption and to perform alternating
phases of bulk and surface diffusion.  The optimum then corresponds to
some fine-tuning of the relative weights of bulk and surface diffusion
by changing the extent of the potential and its value on the boundary.

b) These findings are compatible, in principle, with the prediction of
the non-monotonic behaviour of $T_{\ve}^{(d)}$ as a function of the
desorption rate $\lambda$ made earlier in \cite{greb2}, in which the
NEP with $\kappa = \infty$ has been analysed for an intermittent
diffusion model.  In this model a particle diffuses in the bulk with a
diffusion coefficient $D$ until it hits the boundary of the
micro-domain and switches to surface diffusion of a random duration
(controlled by the desorption rate $\lambda$) with a diffusion
coefficient $D_{\rm surf}$.  This model tacitly presumes that there
are some attractive interactions with the surface, in addition to the
hard-core repulsion, which are taken into account in some effective
way (interactions are replaced by effective contact ones).  In our
settings, the desorption rate $\lambda$ in this intermittent model
should depend on both the strength of the interaction potential $U_0$
at the boundary and also on the gradient of the potential in the
vicinity of the surface, which define the barrier against desorption.
There is, however, some quantitative discrepancy between our
predictions and the predictions made in
\cite{greb2}: In \cite{greb2} (see Fig.~10, right panel), it was
argued that for such an intermittent model with $D=D_{\rm surf}$ (as
in our case) surface diffusion is a preferable search mechanism so
that the global MFET is a monotonic function of $\lambda$.  On
contrary, our analysis demonstrates that $T_{\ve}^{(3)}$ is an
optimisable function even in case of equal bulk and surface diffusion
coefficients, which means that neither the bulk diffusion nor 2D
surface diffusion alone provide an optimal search mechanism, but
rather their combination.  This discrepancy is related to subtle
differences between two models.  For the two-dimensional case,
illustrated in Fig. \ref{fig:denis3}b, the analysis in \cite{greb2}
(see Fig.~10, left panel) suggests that there is an optimum even for
$D=D_{\rm surf}$.  Our analysis agrees with this conclusion.

\begin{figure}
\begin{center}
\includegraphics[width=85mm]{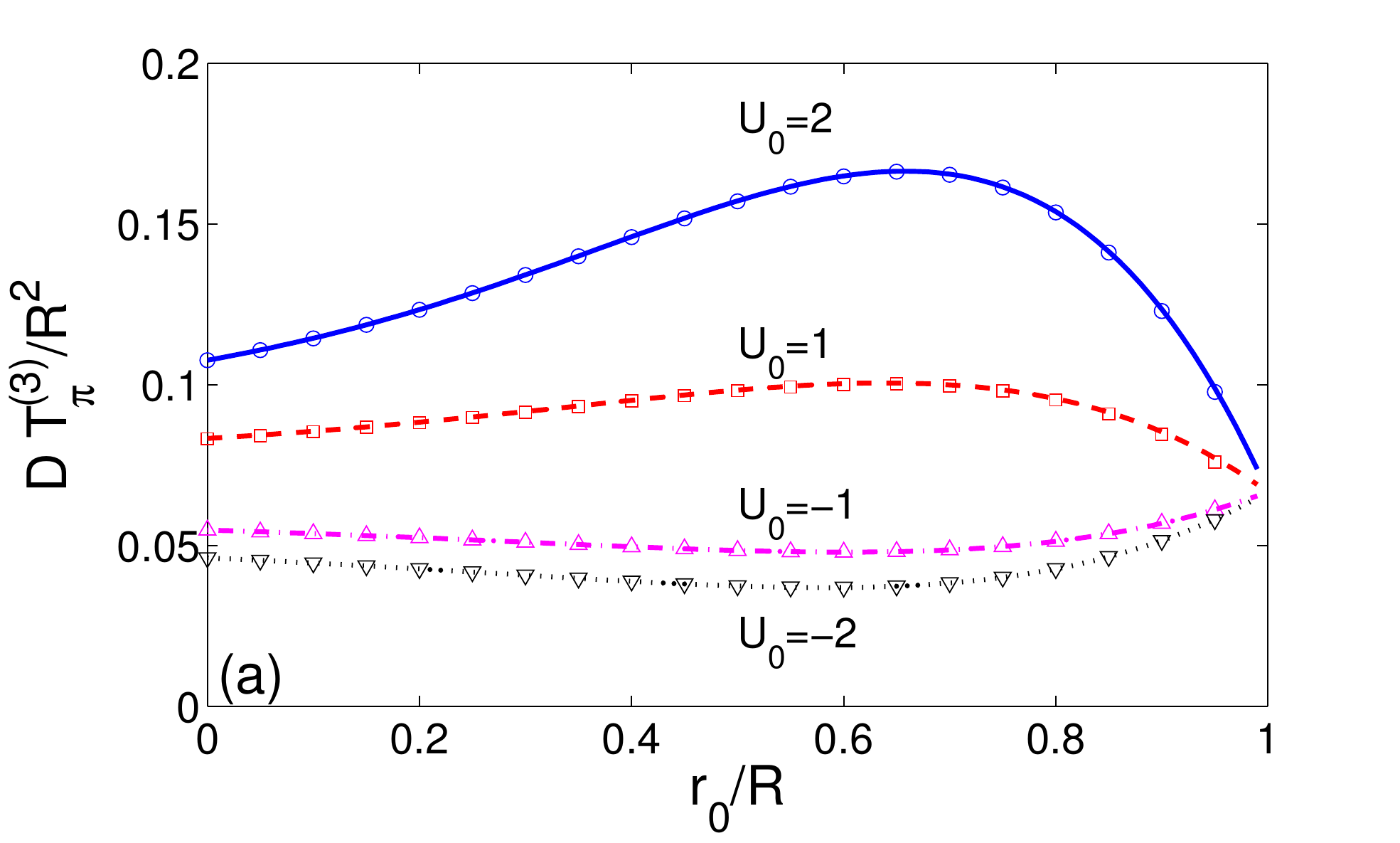} 
\includegraphics[width=85mm]{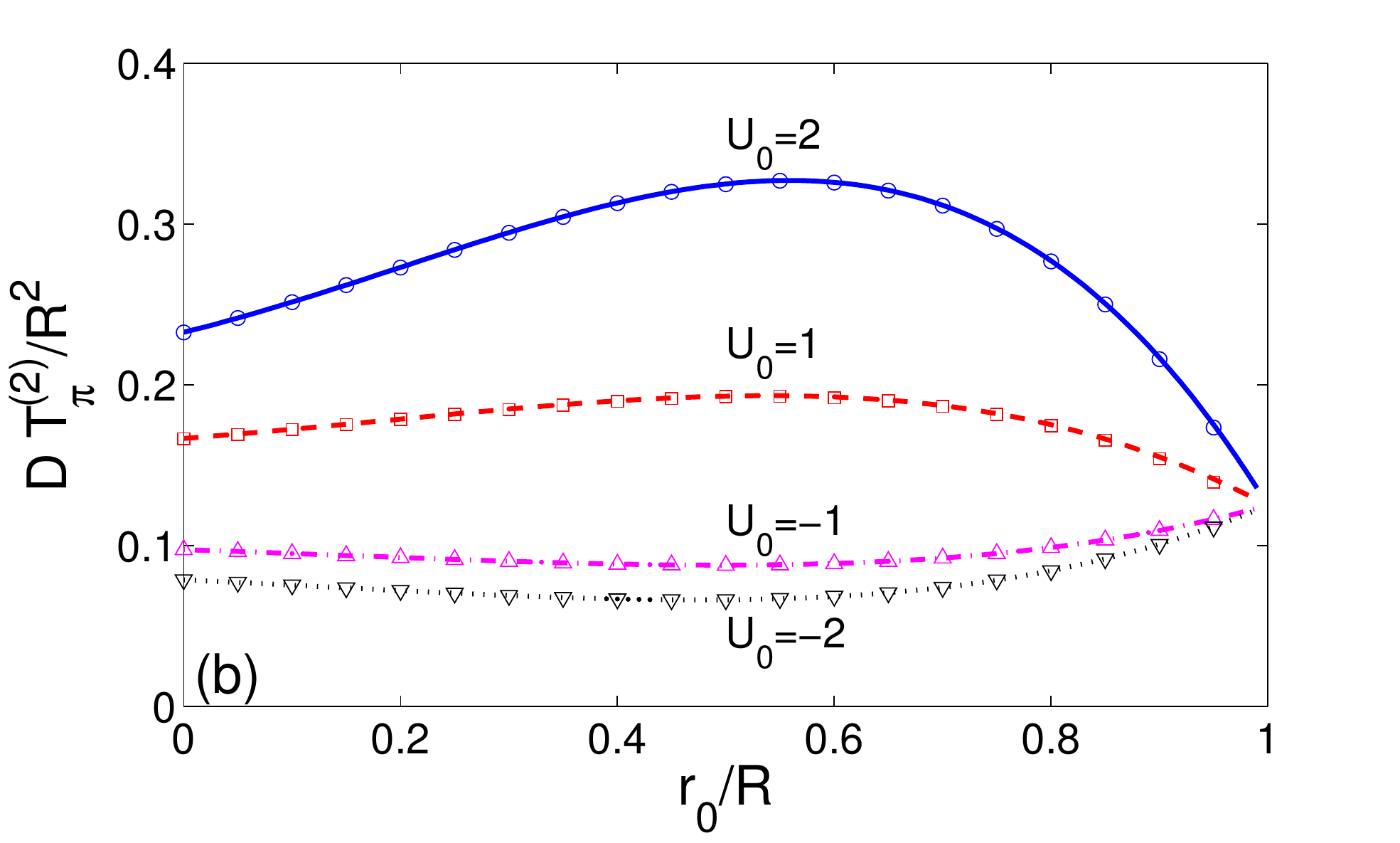} 
\end{center}
\caption{
The dimensionless MFPT, $D T^{(d)}_\pi(\kappa=\infty)/R^2$, for a
triangular-well potential in (\ref{triangular}) versus $r_0/R$ for
several values of $U_0$ and $\kappa = \infty$.  {\bf (a)} 3D case, {\bf (b)} 2D case. }
\label{fig:Tm_pi}
\end{figure}

We close with a brief analysis of the behaviour of
$T_{\pi}^{(d)}(\kappa = \infty)$ -- the MFPT for a diffusive particle,
starting at a random location, to arrive at any point on the boundary
of the micro-domain, in presence of long-range interactions with the
boundary.  This MFPT is included into $T_{\ve}^{(d)}$ but has a little
effect on it since it enters only the constant, $\ve$-independent
terms.  At the same time, it is an interesting quantity in its own
right.  For the triangular-well potential,
$T_{\pi}^{(d)}(\kappa=\infty)$ is defined explicitly by
(\ref{eq:Tpi_3d}) and (\ref{eq:Tpi_2d}) for $d=3$ and $d=2$,
respectively.

A first intuitive guess is that increasing either $U_0$ or the extent
of the potential would make the particle feel the surface stronger, so
that for $U_0 > 0$, the MFPT $T_{\pi}^{(d)}(\kappa = \infty)$ would be
an increasing function of both parameters, while for $U_0 < 0$,
$T_{\pi}^{(d)}(\kappa = \infty)$ would decrease with an increase of
both $U_0$ and $r_0$.  As far as the dependence on $U_0$ is concerned,
this guess appears to be correct.  Indeed, we observe that for a fixed
$r_0$, the MFPTs for both $d=2$ and $d=3$ are monotonic increasing
functions of $U_0$.
Surprisingly enough, this is not the case for the dependence of the
MFPTs $T_{\pi}^{(d)}(\kappa = \infty)$ on the extent $l_{ext}$ of the
potential.  We find that for a fixed $U_0$, $T_{\pi}^{(d)}(\kappa =
\infty)$ exhibits a peculiar non-monotonic behaviour as a function of
$r_0$, with a minimum for $U_0 < 0$ and a maximum for $U_0 > 0$, as
illustrated in Fig. \ref{fig:Tm_pi}.  To the best of our knowledge,
this interesting effect has not been reported earlier.

\section{Conclusion}
\label{conclusion}

To recapitulate, we have presented here some new insights into the
narrow escape problem, which concerns various situations when a
particle, diffusing within a bounded micro-domain, has to escape from
it through a small window (or to bind to some target site) of an
angular size $\ve$ located on the impenetrable boundary.  We have
focused on two aspects of this important problem which had not
received much attention in the past: the effects of an energy or an
entropy barrier at the escape window, always present in realistic
systems, and the effects of long-range potential interactions between
a diffusing particle and the boundary.  Inspired by the
self-consistent approximation developed previously in
\cite{shoup} for calculation of the reaction rates between molecules
with inhomogeneous chemical reactivity, we generalised this approach
to the NEP with long-range potential interactions with the boundary.
In this self-consistent approach, the original mixed boundary
condition is replaced by an effective inhomogeneous Neumann condition,
in which the unknown flux is determined from an appropriate closure
relation.  This modified problem was solved exactly for an arbitrary
radial interaction potential.

We have concentrated on the functional form of the global (or
volume-averaged over the starting point) mean first exit time,
$T_\ve$, for which we derived a general expression analogous to the
celebrated Collins-Kimball relation in chemical kinetics,
incorporating both the contribution due to a finite barrier at the
escape window (or a binding site) and the contribution due to a
diffusive search for its location, for an arbitrary radially-symmetric
potential, any size of the escape window, and a barrier of an
arbitrary height.  We have realised that these two contributions
naturally decouple from each other, which permitted us to study
separately their impact on the MFET.

The accuracy of our analytical results based on the self-consistent
approximation has been confirmed by two independent numerical schemes:
a numerical solution of the backward Fokker-Planck equation by a
finite element method, and Monte Carlo simulations of the diffusive
search for the escape window in presence of particle-boundary
potential interactions.  We have shown that the self-consistent
approximation is very accurate for small escape windows (i.e., in the
true narrow escape limit) but also captures quite well the behaviour
of the global MFET even for rather large escape windows.  In the
latter case, small deviations were observed, related to the fact that
the solution of the modified problem is defined up to a constant.

Turning to the narrow escape limit $\ve \to 0$, we have analysed the
relative weights of each contribution to $T_\ve$.  We have shown that
the contribution due to the passage through the escape window (which
had been ignored in the majority of earlier works) dominates the
global MFET in the narrow escape limit, since it exhibits a stronger
singularity as $\ve \to 0$ than the contribution due to the diffusive
search.  This implies that the kinetics of the narrow escape process
is rather barrier-controlled than diffusion-controlled.  In
consequence, discarding an entropy or an energy barrier at the exit
from the micro-domain can result in strongly misleading estimates in
chemical and biological applications.  Remarkably, the associated
reactivity (or permeability) enters into the global MFET in a very
simple way.

Further on, for the case of radially-symmetric interaction potentials
which possess a bounded first derivative, we have presented an
explicit expression for the contribution to $T_\ve$ due to the
diffusive search for the location of the escape window, in which the
coefficients in front of the terms diverging in the limit $\ve \to 0$
were defined via some integrals and derivatives of the interaction
potential.  The structure of the obtained result suggests that most
likely the general problem considered here can be solved
\textit{exactly} in the narrow escape limit without resorting to any
approximation.

On example of a triangular-well interaction potential, we have
discussed the dependence of the contribution to the MFET due to a
diffusive search for the escape window on the parameters of the
potential. We have shown that this contribution is always a monotonic
function of the value of the potential at the boundary: as expected,
repulsive (resp., attractive) interactions increase (resp., decrease)
the MFET.  Curiously enough, it appeared that for attractive
interactions $T_\ve$ is a non-monotonic function of the extent of the
potential: there exists some optimal extent for which $T_\ve$ has a
minimum. This optimal value corresponds to interactions which are
neither localised near the confining boundary nor extend to deeply
into the bulk. In case of a very small extent (i.e., in the limit of
short-range interactions), with a fixed value of the interactions
potential on the boundary, the force acting on the particle in the
immediate vicinity of the boundary becomes very large and the narrow
escape process proceeds precisely via the Adam-Delbr\"uck
dimensionality reduction scenario: a particle first reaches the
boundary at any point and than, not being able to surmount the barrier
against desorption, continues a diffusive search for the escape window
along the boundary until it finds it.  For more realistic moderate
values of the extent, typical paths consist of alternating,
intermittent bulk diffusion tours followed by diffusion along the
boundary.\\

\section*{Acknowledgments}
DG acknowledges the support from the French National Research Agency
(ANR) under Grant No. ANR-13-JSV5-0006-01.







\pagebreak 
\begin{widetext}

\textbf{\LARGE Supplemental Materials}
\label{SM}
\setcounter{equation}{0}
\setcounter{section}{0}
\setcounter{figure}{0}
\setcounter{table}{0}
\setcounter{page}{1}
\makeatletter
\renewcommand{\theequation}{S\arabic{equation}}
\renewcommand{\thefigure}{S\arabic{figure}}
\renewcommand{\thesection}{SM\arabic{section}}



\vspace{0.2in}

\section{Numerical simulations}
\label{SM1}

In this part of the Supplemental Materials we briefly discuss two
numerical procedures used to check our theoretical predictions.

\subsection{Numerical computation by a finite elements method}
\label{sec:FEM}

To verify the accuracy of the SCA, we solve
the Poisson equation (\ref{eq:Poisson}, \ref{eq:Poisson_2d}) by using
a finite elements method (FEM) implemented in Matlab PDEtool.  This
tool solves the following equation:
\begin{equation}
\label{eq:PDE_Matlab}
-\nabla (c \nabla u) + a u = f ,
\end{equation}
where $c$ is a 2x2 matrix, and $a$ and $f$ are given functions.

In our case, we need to deal with the Laplace operator in radial or
spherical coordinates.  In two dimensions, the original equation
(\ref{eq:Poisson_2d}) can be written in radial coordinates as
\begin{equation}
\frac{1}{r} \partial_r (r e^{-U(r)} \partial_r) u + \frac{e^{-U(r)}}{r^2} \partial^2_\theta u  = - \frac{e^{-U(r)}}{D},
\end{equation}
from which
\begin{equation}
- \left(\begin{array}{c} \partial_r \\ \partial_\theta \\ \end{array} \right)^\dagger
c \left(\begin{array}{c} \partial_r \\ \partial_\theta \\ \end{array} \right) u  = f , 
\end{equation}
with $a = 0$, $f = r e^{-U(r)}/D$, and
\begin{equation}
c = \left(\begin{array}{c c} re^{-U(r)} & 0 \\ 0 & e^{-U(r)}/r \\ \end{array} \right) .
\end{equation}

In three dimensions, the Poisson equation (\ref{eq:Poisson}) reads in
spherical coordinates as
\begin{align}
& \frac{1}{r^2} \partial_r (r^2 e^{-U(r)} \partial_r) u + \frac{e^{-U(r)}}{r^2} \frac{1}{\sin\theta} \partial_\theta (\sin \theta \partial_\theta) u   \nonumber\\
& + \frac{e^{-U(r)}}{r^2 \sin^2\theta} \partial^2_\varphi u  = - \frac{e^{-U(r)}}{D}.  
\end{align}
Since our solution does not depend on $\varphi$, the last term on the
left hand side can be omitted so that
\begin{equation}
- \left(\begin{array}{c} \partial_r \\ \partial_\theta \\ \end{array} \right)^\dagger c
\left(\begin{array}{c} \partial_r \\ \partial_\theta \\ \end{array} \right) u  = f,
\end{equation}
with $a = 0$, $f = r^2 e^{-U(r)}\sin\theta /D$, and
\begin{equation}
c = \left(\begin{array}{c c} r^2 e^{-U(r)}\sin\theta & 0 \\ 0 & e^{-U(r)} \sin\theta \\ \end{array} \right).
\end{equation}

We set the rectangular domain $V = [0,1]\times [0,\pi]$ with
mixed boundary conditions (\ref{eq:BC_flux}), i.e., a zero flux
condition for $\partial V \backslash \Gamma_0$, except for the
segment $\Gamma_0 = \{1\}\times [0,\ve]$ representing the EW (see
Fig. \ref{fig:domain}):
\begin{equation}
\begin{split} 
\biggl(\partial_r u + \frac{k}{D} u \biggr)_{r=1} & = 0  \qquad (\textrm{on}~\Gamma_0), \\
(\partial_r u)_{r=1} & = 0  \qquad  (\textrm{on}~\Gamma_1), \\
(\partial_r u)_{r=0} & = 0  \qquad  (\textrm{on}~\Gamma_3), \\
(\partial_\theta u)_{\theta=0} & = 0  \qquad  (\textrm{on}~\Gamma_4), \\
(\partial_\theta u)_{\theta=\pi} & = 0  \qquad  (\textrm{on}~\Gamma_2). \\
\end{split}
\end{equation} 
In Matlab, the generalised Neumann boundary condition has the form
\begin{equation}
\vec n \cdot (c \nabla u) + q u = g,
\end{equation}
where the matrix $c$ is the same as in the PDE (\ref{eq:PDE_Matlab}).
We set $g = 0$ and
\begin{equation}
\begin{split}
q & = R e^{-U(R)} \kappa/D  \qquad (\textrm{on}~\Gamma_0), \\
q & = 0  \qquad (\textrm{on}~\Gamma_1\cup \Gamma_2\cup \Gamma_3 \cup \Gamma_4) \\
\end{split}
\end{equation}
in two dimensions, and 
\begin{equation}
\begin{split}
q & = R^2 \sin\theta ~e^{-U(R)} \kappa/D  \qquad (\textrm{on}~\Gamma_0), \\
q & = 0  \qquad (\textrm{on}~\Gamma_1\cup \Gamma_2\cup \Gamma_3\cup \Gamma_4) \\
\end{split}
\end{equation}
in three dimensions.  For fully reactive EW ($\kappa = \infty$), the
Dirichlet boundary condition is imposed on $\Gamma_0$.

\begin{figure}
\begin{center}
\includegraphics[width=80mm]{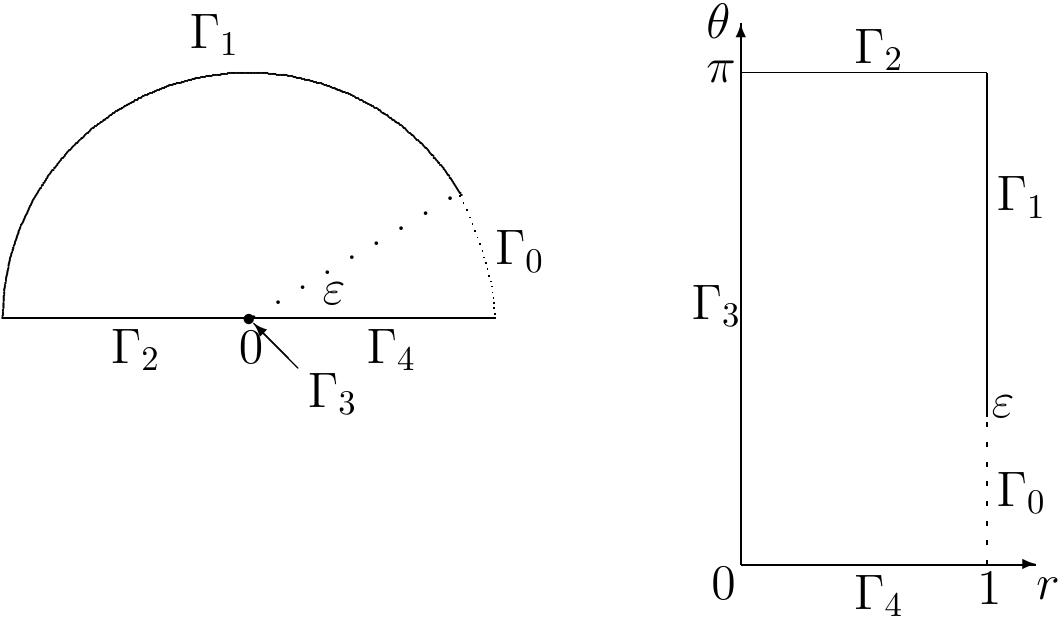} 
\end{center}
\caption{
The original domain (half-disc) and the associated computational
domain (rectangle).  Similar in 3D case.}
\label{fig:domain}
\end{figure}

\subsection{Monte Carlo simulations}

We also compute the distribution of first passage times to the EW by
simulating diffusion trajectories which end up at the EW.  In
practice, we solve a Langevin equation by the following iterative
procedure: after generating a uniformly distributed starting point
$\r_0$, one re-iterates
\begin{equation}
\r_{n+1} = \r_n + D\delta~ \f(\r_n) + \sqrt{2D\delta}~ \xi_n \,,
\end{equation}
where $\delta$ is a one-step duration, $\r_n$ is the position after
$n$ steps, $\f = - \partial_r U(r) \e_r$ is the normalised applied
force in the radial direction $\e_r$, and $\xi$ is the normalised
random thermal force.  For instance, we have in two dimensions:
\vspace{0.2in}
\begin{equation}
\begin{split}
x_{n+1} & = x_n + D\delta~ f(|\r_n|)~ x_n/|\r_n| + \sqrt{2D\delta}~ \xi_{x,n} \,,  \\
y_{n+1} & = y_n + D\delta~ f(|\r_n|)~ y_n/|\r_n| + \sqrt{2D\delta}~ \xi_{y,n} \,, \\
\end{split}
\end{equation}
where $x_n/|\r_n|$ and $y_n/|\r_n|$ represent $\cos(\theta)$ and
$\sin(\theta)$ in the projection of the radial force, and $\xi_{x,n}$,
$\xi_{y,n}$ are independent normal variables with zero mean and unit
variance.

At each step, one checks whether the new position $(x_{n+1},y_{n+1})$
remains inside the disk: $x_{n+1}^2 + y_{n+1}^2 < R^2$.  If this
condition is not satisfied, the particle is considered as being on the
boundary.  If the particle hits the EW, the trajectory simulation is
stopped and $n\delta$ is recorded as the generated exit time.
Otherwise, the particle is reflected back and continues to diffuse.
The Monte Carlo simulations in three dimensions are similar.  Finally,
the partial reactivity of the EW can be introduced by partial
reflections \cite{Grebenkov06,Grebenkov14}.

\section{Asymptotic behaviour of the series in
(\ref{eq:Rd_3d}) and (\ref{rv2d})}
\label{SM2}

We focus on the asymptotic behaviour of the infinite series
$\Rve^{(3)}$ and $\Rve^{(2)}$, defined in (\ref{eq:Rd_3d}) and
(\ref{rv2d}), for potentials $U(r)$ which have a bounded first
derivative for any $r \in (0,R)$.  Our aims here are two-fold: first
we establish the exact asymptotic expansions for these infinite series
in the narrow escape limit $\ve \to 0$, and second, we derive
approximate explicit expressions for $\Rve^{(3)}$ and $\Rve^{(2)}$
which permit us to investigate their asymptotic behaviour in the limit
$R |U'(R)| \to \infty$.

Our analysis is based on two complementary approaches.  In the first
approach we take advantage of the following observation.  When $\ve =
0$, there is no EW, and the MFET is infinite, whatever the potential
is.  Since $\l_U^{(d)}$ does not depend on $\ve$, the divergence of
the MFET as $\ve\to 0$ should be ensured by the divergence of
$\Rve^{(d)}$.  Suppose that we truncate the infinite series in
(\ref{eq:Rd_3d}) and (\ref{rv2d}) at some arbitrary $n= N^*$.  Then,
turning to the limit $\ve \to 0$, we find that both $\Rve^{(3)}$ and
$\Rve^{(2)}$ attain some constant values, which depend on the upper
limit $N^*$ of summation.  As a consequence, these truncated sums
should diverge as $N^*\to \infty$ while their small-$\ve$ behaviour is
dominated by the terms with $n \to \infty$.  One needs therefore to
determine the asymptotic behaviour of $g_n(R)/g'_n(R)$ in this limit
and to evaluate the corresponding small-$\ve$ asymptotics for
$\Rve^{(3)}$ and $\Rve^{(2)}$. This will be done in the subsection
\ref{small}.

Next, in subsection \ref{large} we will pursue a different approach
based on the assumption that, once we are interested in the behaviour
of the ratio $g_n/g'_n$ at $r = R$ only, we may approximate the
coefficients in the differential equations (\ref{eq:gn}) and
(\ref{eq:gn2d}), which are functions of $r$, by taking their values at
the confining boundary.  This will permit us to derive an explicit
expression for $g_n(R)/g'_n(R)$ valid for arbitrary $n$, not
necessarily large, and arbitrary $|U'(R)| < \infty$.  This expression
will be checked subsequently against an exact solution obtained for a
triangular-well potential (see \ref{SM4} and \ref{SM5}).  We set out
to show that an approximate expression for $g_n(R)/g'_n(R)$ and an
exact result for such a choice of the potential agree very well
already for quite modest values of $n$ and the agreement becomes
progressively better with an increase of $|U'(R)|$.  On this basis, we
also determine the small-$\ve$, as well large-$R U'(R)$ asymptotic
behaviour of $\Rve^{(3)}$ and $\Rve^{(2)}$, which agrees remarkably
well with the expressions obtained within the first approach, and the
exact solution derived for the particular case of a triangular-well
potential.

\subsection{Large-$n$ asymptotics  of $g_n(R)/g'_n(R)$  and the corresponding small-$\ve$ behaviour of the infinite series $\Rve^{(3)}$ and $\Rve^{(2)}$.}
\label{small}

We introduce an auxiliary function $\psi=\psi_n(r) = g_n(r)/g'_n(r)$,
which is the inverse of the logarithmic derivative of $g_n(r)$ and
obeys, in virtue of (\ref{eq:gn}) and (\ref{eq:gn2d}), the following
equations:
\begin{equation}
\label{1}
r^2 \left(1- \psi'\right) + r \left(2 - r U'(r)\right) \psi - n(n+1) \psi^2 =0 
\end{equation} 
for the 3D case, and 
\begin{equation}
\label{2}
r^2 \left(1- \psi'\right) + r\left(1 - r U'(r)\right) \psi - n^2 \psi^2 =0 
\end{equation} 
for the 2D case, respectively.  We will seek the solutions of these
non-linear Riccati-type differential equations in form of the
asymptotic expansion in the inverse powers of $n$ in the limit $n \to
\infty$.

Supposing that $U'(r)$ does not diverge at any point within the
domain, we find that the leading term of $\psi$ in the limit $n \to
\infty$ is given by $\psi \sim r/n$ for both 2D and 3D cases, which is
completely \textit{independent} of the potential $U(r)$.  Pursuing
this approach further, we make no other assumption to get the second
term in this large-$n$ expansion, while for the evaluation of the
third term we stipulate that $|U''(r)| <
\infty$.  We have then for the 3D case 
\begin{align}
\label{c}
&\psi = \frac{r}{n} - \frac{r^2 U'(r)}{2 n^2} + \frac{r^2 \left(U'(r) \left(4 + r \, U'(r)\right) + 2 \, r \, U''(r)\right)}{8 n^3} + O\left(\frac{1}{n^4}\right) \,,
\end{align} 
and hence, 
\begin{align}
\label{cc}
&\frac{g_n(R)}{R g'_n(R)} = \frac{1}{n} - \frac{R U'(R)}{2 n^2} + \frac{R U'(R) \left(4 + R \, U'(R)\right) + 2 \, R^2 \, U''(R)}{8 n^3} + O\left(\frac{1}{n^4}\right) \,,
\end{align} 
where the omitted terms decay in the leading order as $1/n^4$.
Similarly, for the 2D case we find
\begin{align}
\label{c2D}
&\psi = \frac{r}{n} - \frac{r^2 U'(r)}{2 n^2} + \frac{r^2 \left(U'(r) \left(2 + r \, U'(r)\right) + 2 \, r \, U''(r)\right)}{8 n^3} + O\left(\frac{1}{n^4}\right) \,,
\end{align} 
and hence, 
\begin{align}
\label{cc2D}
&\frac{g_n(R)}{R g'_n(R)} = \frac{1}{n} - \frac{R U'(R)}{2 n^2} + \frac{R U'(R) \left(2 + R \, U'(R)\right) + 2 \, R^2 \, U''(R)}{8 n^3} + O\left(\frac{1}{n^4}\right) \,.
\end{align} 
We observe that the asymptotic expansions (\ref{cc}) and (\ref{cc2D})
for the 3D and the 2D cases become different from each other only
starting from the third term; first two terms are exactly the same.
As it will be made clear below, we do not have to proceed further with
this expansion and, as an actual fact, just two first terms will
suffice us to determine the leading asymptotic behaviour of
$\Rve^{(3)}$ and just the first one will be enough to determine the
analogous behaviour of $\Rve^{(2)}$.  In what follows, in \ref{SM4}
and \ref{SM5} we will also check these expansions against the exact
results obtained for the triangular-well potential, in which case the
radial functions $g_n$ and hence, the ratio $g_n(R)/g'_n(R)$ can be
calculated exactly. We proceed to show that the asymptotic forms in
(\ref{cc}) and (\ref{cc2D}) coincide with the exact asymptotic
expansions at least for the first three terms.

Focusing first on the 3D case, we formally write
\begin{equation}
\label{d}
\frac{g_n(R)}{R g'_n(R)} \equiv \frac{1}{n} - \frac{R \, U'(R)}{2 n^2} + \left(\frac{g_n(R)}{R g'_n(R)} - \frac{1}{n} + \frac{R \, U'(R)}{2 n^2}\right) \,,
\end{equation}
where the terms in brackets, in virtue of (\ref{c}) and (\ref{cc}),
decay as $1/n^3$ when $n \to \infty$.  Inserting (\ref{d}) into
(\ref{eq:Rd_3d}), we have
\begin{align}
\label{eq:Rd_3d_SM}
\Rve^{(3)} & = \Sigma_1 - R \, U'(R) \, \Sigma_2 
+ \sum\limits_{n=1}^\infty \left(\frac{g_n(R)}{R g'_n(R)} - \frac{1}{n} + \frac{R U'(R)}{2 n^2}\right) \frac{\phi_n^2(\ve)}{(2n+1)} \,,
\end{align}
where 
\begin{align}
\label{series1}
\Sigma_1 &= \sum\limits_{n=1}^\infty \frac{\phi_n^2(\ve)}{n (2n+1)}   , \\
\label{series}
\Sigma_2 &=  \frac{1}{2} \sum_{n=1}^{\infty} \frac{\phi_n^2(\ve)}{n^2 (2n+1)} \,,
\end{align}
and $\phi_n(\ve)$ is defined by (\ref{eq:phin3d}).

Next, we find that in the limit $\ve \to 0$,
\begin{eqnarray}
\label{sigma1}
&& \Sigma_1 = \frac{32}{3 \pi} \ve^{-1} + \ln\left(1/\ve\right) - \frac{7}{4} + \ln 2 + O(\ve) \,, \\
\label{sigma2}
&&\Sigma_2 = \ln\left(1/\ve\right) + \frac{1}{4} + \ln 2 + \frac{\pi^2}{12} + O(\ve) \,, \\
\label{sigma3}
&& \sum\limits_{n=1}^\infty \left(\frac{g_n(R)}{R g'_n(R)} - \frac{1}{n} + \frac{R U'(R)}{2 n^2}\right) \frac{\phi_n^2(\ve)}{2n+1}  = O(1) \,. 
\end{eqnarray}
Combining (\ref{eq:Rd_3d_SM}) and (\ref{sigma1}, \ref{sigma2},
\ref{sigma3}) renders our central result in (\ref{asympR3D}) for the
3D case.  The derivation of the asymptotic forms in (\ref{sigma1},
\ref{sigma2}) is straightforward but rather lengthy and we relegate it
to the end of this subsection.  Here we only briefly comment on the
term in (\ref{sigma3}).  We have
\begin{align}
\lim_{\ve \to 0} \phi_n^2(\ve) = (2 n + 1)^2 \,,
\end{align}
so that the sum in (\ref{sigma3}) converges as $\ve \to 0$ to
\begin{align}
\sum\limits_{n=1}^\infty \left(\frac{g_n(R)}{R g'_n(R)} - \frac{1}{n} + \frac{R U'(R)}{2 n^2}\right) (2 n +1) \,.
\end{align}
In view of the discussion above and (\ref{cc}), the terms in brackets
decay as $1/n^3$, which implies that this series converges.  In turn,
it means that the expression in (\ref{sigma3}) contributes in the
limit $\ve \to 0$ only to a constant, $\ve$-independent term in the
small-$\ve$ expansion of $\Rve^{(3)}$.

For the 2D case we use only the first term in the expansion in
(\ref{c}) to formally represent the ratio $g_n(R)/g'_n(R)$ as
\begin{align}
\label{d2D}
\frac{g_n(R)}{R g'_n(R)} \equiv \frac{1}{n}  + \left(\frac{g_n(R)}{R g'_n(R)} - \frac{1}{n}\right) \,,
\end{align}
where the terms in the brackets vanish as $1/n^2$. 
Then, inserting the latter identity into (\ref{rv2d}), we have
the following expression for $\Rve^{(2)}$:
\begin{align}
\label{label}
\Rve^{(2)} & = 2 \sum_{n=1}^{\infty} \frac{1}{n} \left(\frac{\sin(n \ve)}{n \ve}\right)^2 
+ 2 \sum_{n=1}^{\infty} \left(\frac{\sin(n \ve)}{n \ve}\right)^2 \left(\frac{g_n(R)}{R g'_n(R)} - \frac{1}{n}\right) .
\end{align}
The first sum evidently diverges as $\ve \to 0$, since
\begin{equation}
\lim_{\ve \to 0} \left(\frac{\sin(n \ve)}{n \ve}\right) = 1 \, \quad \text{and}  \quad \sum_{n=1}^{\infty} \frac{1}{n} = \infty \,.
\end{equation}
As a matter of fact, this sum can be calculated in an explicit form
(see (\ref{eq:xi}) in the main text) for an arbitrary $\ve$.  In the
small-$\ve$ limit it is given by
\begin{align}
\label{eq:Sigma1_2d}
2 \sum_{n=1}^{\infty} \frac{1}{n} \left(\frac{\sin(n \ve)}{n \ve}\right)^2 = 2 \ln(1/\ve) + 3 - 2 \ln 2 + O\left(\ve^2\right) .
\end{align}
On the other hand, for the sum in the last line in (\ref{label}) we
have
\begin{align}
\label{label1}
&\lim_{\ve \to 0} \sum_{n=1}^{\infty} \left(\frac{\sin(n \ve)}{n \ve}\right)^2 \left(\frac{g_n(R)}{R g'_n(R)} 
- \frac{1}{n}\right) = \sum_{n=1}^{\infty} \left(\frac{g_n(R)}{R g'_n(R)} - \frac{1}{n}\right) .
\end{align}
Since the terms in brackets decay as $1/n^2$ according to
(\ref{cc2D}), we infer that the sum in (\ref{label1}) converges, so
that the second term in (\ref{label}) contributes only to the
constant, $\ve$-independent term in the small-$\ve$ expansion of
$\Rve^{(2)}$.  Collecting (\ref{label}) to (\ref{label1}) we arrive at
the asymptotic expansion in (\ref{asympR2D}).

Lastly, we outline the derivation of the asymptotic forms in
(\ref{sigma1}, \ref{sigma2}).  For this purpose, we represent the
difference of two Legendre polynomials of orders $n-1$ and $n+1$ as
\begin{equation}
\label{eq:Leg_diff}
P_{n-1}(x) - P_{n+1}(x) = \frac{(2n+1)}{n(n+1)}  (1-x^2) \frac{d}{dx} P_n(x) \,.
\end{equation}
Using next the standard integral representation of the Legendre
polynomials,
\begin{equation}
\label{nu}
P_n(x) = \frac{1}{\pi} \int\limits_0^\pi dz_1 \, \nu^n(z_1) \,,  \quad  \nu_\ve(z_1) = x + i\sqrt{1-x^2} \cos(z_1) \,,
\end{equation}
we have
\begin{align}
\label{mu}
&P_{n-1}(x) - P_{n+1}(x) = \frac{(2n+1)}{(n+1)} ~ \frac{\sqrt{1-x^2}}{\pi}  \int\limits_0^\pi dz_1 \, \mu_\ve(z_1) \, \nu_\ve^{n-1}(z_1) \,, \,\,\, \mu_\ve(z_1) =  \sqrt{1-x^2} - ix \cos(z_1)  \,.
\end{align}
Plugging the latter representation into (\ref{series1}), performing the
summation over $n$, and setting $x = \cos \ve$, we cast $\Sigma_1$
into the form of the following double integral:
%
\begin{align}
\label{eq:xi_int}
&\Sigma_1 =   \int\limits_0^\pi dz_1 \int\limits_0^\pi dz_2 \, \Phi^{(1)}_\ve\left(z_1,z_2\right) ,
\end{align}
with
\begin{align}
& \Phi^{(1)}_\ve\left(z_1,z_2\right) = \frac{1}{\pi^2} \, \left(\frac{1+x}{1-x}\right) \, 
\frac{ \mu_\ve(z_1) \, \mu_\ve(z_2) }{\nu_\ve^2(z_1) \nu_\ve^2(z_2)}   \biggl(\bigl(1-\nu_\ve(z_1) \nu_\ve(z_2) \bigr)\ln\bigl(1- \nu_\ve(z_1) \nu_\ve(z_2)\bigr) 
+ \Li_2\bigl(\nu_\ve(z_1) \nu_\ve(z_2)\bigr) \biggr) \,,
\end{align}
where $\Li_2(y)$ is the dilogarithm: $\Li_2(y) =
\sum\limits_{n=1}^\infty y^n/n^2$.

We focus next on the small-$\ve$ behaviour of
$\Phi^{(1)}_{\ve}(z_1,z_2)$.  After straightforward but lengthy
calculations, we find that in this limit $\Phi^{(1)}_{\ve}(z_1,z_2)$
admits the following expansion
\begin{align}
\label{expansion}
 \Phi^{(1)}_{\ve}(z_1,z_2)  &= B^{(1)}_1 \, \ve^{-2} + B^{(1)}_2 \, \ve^{-1} \ln(\ve) + B^{(1)}_3 \, \ve^{-1}  + B^{(1)}_4 \, \ln(\ve) + B^{(1)}_5 + O(\ve) \,,
\end{align}
where $B_j^{(1)}$ are functions of both $z_1$ and $z_2$:
\begin{align*}
B^{(1)}_1 &= - \frac{2}{3} \cos z_1 \cos z_2 \,, \quad  B^{(1)}_2 = \frac{8 i}{\pi^2} \cos z_1 \cos z_2 \left(\cos z_1 +  \cos z_2\right) \,, \nonumber\\
 B^{(1)}_3  & =\frac{2 i}{3 \pi^2}  \Bigg( 2 \cos z_1 \cos z_2
  \bigg[\pi^2 - 3 - 3 i \pi +  6 \ln\left(\cos z_1 + \cos z_2\right)\bigg] - \pi^2\Bigg) \left(\cos z_1 + \cos z_2\right)\,, \nonumber\\
B^{(1)}_4 &= - \frac{2}{\pi^2} \Bigg( \left(4 - \left(\cos z_1 - \cos z_2\right)^2\right) \cos z_1 \cos z_2  + 
4 \left(\cos z_1 + \cos z_2 \right)^2 \left(1 - 2 \cos z_1 \cos z_2 \right) \Bigg)  \,, \nonumber\\
\end{align*}
and 
\begin{align}
B^{(1)}_5 &= \frac{1}{9} \Bigg(6 - 29 \cos z_1 \cos z_2 + 24 \cos^2 z_1 \cos^2 z_2  + 6 \left(3 \cos z_1 \cos z_2 -2\right)\left(\cos^2 z_1 + \cos^2 z_2\right) \Bigg)  \nonumber\\
&+\frac{1}{\pi^2} \Bigg[4 \left(\cos z_1 + \cos z_2\right)^2 \left(1 - 2 \cos z_1 \cos z_2\right)  -\cos z_1 \cos z_2 \bigg(4 + \cos^2 z_1 + \cos^2 z_2 + 6 \cos z_1 \cos z_2\bigg)\Bigg] \nonumber \\
& + \frac{i}{\pi} \Bigg[  \cos z_1 \cos z_2 \bigg(4 - (\cos z_1 - \cos z_2)^2\bigg) + 4  \left(\cos z_1 + \cos z_2\right)^2   \left(1 - 2 \cos z_1 \cos z_2\right)  \Bigg] \nonumber\\
&- \frac{2}{\pi^2} \Bigg\{\cos z_1 \cos z_2 \bigg(4 - \left(\cos z_1 - \cos z_2\right)^2\bigg) + 4 \left(\cos z_1 + \cos z_2\right)^2 \left(1 - 2 \cos z_1 \cos z_2\right)\Bigg\} 
\ln\left(\cos z_1 + \cos z_2\right)  \,.
\end{align}
Integrating $B_j^{(1)}$ over $z_1$ and $z_2$, we get
\begin{align}
\label{integrals1}
& \int^{\pi}_0 \int^{\pi}_0 dz_1 dz_2\, B_1^{(1)} = \int^{\pi}_0 \int^{\pi}_0 dz_1 dz_2\, B_2^{(1)} = 0  \,,  \nonumber\\
& \int^{\pi}_0 \int^{\pi}_0 dz_1 dz_2\, B_3^{(1)} = \frac{32}{3 \pi} \,, \qquad
  \int^{\pi}_0 \int^{\pi}_0 dz_1 dz_2\, B_4^{(1)} = - 1 \,, \nonumber\\
&\int^{\pi}_0 \int^{\pi}_0 dz_1 dz_2\, B_5^{(1)} = \ln 2 - \frac{7}{4} \,. 
\end{align}
Collecting the expressions in (\ref{integrals1}) we get the expansion
in (\ref{sigma1}).  Note that the coefficient $32/(3 \pi)$ in front of
the leading term in (\ref{sigma1}) was obtained earlier in
\cite{shoup}.

Similarly, using (\ref{mu}), we represent the infinite series
$\Sigma_2$ in (\ref{series}) as
\begin{equation}
\label{eq:xi2_int}
\Sigma_2  =  \int^{\pi}_0 \int_0^{\pi} dz_1 dz_2 \, \Phi_{\ve}^{(2)}(z_1,z_2) ,
\end{equation}
where
\begin{align}
&\Phi_{\ve}^{(2)}(z_1,z_2) = \frac{1}{2\pi^2} ~ \left(\frac{1+x}{1-x}\right) ~ 
 \frac{\mu(z_1) \mu(z_2)}{\nu^2(z_1) \nu^2(z_2)}  \Bigg((\nu(z_1) \nu(z_2) -1)\Li_2\bigl(\nu(z_1) \nu(z_2)\bigr) + \nu(z_1) \nu(z_2)\Bigg) \,,
\end{align}
with $\nu(z_{1,2})$ defined in (\ref{nu}).  The small-$\ve$ behaviour
of $\Phi^{(2)}_{\ve}(z_1,z_2)$ follows
\begin{align}
\label{expansion2}
\Phi^{(2)}_{\ve}(z_1,z_2) & = B^{(2)}_1 \, \ve^{-2} + B^{(2)}_3 \, \ve^{-1}  + B^{(2)}_4 \, \ln(\ve) + B^{(2)}_5 + O(\ve) \,,
\end{align}
where $B^{(2)}_1$, $B^{(2)}_3$, $B^{(2)}_4$ and
$B^{(2)}_5$ are given explicitly by
\begin{align*}
B^{(2)}_1 &= - \frac{2}{\pi^2} \cos z_1 \cos z_2 \,, \,\,\,
B^{(2)}_3 = - \frac{i}{3 \pi^2}  \bigg(6 + (\pi^2 - 6) \cos z_1 \cos z_2 \bigg)  
 (\cos z_1 +\cos z_2) \,, \,\,\, \nonumber\\
 B^{(2)}_4 & = - \frac{2}{\pi^2} \cos z_1 \cos z_2 \left(\cos z_1 + \cos z_2\right)^2 \,,   \nonumber\\
\end{align*}
and
\begin{align}
B^{(2)}_5 &= \frac{1}{3} \Bigg(\left(\cos^2 z_1 + \cos^2 z_2\right)  \left(1 - 2 \cos z_1 \cos z_2\right) + 3 \cos z_1 \cos z_2 \left(1  - \cos z_1 \cos z_2\right) \Bigg)  \nonumber\\
&+ \frac{1}{3 \pi^2} \Bigg[ 6 - 6 \left(\cos^2 z_1 + \cos^2 z_2\right) \left(1 - 2 \cos z_1 \cos z_2\right) - 11 \cos z_1 \cos z_2 + 18 \cos^2 z_1 \cos^2 z_2 \Bigg] \nonumber\\
&+\frac{i}{\pi} \cos z_1 \cos z_2 \left(\cos z_1 + \cos z_2\right)^2 - \frac{2}{\pi^2}  \cos z_1 \cos z_2 \left(\cos z_1 + \cos z_2\right)^2  \ln \left(\cos z_1 + \cos z_2\right) \,.
\end{align}
Integrating $B_j^{(2)}$ over $z_1$ and $z_2$, we obtain
\begin{align}
\label{integrals2}
& \int^{\pi}_0 \int^{\pi}_0 dz_1 dz_2\, B_1^{(2)} = \int^{\pi}_0 \int^{\pi}_0 dz_1 dz_2\, B_3^{(2)} = 0  \,, \nonumber\\
& \int^{\pi}_0 \int^{\pi}_0 dz_1 dz_2\, B_4^{(2)} = - 1 \,, \nonumber\\
&  \int^{\pi}_0 \int^{\pi}_0 dz_1 dz_2\, B_5^{(2)} = \frac{1}{4} + \ln 2 + \frac{\pi^2}{12} \,. 
\end{align}
Collecting these results, we obtain eventually the asymptotic
expansion in (\ref{sigma2}).

\subsection{Approximation for $g_n(R)/g'_n(R)$ and its limiting behavior for sufficiently large 
$ |U'(R)|$.}
\label{large}

We pursue next a different approach for calculation of the logarithmic
derivative of the radial functions at the boundary, and of the
corresponding expressions for the infinite series $\Rve^{(3)}$ and
$\Rve^{(2)}$.  This approach is based on the assumption that, once we
are only interested in the behaviour on the boundary only, we may
replace the coefficients in the differential equations (\ref{eq:gn})
and (\ref{eq:gn2d}) by their values at the boundary.  Such an
approximation is legitimate, of course, only for the potentials for
which $U'(R)$ exists.  In doing so, we will be able to derive an
explicit, albeit an approximate expression for $g_n(R)/g'_n(R)$ which
is valid, in principle, for arbitrary $n$ and arbitrary\footnotemark\footnotetext{Note
that the condition that $U'(R)$ is bounded does not prevent us to
study the behaviour of $g_n(R)/g'_n(R)$ in the limit $|U'(R)|
\to \infty$.} $|U'(R)| < \infty$.  This approximate expression will be
subsequently checked against exact results obtained for the
triangular-well potential (see \ref{SM4} and \ref{SM5}) and the
asymptotic forms in (\ref{cc}) and (\ref{cc2D}).

We turn to the differential equations (\ref{eq:gn}) and
(\ref{eq:gn2d}) and replace the coefficients in these equations (which
are functions of $r$) by their values at the boundary.  This gives the
following differential equations with constant coefficients:
\begin{equation}
\label{apprx3}
g''_n + \left(\frac{2}{R} - U'(R)\right) g'_n - \frac{n (n+1)}{R^2} g_n =0 \,
\end{equation} 
and 
\begin{equation}
\label{apprx2}
g''_n + \left(\frac{1}{R} - U'(R)\right) g'_n - \frac{n^2}{R^2} g_n = 0
\end{equation}
for the 3D and the 2D cases, respectively.  Using the notation
\begin{equation}
\etamm = R \, U'(R) ,
\end{equation}
we write down a general solution of  (\ref{apprx3}):
\begin{align}
\label{apprx4}
g_n &= c_1 \exp\left(\frac{r}{2 R} \left(\left(\etamm - 2\right) 
- \sqrt{\left(2 n + 1\right)^2 + \left(\etamm -2\right)^2 - 1} \right)\right) \nonumber\\
&+ c_2  \exp\left(\frac{r}{2 R} \left(\left(\etamm - 2\right) 
+ \sqrt{\left(2 n + 1\right)^2 + \left(\etamm - 2\right)^2-1} \right)\right) ,
\end{align}
where $c_1$ and $c_2$ are adjustable constants.  We note that since
$\left(2 n + 1\right)^2-1 > 0$ for any $n > 0$, the expression under
the square root is always positive.  Further, differentiating
(\ref{apprx4}) and setting $r=R$, we get the following approximate
expression for the inverse of the logarithmic derivative at the
boundary:
\begin{align}
\label{apprx5}
&\dfrac{g_n(R)}{R g'_n(R)} \approx 2 \Bigg(\etamm - 2 + \sqrt{(2 n + 1)^2 + (\etamm - 2)^2-1}  - \dfrac{2 c_1 \sqrt{(2 n + 1)^2 + (\etamm- 2)^2-1} }{c_1 + c_2 \exp\bigl(\sqrt{(2 n + 1)^2 + (\etamm - 2)^2-1} \bigr)}\Bigg)^{-1} \,.
\end{align}
We notice that the last term in brackets in (\ref{apprx5}), which is
the ratio of an algebraic and an exponential function, can be safely
neglected because the exponential function becomes large when either
(or both) $n$ and/or $|\etamm|$ are large.  This yields the following
approximation for the inverse logarithmic derivative, which is
independent of the constants $c_1$ and $c_2$:
\begin{align}
\label{apprx7}
\dfrac{g_n(R)}{R g'_n(R)} &\approx \dfrac{2}{\etamm  - 2 + \sqrt{\left(2 n + 1\right)^2 + \left(\etamm - 2\right)^2-1} }  = \dfrac{\sqrt{\left(2 n + 1\right)^2 + \left(\etamm - 2\right)^2-1} - \etamm + 2}{2 n (n+1) } \, .
\end{align} 
The same arguments yield an analogous approximation for the 2D case:
\begin{equation}
\label{apprx8}
\dfrac{g_n(R)}{R g'_n(R)} \approx \dfrac{\sqrt{4 n^2 + \left(\etamm - 1\right)^2} - \etamm + 1}{2 n^2} \,.
\end{equation} 
In Figs. \ref{denis7} and \ref{denis8} in the following sections
\ref{SM4} and \ref{SM5} we compare the expressions in (\ref{apprx7})
and (\ref{apprx8}) with the exact results for the ratio $g_n(R)/(R
g'_n(R))$ derived for the special case of a triangular-well potential
in (\ref{triangular}).  We observe a fairly good agreement between the
approximate forms in (\ref{apprx7}) and (\ref{apprx8}) and the exact
results in (\ref{eq:gn_Rgnprime2}) and (\ref{eq:gn_Rgnprime2_2d}) even
for very modest values of $n$ (say, for $n \geq 10$).  For smaller $n$
there are some apparent deviations which however become smaller the
larger $|\etamm|$ is.

We turn to the limit $n \to \infty$.  We find that in this limit the
expressions in $(\ref{apprx7})$ and (\ref{apprx8}) exhibit the
following asymptotic behaviour
\begin{align}
\label{apprx11}
&\dfrac{g_n(R)}{R g'_n(R)} \approx \frac{1}{n} - \frac{\etamm - 1}{2 n^2} + \frac{\etamm^2-1}{8 n^3} + O\left(\frac{1}{n^4}\right)
\end{align}
and
\begin{align}
\label{apprx12}
&\dfrac{g_n(R)}{R g'_n(R)} \approx \frac{1}{n} - \frac{\etamm - 1}{2 n^2} + \frac{\etamm^2- 2 \etamm +1}{8 n^3} + O\left(\frac{1}{n^4}\right)
\end{align}
for the 3D and the 2D cases, respectively.  Comparing these expansions
with the asymptotic forms in (\ref{cc}) and (\ref{cc2D}), we observe
that they are identical in the leading terms for large $|\etamm|$.
This suggests, in turn, that the approximate expressions for the
inverse of the logarithmic derivatives in (\ref{apprx7}) and
(\ref{apprx8}) are reliable (as well as the assumptions underlying
their derivation) for $|\etamm|$ large enough.

Further, using (\ref{apprx7}) and (\ref{apprx8}), we evaluate
approximate expressions for the infinite series $\Rve^{(3)}$ and
$\Rve^{(2)}$, and the corresponding small-$\ve$ expansions.  To this
end, it is expedient to use an auxiliary integral identity
\begin{equation}
\label{aa}
\sqrt{A^2 + B^2} = A + B  \int^{\infty}_0 \frac{d\xi}{\xi} e^{- A \xi} \, J_1\left(B \xi \right) \,,
\end{equation} 
where $J_1(\cdot)$ is the Bessel function.  This identity is valid for
$A$ and $B$ such that $| {\rm Im} \, B| < {\rm Re} \, A$.  

\subsubsection*{2D case}

We start with the 2D case, which is simpler than the 3D one, and set
$A = 2 n$ and $B = \etamm - 1$.  Such a choice evidently fulfils the
condition of the applicability of the identity in (\ref{aa}).  We have
then
\begin{align}
\label{bb}
\Rve^{(2)} & \approx 2 \sum_{n=1}^{\infty} \frac{\sin^2(n \ve)}{n^3 \, \ve^2} - 
B \sum_{n=1}^{\infty} \frac{\sin^2(n \ve)}{n^4 \, \ve^2}  + B \int^{\infty}_0 \frac{d\xi}{\xi} J_1(B \xi) \sum_{n=1}^{\infty} \frac{\sin^2(n \ve)}{n^4 \, \ve^2} e^{- 2 n \xi}  \,,
\end{align}
where the symbol $\approx$ signifies that this expression is obtained
via an approximate approach.  The asymptotic small-$\ve $ behaviour of
the first sum is given by (\ref{eq:Sigma1_2d}), while the second and
the third terms converge to $\ve$-independent constants :
\begin{align}
\label{bb1}
\Rve^{(2)} &\approx \underbrace{2\ln(1/\ve) + 3-2\ln 2 + O(\ve^2)}_{\textrm{first sum}} - 
\underbrace{B \, \frac{\pi^2}{6}}_{\textrm{second sum}}  + \underbrace{B \, \int^{\infty}_0 \frac{d\xi}{\xi} J_1(B \xi) \Li_2(e^{-2\xi})}_{\textrm{third sum}} \,.
\end{align}
Since $J_1(z)$ is an odd function, $J_1(-z)= -J_1(z)$, the integral in
the last line in (\ref{bb1}) is an even function of $B$, i.e., it
depends only on $|B|$.  For large $|B|$ (or large $|\etamm|$), the
major contribution to this integral comes from small values of $\xi$,
$\xi \ll 1$, so that this integral is given approximately by
\begin{equation}
B \int^{\infty}_0 \frac{d\xi}{\xi} J_1(B \xi) \Li_2(e^{-2\xi}) \approx |B| \frac{\pi^2}{6} + 2\ln |B| + O(1) \,,
\end{equation}
where the omitted terms $O(1)$ are $B$-independent constants.
We therefore obtain
\begin{align}
\label{eq:Rve_asympA}
\Rve^{(2)} \approx 2 \ln(1/\ve) + \frac{\pi^2}{6} R \bigl(|U'(R)| - U'(R)\bigr) + 2\ln (R|U'(R)|) + O(1) \,.
\end{align}
We conclude that in the 2D case, the leading in the limit $\ve\to 0$
term in $\Rve^{(2)}$ is independent of the interaction potential and
is identical to the result in (\ref{asympR2D}) based on the large-$n$
expansions.  Remarkably, the second term in (\ref{eq:Rve_asympA}) is
non-zero for attractive potentials (negative $U'(R)$) only, and
becomes identically equal to zero in case of repulsive potentials
(positive $U'(R)$).  As a matter of fact, this term provides the major
contribution in the limit of infinitely strong attractive potentials.
For instance, in the case of a triangular-well potential, one has
$\l^{(2)}_U \sim 2/|\etamm|$ from (\ref{attr}) for negative $U'(R)$ of
very large amplitude, so that the MFET in the limit $\etamm \to
-\infty$ becomes
\begin{equation}
T^{(2)}_\ve \simeq \left(\frac{r_0}{R}\right)^2 \, \frac{r_0^2}{8D} + \frac{\pi^2 R^2}{3D} \, .
\end{equation}
As discussed in the main text, the first term is the time for a
particle started uniformly to reach the boundary (in presence of an
infinitely strong attractive potential in the region $r_0 < r < R$),
whereas
the second term represents the MFPT from a uniform starting point on a
circle of radius $R$ to a point-like target ($\ve = 0$).

\subsubsection*{3D case}

In the 3D case we set $A = 2 n +1$, which is real and positive, and $B
= \sqrt{(\etamm - 2)^2 - 1}$.  Note that the maximum imaginary value
of $B$ is $1$, and it is less than the minimal value of $A = 3$,
attained for $n=1$, so that the identity in (\ref{aa}) is valid for
any $n$ and $\etamm$.  Using this identity, we can cast $\Rve^{(3)}$
into the following form
\begin{align}
\label{sum}
\Rve^{(3)} \approx - \frac{\etamm - 2}{2} \sum_{n=1}^{\infty} \frac{\phi^2_n(\ve)}{n (n+1) (2 n+1)} 
+ \frac{1}{2} \sum_{n=1}^{\infty} \frac{\phi_n^2(\ve)}{n (n+1)} + F_\ve(B)  \,,
\end{align}
\vspace{0.1in}
where
\begin{equation}
\label{eq:FveB}
F_\ve(B) = \frac{B}{2} \int^{\infty}_0 \frac{d\xi}{\xi} \, e^{-\xi} \, J_1\left(B \xi \right) \, 
\sum_{n=1}^{\infty} \frac{\phi^2_n(\ve)}{n (n+1) (2 n + 1)} e^{- 2 n \xi} \,.
\end{equation}
For the infinite series entering the first term on the right-hand-side
of (\ref{sum}) we have
\begin{align}
\label{13}
\frac{1}{2} \sum_{n=1}^{\infty} \frac{\phi^2_n(\ve)}{n (n+1) (2 n+1)}  = \Sigma_2 + 
\frac{1}{2} \sum_{k=1}^{\infty} (-1)^k \sum_{n=1}^{\infty} \frac{\phi^2_n(\ve)}{n^{2 + k} (2 n+1)} \,,
\end{align}
where $\Sigma_2$ and its asymptotic behaviour are defined in
(\ref{series}) and (\ref{sigma2}).  Noticing that the second term on
the right-hand-side of (\ref{13}) converges to an $\ve$-independent
constant as $\ve \to 0$, i.e.,
\begin{align}
\lim_{\ve \to 0} \sum_{k=1}^{\infty} (-1)^k \sum_{n=1}^{\infty} \frac{\phi^2_n(\ve)}{n^{2 + k} (2 n+1)} = 
\sum_{k=1}^{\infty} (-1)^k \sum_{n=1}^{\infty} \frac{(2 n+1 )}{n^{2 + k}} = - 1 - \frac{\pi^2}{6} \,,
\end{align}
we infer that 
\begin{align}
\label{14}
\frac{1}{2} \sum_{n=1}^{\infty} \frac{\phi^2_n(\ve)}{n (n+1) (2 n+1)}  = \ln(1/\ve) + O(1) \,.
\end{align} 
The sum in the second term on the right-hand-side of (\ref{sum}) can
be formally rewritten as
\begin{align}
\frac{1}{2} \sum_{n=1}^{\infty} \frac{\phi_n^2(\ve)}{n (n+1)} &= \sum_{n=1}^{\infty} \frac{\phi_n^2(\ve)}{n (2 n+1)}
 \left(1 - \frac{1}{2 (n+1)}\right) = \Sigma_1 - \frac{1}{2} \sum_{n=1}^{\infty} \frac{\phi^2_n(\ve)}{n (n+1) (2 n+1)} \,,
\end{align}
where $\Sigma_1$ and its asymptotic behaviour are defined in
(\ref{series1}) and (\ref{sigma1}).  Consequently, we have
\begin{align}
\label{15}
\frac{1}{2} \sum_{n=1}^{\infty} \frac{\phi_n^2(\ve)}{n (n+1)} = \frac{32}{3 \pi} \ve^{-1} + O(1) \,.
\end{align}

Lastly, we consider the contribution in (\ref{eq:FveB}).  For large
$|B|$, the major contribution to the integral comes from $\xi$ close
to $0$.  Since $\phi_n(\ve) \to (2n+1)$ as $\ve \to 0$, the sum would
logarithmically diverge if both $\ve$ and $\xi$ were set to $0$.  This
simple observation suggests that this sum may exhibit a logarithmic
dependence either on $\ve$, or on $\xi$.  In order to evaluate the
contribution $F_\ve(B)$, we adopt the summation technique used in the
previous subsection.  Recalling the integral representations in
(\ref{nu}, \ref{mu}), we have
\begin{equation}
\label{eq:FveB2}
F_\ve(B) = \frac{B}{2} \int^{\infty}_0 \frac{d\xi}{\xi} \, J_1\left(B \xi \right) \, G_\ve(\xi),  
\end{equation}
where, explicitly, 
\begin{equation}
G_\ve(\xi) = e^{-\xi} \sum_{n=1}^{\infty} \frac{\phi^2_n(\ve) \, e^{- 2 n \xi} }{n (n+1) (2 n + 1)} 
= \int\limits_0^\pi dz_1 \int\limits_0^\pi dz_2 \, \Phi^{(3)}_\ve(z_1,z_2,\xi) 
\end{equation}
and
\begin{align}
\Phi^{(3)}_\ve(z_1,z_2,\xi) &= \frac{e^{-\xi}}{\pi^2} \, \left(\frac{1+\cos \ve}{1-\cos \ve}\right) \mu_\ve(z_1) \mu_\ve(z_2)  \sum_{n=1}^{\infty} \frac{2n+1}{n (n+1)^3} \bigl[\nu_\ve (z_1)\nu_\ve(z_2)\bigr]^{n-1} e^{- 2 n \xi}  \,,
\end{align}
with $\mu_\ve(z)$ and $\nu_\ve(z)$ defined in (\ref{nu}, \ref{mu}).
Denoting $\zeta = \nu_\ve(z_1) \nu_\ve(z_2) e^{-2\xi}$, we get
\begin{align}
\label{phi3}
\Phi^{(3)}_\ve(z_1,z_2,\xi) & = \frac{1}{\pi^2} \, \left(\frac{1+\cos \ve}{1-\cos \ve}\right) \mu_\ve(z_1) \mu_\ve(z_2) e^{-3\xi}   \frac{\left(\Li_3(\zeta) - \Li_2(\zeta) + (1-\zeta) \ln(1-\zeta) + \zeta\right)}{\zeta^2} \,.
\end{align}
Now, we have two options, either to expand this function first in
powers of $\ve$ and then in powers of $\xi$, or to expand it first in
powers of $\xi$ and then in powers of $\ve$.  These two options
correspond to two possible orders of limits: $\ve\to 0$ and $|B| \to
\infty$.

\vspace{0.1in}
(i) {\bf Limit $\ve\to 0$ for a fixed $|B|$}.  

For a fixed $\xi > 0$, we expand $\Phi^{(3)}_\ve(z_1,z_2,\xi)$ in
powers of $\ve$ to get
\begin{align}
\Phi^{(3)}_\ve(z_1,z_2,\xi) & = C_{-2}(z_1,z_2,\xi) \ve^{-2} + C_{-1}(z_1,z_2,\xi) \ve^{-1} +C_{0}(z_1,z_2,\xi) + O(\ve). 
\end{align}
Note that this expansion does not contain a term, which
logarithmically diverges as $\ve \to 0$.  Next, each coefficient
$C_{j}(z_1,z_2,\xi)$ has to be expanded in powers of $\xi$.  After
integration over $z_1$ and $z_2$, the contributions from $C_{-2}$ and
$C_{-1}$ vanish (as expected), and the leading terms are given by
\begin{align}
F_0(B) & = \frac{B}{2} \int^{\infty}_0 \frac{d\xi}{\xi} \, J_1\left(B \xi \right) \, \biggl[ -2\ln \xi - 2\ln 2 - 1 + 3\xi + O(\xi^2)\biggr] \nonumber\\
& = |B| \ln |B| + |B| (\gamma - 3/2) + O(1) \nonumber\\
& = |\etamm| \ln |\etamm| + |\etamm| (\gamma - 3/2) + O(\ln|\etamm|), 
\end{align}
where $\gamma \approx 0.5772$ is the Euler-Mascheroni constant.
Combining this contribution with (\ref{14}, \ref{15}), we obtain the
small-$\ve$ asymptotic behaviour of $\Rve^{(3)}$ for sufficiently large $|\etamm|$:
\begin{align}
\label{eq:Rve3_1}
\Rve^{(3)} &\approx \frac{32}{3 \pi} \ve^{-1} - (\etamm - 2)  \, \ln(1/\ve) + |\etamm| \, \ln |\etamm| + (\gamma - 3/2) |\etamm|  + O(\ln|\etamm|) \,.
\end{align}
%
%
%
We note that despite the fact that this small-$\ve$ asymptotics is
formally valid for sufficiently large $|\etamm|$, it predicts a
spurious logarithmic divergence of the MFET in the limit $|\etamm| \to
\infty$.  This divergence is clearly unphysical (a stronger attractive
potential should reduce the MFET, instead of increasing it) and
indicates that (\ref{eq:Rve3_1}) holds for large but bounded $\etamm$.
Upon a more detailed analysis, we infer that (\ref{eq:Rve3_1}) is only
applicable for $1 \ll |\etamm| \ll 1/\ve$.

\vspace{0.1in}
(ii) {\bf Limit $|B|\to \infty$ for a fixed small $\ve$}.

Expanding $\Phi^{(3)}_\ve(z_1,z_2,\xi)$ in (\ref{phi3}) in powers of
$\xi$, we have
\begin{equation}
\Phi^{(3)}_\ve(z_1,z_2,\xi) = \Phi^{(3),0}_\ve (z_1,z_2) + \xi \, \Phi^{(3),1}_\ve (z_1,z_2) + O(\xi^2) \,,
\end{equation}
which yields, upon inserting the latter expansion into (\ref{eq:FveB2}),
\begin{equation}
\label{eq:FveB3}
F_\ve(B) = \frac{1}{2} \int\limits_0^\pi dz_1 \int\limits_0^\pi dz_2 \biggl[|B| \, \Phi^{(3),0}_\ve (z_1,z_2) + \Phi^{(3),1}_\ve (z_1,z_2) \biggr].
\end{equation}
Concentrating next on the narrow escape limit $\ve\to 0$, we expand
$\Phi^{(3),0}_\ve (z_1,z_2)$ and $\Phi^{(3),1}_\ve (z_1,z_2)$ in
powers of $\ve$ to get
\begin{align}
\label{formal}
&\Phi^{(3),j}_\ve (z_1,z_2) = B^{(3),j}_1(z_1,z_2) \, \ve^{-2} + B^{(3),j}_2(z_1,z_2) \, \ve^{-1} \ln(\ve)  \nonumber\\
&+B^{(3),j}_3(z_1,z_2) \, \ve^{-1} + B^{(3),j}_4(z_1,z_2) \, \ln(\ve) + B^{(3),j}_5(z_1,z_2) + O(\ve),
\end{align}
where $B^{(3),j}_1(z_1,z_2)$ with $j=0,1$ are given explicitly by
\begin{align}
B_1^{(3),0} & = -\frac{2\left(6\zeta(3)-\pi^2+6\right)}{3\pi^2} \cos z_1 \cos z_2 \,, \quad B_2^{(3),0}  \equiv 0 \,, \nonumber\\
B_3^{(3),0} & = \frac{2i}{3\pi^2} (\cos z_1 + \cos z_2) \biggl( 3\left(4\zeta(3) - \pi^2 + 4 \right)\cos z_1 \cos z_2  - \left(6\zeta(3) - \pi^2 + 6\right) \biggr) \,,  \nonumber\\
B_4^{(3),0}  &= - \frac{4}{\pi^2} \cos z_1 \cos z_2 \bigl(\cos z_1 + \cos z_2 \bigr)^2 \,, \nonumber\\
B_5^{(3),0}  &= - \frac{4}{\pi^2} \cos z_1 \cos z_2 \bigl(\cos z_1 + \cos z_2\bigr)^2 \ln\bigl(\cos z_1 + \cos z_2\bigr) + \nonumber\\
&+ \frac{1}{9\pi^2} \biggl(36+36\zeta(3) - 6\pi^2 + \bigl(144 -33\pi^2 +18\pi i + 108\zeta(3)\bigr)  \cos z_1 \cos z_2 \bigl(\cos^2 z_1 + \cos^2 z_2 \bigr)  \nonumber\\
& + \bigl(18\pi^2 - 72 - 72\zeta(3)\bigr) \bigl(\cos^2 z_1 + \cos^2 z_2\bigr)+ \bigl(47\pi^2 -174 - 174\zeta(3)\bigr) \cos z_1 \cos z_2 \nonumber\\
& + \bigl(216-48\pi^2+36\pi i\bigr)\cos^2 z_1 \cos^2 z_2 \biggr) \,,
\end{align}
and
\begin{align}
B_1^{(3),1} & = -\frac{2(2\zeta(3)-\pi^2+2)}{\pi^2} \cos z_1 \cos z_2 \,, \,\,\, B_2^{(3),1}  = -\frac{16i}{\pi^2} \cos z_1 \cos z_2 \bigl(\cos z_1 + \cos z_2\bigr) \,, \nonumber\\
B_3^{(3),1} & = -\frac{16i}{\pi^2} \cos z_1 \cos z_2 \bigl(\cos z_1 + \cos z_2\bigr) \ln\bigl(\cos z_1 + \cos z_2 \bigr) + \frac{2i}{3\pi^2} \bigl(\cos z_1 + \cos z_2\bigr) \nonumber\\
& \times \biggl(\cos z_1 \cos z_2 \bigl(12\zeta(3)+24-7\pi^2+ 12i\pi\bigr) - 3\bigl(2\zeta(3)-\pi^2+2\bigr)\biggr) \,, \nonumber\\
B_4^{(3),1} & = \frac{8}{\pi^2} \biggl( \bigl(\cos^2 z_1 +\cos^2 z_2\bigr) \bigl(2-5\cos z_1 \cos z_2\bigr) +2\cos z_1 \cos z_2 \bigl(3-4\cos z_1 \cos z_2\bigr) \biggr) \,. 
\end{align}
To find an explicit expression for $F_\ve(B)$ in (\ref{eq:FveB3}), we
now have to integrate all the coefficients $B_1^{(3),j}$ over $z_1$
and $z_2$.  This can be done rather straightforwardly and we find that
the double integrals
\begin{equation}
b_k^j = \frac{1}{2} \int\limits_0^\pi  \int\limits_0^\pi dz_1 \, dz_2 \, B^{(3),j}_k(z_1,z_2) \,,
\end{equation}
are given explicitly by
\begin{equation}
b_1^0 = b_2^0 = b_3^0 = 0, \quad b_4^0 = -1, \quad  b_5^0 = \ln 2 - \frac{1}{4} \,, 
\end{equation}
for $j=0$, and
\begin{equation}
b_1^1 = b_2^1 = b_4^1 = 0 \,,  \quad  b_3^1 = - \frac{32}{3\pi} \,,
\end{equation}
for $j=1$, respectively.  Collecting these explicit expressions for
the coefficients $b_k^j$, we get
\begin{equation}
\label{eq:FveB4}
F_\ve(B) = |B|\, \ln(1/\ve) + (\ln 2 - 1/4) \, |B|  - \frac{32}{3\pi} \, \ve^{-1} + O(1) \,.
\end{equation}
Note that the coefficient in front of the term which diverges as
$1/\ve$ is negative and is equal by the absolute value to the
coefficient of the leading diverging term in (\ref{14}), so that these
two terms cancel each other.  Recalling next the definition of $B$ for
the 3D case, and combining (\ref{eq:FveB3}) with (\ref{14}, \ref{15}),
we obtain the asymptotic behaviour of $\Rve^{(3)}$ for very large
$|\etamm|$ and small fixed $\ve$:
\begin{equation}
\label{eq:Rve3_2}
\Rve^{(3)} \approx (|\etamm| - \etamm) \, \ln(1/\ve) + (\ln 2 - 1/4) |\etamm| + O(1)  \,.
\end{equation}
Comparing the latter expression with (\ref{eq:Rve3_1}), we note that
(\ref{eq:Rve3_2}) does not contain the term $|\etamm| \ln|\etamm|$
(that caused an unphysical divergence of the MFET in the limit $\etamm
\to -\infty$), and includes an extra term $(|\etamm|-\etamm)
\ln(1/\ve)$ so that the logarithmically diverging term in
(\ref{eq:Rve3_2}) is twice larger than the one in (\ref{eq:Rve3_1}) in
case of negative $\etamm$.  We note that due to this additional
numerical factor, the expression in (\ref{eq:Rve3_2}) reproduces
correctly, in the limit $\etamm \to - \infty$, the exact result
obtained in \cite{Sano79} for the MFPT to the EW solely due to
diffusion along the surface of the 3D spherical micro-domain.  We also
emphasise that the coefficient in front of $\ln(1/\ve)$ is non-zero
only for negative $\etamm$ (attractive interactions), and vanishes for
positive $\etamm$ (repulsive interactions).

\section{Systems without long-range interactions}
\label{SM3} 

We examine next the simplest case without long-range interactions,
$U(r) \equiv 0$, so that a particle diffuses freely with a bounded
micro-domain.

\subsection{3D case}

The general solution of equation (\ref{eq:gn}) for the radial
functions $g_n(r)$ reads
\begin{equation}
g_n(r) = c_1 r^n + c_2 r^{-n-1} .
\end{equation}
We set $c_1 = 1$ for convenience, and choose $c_2 = 0$ to ensure the
regularity at the origin.  Then, the particular solution $t_0(r)$ is
\begin{equation}
t_0(r) = \frac{R^2 - r^2}{6D} ,
\end{equation}
so that $t'_0(R) = -R/(3D)$.  We therefore obtain
\begin{equation}
\label{eq:t3d}
t(r,\theta) = \frac{R^2 - r^2}{6D} + a_0 - \frac{R^2}{3D} \sum\limits_{n=1}^\infty \dfrac{\phi_n(\ve)}{n} \left(\dfrac{r}{R}\right)^{n} P_n(\cos\theta),
\end{equation}
where the coefficient $a_0$ is fixed by the self-consistent condition
in (\ref{eq:a0}).  This gives
\begin{equation}
\label{eq:a0_3d_nopot}
a_0 = \frac{2R}{3\kappa (1-\cos \ve)} + \frac{R^2 \Rve^{(3)}}{3D}  ,
\end{equation}
with $\Rve^{(3)}$ defined in (\ref{eq:Rve_3d_U0}).  By integrating
(\ref{eq:t3d}) over the volume of the sphere, we find that the global
MFET $T_{\ve}$ from a random location is given by (\ref{ckI2}).

We note finally that for a perfect EW (no barrier, $\kappa = \infty$),
such that any arrival of the particle to the EW location will result
in the escape from the sphere, the condition
(\ref{eq:a0_selfconsistent}) reduces to
\begin{equation}
\label{eq:BC_weak}
\int\limits_0^{\ve} d\theta \sin\theta~ t(R,\theta) = 0 .
\end{equation}
In other words, the original Dirichlet boundary condition at each
point of the EW, $t(R,\theta) = 0$ for $0\leq \theta \leq \ve$, is
replaced by a weaker condition requiring that the MFET vanishes on the
EW {\it on average}.  Hence, the condition (\ref{eq:BC_weak}) implies
that some values of $t(R,\theta)$ can become negative.  As a
consequence, the approximation is not expected to yield accurate
results for the MFET with the starting point $(r,\theta)$ close to the
EW.  One can check numerically (not shown) that the approximation is
nonetheless very accurate when the starting point is far from the EW.
In general, the SCA is expected to be more accurate for small targets,
as well as for weak reactivities $\kappa$.

\subsection{2D case}

In 2D case, the radial functions $g_n(r)$ are given by $g_n(r) = c_1
r^n + c_2 r^{-n} = r^n$, where we set $c_1 = 1$ and $c_2 = 0$.  We
also have
\begin{equation}
t_0(r) = \frac{R^2-r^2}{4D} ,
\end{equation}
from which $t'_0(R) = -R/(2D)$ follows.  We therefore obtain
\begin{equation}
\label{eq:t_2d_approx}
t(r,\theta) = \frac{R^2-r^2}{4D} + a_0 - \frac{R^2}{D} \sum\limits_{n=1}^\infty \frac{\sin(n \ve)}{n^2 \ve}  (r/R)^n   \cos(n\theta),
\end{equation}
with
\begin{equation}
\label{eq:a0_2d_U0}
a_0 = \frac{\pi R}{2\kappa \ve} + \frac{R^2 \Rve^{(2)}}{D} ,
\end{equation}
and $\Rve^{(2)}$ defined in (\ref{r2}).  Integrating
(\ref{eq:t_2d_approx}) over the area of the circular micro-domain, we
arrive at our result in (\ref{ckII2}).

Lastly, we note that the problem of finding the MFET through the fully
reactive arc $(-\ve,\ve)$ of a disk, without LRI potential
($U(r)\equiv 0$) and without a barrier at the EW ($\kappa =
\infty$) was solved analytically by Singer {\it et al.}
\cite{Singer06b} (see also \cite{Caginalp12,Rupprecht15}).

\section{Triangular-well potential in 3D case}
\label{SM4}

We now make a particular choice of the interaction potential between
the diffusive particle and the boundary -- a triangular-well radial
potential defined in (\ref{triangular}) (see Fig. \ref{Fig2}).  An
advantage of such a choice is that i) it is simple but physically
meaningful (see the discussion in \cite{pop}), ii) it permits to
obtain an exact solution of the modified boundary-value problem and
hence to check the accuracy of our predictions in (\ref{gen3}) and
(\ref{gen2}), iii) it allows to verify our arguments behind the
derivation of the asymptotic series in (\ref{cc}) and (\ref{cc2D}),
and finally, iv) it helps to highlight some spectacular effects of the
long-range particle-boundary interactions on the MFET.

\begin{figure}
\begin{center}
\includegraphics[width=80mm]{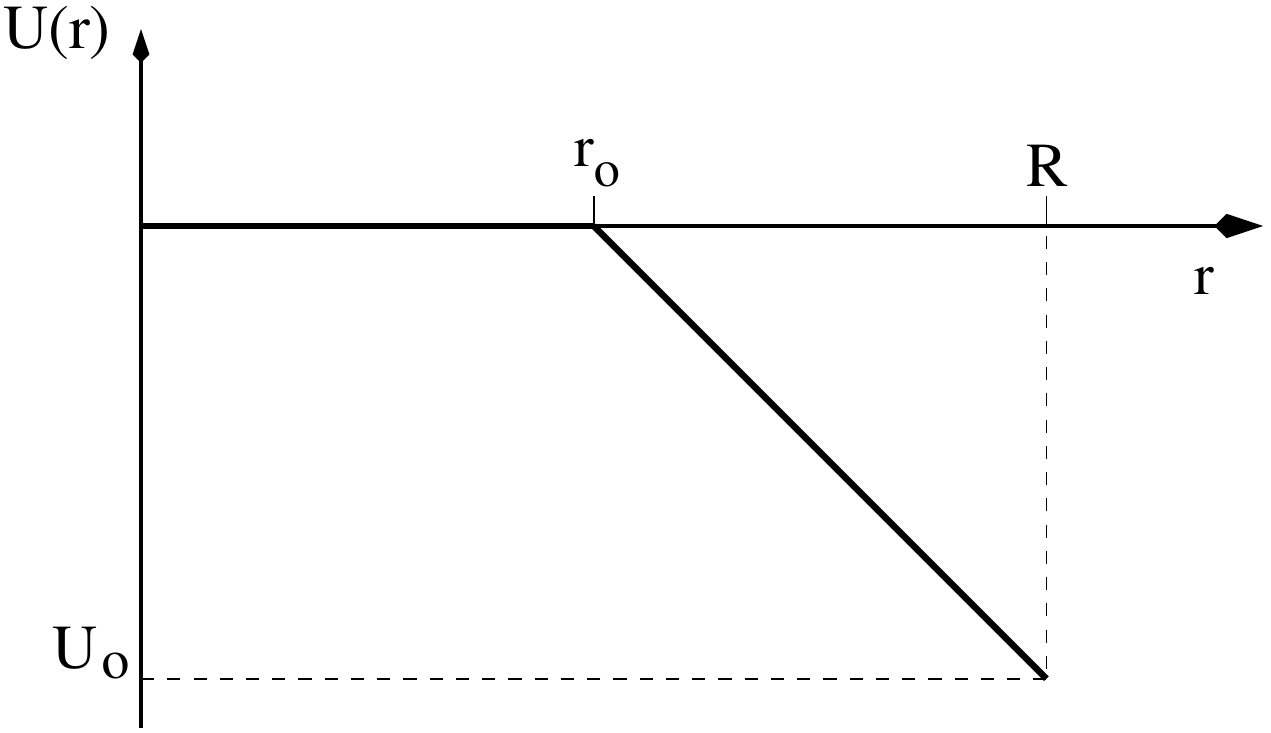} 
\end{center}
\caption{
A sketch of the triangular-well interaction potential in
(\ref{triangular}). }
\label{Fig2}
\end{figure}

\subsection{Solution of the inhomogeneous problem (\ref{eq:u0}).}

First, we compute $t_0(r)$ by direct integration of the expression in
(\ref{eq:u0}) to get
\begin{equation}
\label{eq:u0_lin11}
t_0(r) = \begin{cases} \displaystyle \frac{r_0^2 - r^2}{6D} + H^{(3)}(\etam_0)  ,\quad 0 \leq r \leq r_0 , \cr
H^{(3)}\left(\dfrac{\etam \, r}{R}\right) , \hskip 12mm r_0 < r \leq R, \end{cases}
\end{equation}
where
\begin{align}
\label{eq:H}
H^{(3)}(z) &= \frac{R^2}{D \etam^2} \biggl[\bigl(\etam_0^3/3 + \etam_0^2 + 2 \etam_0 + 2\bigr) e^{-\etam_0} 
\int\limits_{z}^{\etam} dx \frac{e^x}{x^2}  - \etam (1-x_0) - \frac{2 (1-x_0)}{\etam_0}  + 2\ln x_0 \biggr],
\end{align}
with the dimensionless parameters 
\begin{equation*}
x_0 = \frac{r_0}{R} \,,  \quad \etam = \frac{R U_0}{R - r_0} = \frac{U_0}{1-x_0} \,, 
\quad \etam_0 = \frac{r_0 U_0}{R - r_0} = x_0 \etam \,.  
\end{equation*}
Next, the derivative of $t_0(r)$ reads
\begin{equation}
\label{eq:u0_prime_lin}
t'_0(R) = \frac{R}{D\etam^3} \biggl(\etam^2 + 2 \etam + 2 - \bigl(\etam_0^3/3 + \etam_0^2 + 2 \etam_0 + 2\bigr)e^{\etam -\etam_0}\biggr).
\end{equation}
Integrating (\ref{eq:u0_lin11}), one finds the result in  (\ref{eq:Tpi_3d}).

\subsection{Radial functions $g_n(r)$}

In order to solve (\ref{eq:gn}), one finds solutions on each
of the subintervals
\begin{equation}
g_n(r) = \begin{cases} A^- r^n + B^- r^{-n-1},  \hskip 17mm   0 \leq r \leq r_0, \cr
A^+ u_n\left(\dfrac{\etam \, r}{R}\right) + B^+ v_n\left(\dfrac{\etam \, r}{R}\right),  \quad r_0 < r \leq R, \end{cases}
\end{equation}
where $A^\pm$ and $B^\pm$ are unknown coefficients to be determined,
and
\begin{equation}
\begin{split}
u_n(z) & = z^n M(n,2n+2, z) , \\  
v_n(z) & =  z^{-n-1} U(-n-1,-2n, z) \\
\end{split}
\end{equation}
are two independent solutions in the presence of a triangular-well
potential (\ref{triangular}), $M(a,b,z)$ and $U(a,b,z)$ being Kummer's
and Tricomi's confluent hypergeometric functions, respectively.  The
regularity of $g_n(r)$ at $r=0$ requires $B^- = 0$.  Requiring the
continuity of $g_n(r)$ and of its derivative $g'_n(r)$ at $r = r_0$,
one relates $A^+$ and $B^+$ to $A^-$:
\begin{equation}
\begin{split}
A^- \, r_0^n & = A^+ \, u_n\left(\etam_0\right) + B^+ \, v_n(\etam_0), \\
n \, A^- R \, r_0^{n-1} & = A^+ \, \etam \, u'_n(\etam_0) + B^+ \, \etam \, v'_n(\etam_0). \\
\end{split}
\end{equation}
These relations can be inverted to get
\begin{equation}
\begin{split}
A^+ & = A^- \, \frac{\left(r_0^n \, v'_n(\etam_0) - n \, R \, r_0^{n-1} v_n(\etam_0)/\etam\right)}
{u_n(\etam_0) \, v'_n(\etam_0) - v_n(\etam_0) \, u'_n(\etam_0)}  , \\
B^+ & = A^- \, \frac{\left(-r_0^n \, u'_n(\etam_0) + n \, R \, r_0^{n-1} \, u_n(\etam_0)/\etam\right)}
{u_n(\etam_0) \, v'_n(\etam_0) - v_n(\etam_0) \, u'_n(\etam_0)} \,. \\
\end{split}
\end{equation}
The denominator in the latter expressions is the Wronskian of the
solution, which can be calculated explicitly:
\begin{equation}
u_n(z) v'_n(z) - v_n(z) u'_n(z) = - \frac{(2n+1)!}{(n-1)!} ~ \frac{e^{z}}{z^2} .
\end{equation}
Note that the Wronskian can be ``absorbed'' into a
prefactor, which will then be factored out.  We write then
\begin{equation}
g_n(r) = A^* \biggl[u_n\left(\dfrac{\etam \, r}{R}\right) - v_n\left(\dfrac{\etam \, r}{R}\right) w_n\left(\etam_0\right)\biggr] ,
\end{equation}
where
\begin{equation}
w_n(z) = \frac{z u'_n(z) - n u_n(z)}{z v'_n(z) - n v_n(z)} .
\end{equation}
We therefore obtain
\begin{equation}
\label{eq:gn_Rgnprime}
\frac{g_n(R)}{R g'_n(R)} = \frac{u_n(\etam) - v_n(\etam) ~w_n(\etam_0)}
{\etam ~u'_n(\etam) -  \etam ~v'_n(\etam) ~w_n(\etam_0)}. 
\end{equation}
Next, using the relations
\begin{align*}
&zu'_n(z) = n z^n M(n+1,2n+2,z) , \nonumber\\
&zv'_n(z) = - \frac{n (n+1)}{z^{n+1}} U(-n,-2n,z)  ,  \nonumber 
\end{align*}
one can represent
\begin{align*} 
zu'_n(z) - n u_n(z) &= n z^n \bigl[M(n+1,2n+2,z) - M(n,2n+2,z)\bigr]  \nonumber\\
&=\frac{n z^{n+1}}{2n+2} M(n+1,2n+3,z) , \nonumber\\
zv'_n(z) - n v_n(z) &= -n z^{-n-1} \bigl[(n+1)U(-n,-2n,z) + U(-n-1,-2n,z)\bigr] \nonumber\\
&= -n z^{n+1} U(n+1,2n+3,z) ,
\end{align*}
so that
\begin{equation}
\label{w}
w_n(z) = - \frac{1}{2(n+1)}~ \frac{M(n+1,2n+3,z)}{U(n+1,2n+3,z)} .
\end{equation}
Taking together (\ref{eq:gn_Rgnprime}) to (\ref{w}),  we obtain
\begin{align}
\label{eq:gn_Rgnprime2}
\frac{g_n(R)}{R g'_n(R)}  &= \frac{1}{n} ~ \frac{M(n,2n+2,\etam)}{M(n+1,2n+2,\etam)}  \, 
\left(1 - \frac{U(n,2n+2,\etam)}{M(n,2n+2,\etam)}~ w_n(\etam_0)\right)  \nonumber\\
& \times \left(1 + \frac{(n+1) U(n+1,2n+2,\etam)}{M(n+1,2n+2,\etam)}~ w_n(\etam_0)\right)^{-1} \,,  
\end{align}
which is the desired exact expression for the ratio $g_n(R)/(R
g'_n(R))$ for the triangular-well potential.  

For numerical computations, another representation in terms of the
modified Bessel functions $I_{n+1/2}(z)$ and $K_{n+1/2}(z)$ can be
convenient.  Starting from the identities
\begin{align}
\label{eq:bessels}
M(n+1,2n+2,x) & = \Gamma(n+3/2) \left(\frac{4}{x}\right)^{n+1/2} e^{x/2} I_{n+1/2}(x/2) \,, \nonumber\\
U(n+1,2n+2,x) & = \frac{e^{x/2}}{\sqrt{\pi} x^{n+1/2}} K_{n+1/2}(x/2) \,, 
\end{align}
one can use the recurrence relations between Kummer's and Tricomi's
functions to represent all the entries in (\ref{eq:gn_Rgnprime2}) in
terms of $I_{n+1/2}(z)$ and $K_{n+1/2}(z)$. This gives
\begin{align}
\frac{g_n(R)}{Rg'_n(R)} &= \frac{1}{n} \left(1 + \frac{\etam\, i_n(\etam)}{2(n+1)}\right) \bigl(1 + j_n(\etam,\etam r_0/R) \bigr)^{-1}   \nonumber\\
&\times \left(1 - j_n(\etam, \etam r_0/R) \, \frac{k_n(\etam) - 2 \frac{n+1}{\etam}}{i_n(\etam) + 2 \frac{n+1}{\etam}}\right) \,, 
\end{align}
with
\begin{align}
&i_n(z) = \frac{I_{n+3/2}(z/2)}{I_{n+1/2}(z/2)} - 1 \,, \quad 
k_n(z) = \frac{K_{n+3/2}(z/2)}{K_{n+1/2}(z/2)} + 1 \,, \nonumber\\
&j_n(z,z_0) = \frac{K_{n+1/2}(z/2)}{K_{n+3/2}(z_0/2)} \, \frac{I_{n+3/2}(z_0/2)}{I_{n+1/2}(z/2)} \, .
\end{align}

Before we proceed with the analysis of the asymptotic large-$n$
behaviour of $g_n(R)/(R g'_n(R))$, it might be expedient to note that
in the particular case $r_0 = 0$, the expression in
(\ref{eq:gn_Rgnprime2}) simplifies to give
\begin{equation}
\frac{g_n(R)}{R g'_n(R)} = \frac{u_n(\etam)}{\etam ~ u'_n(\etam )} = \frac{1}{n} ~ \frac{M(n,2n+2,\etam )}{M(n+1,2n+2,\etam )} \,,
\end{equation}
which is just the first factor in (\ref{eq:gn_Rgnprime2}) since
$w_n(0)$ appears to be equal identically to zero.  In this particular
case, we have
\begin{equation}
\label{eq:H_r0}
H^{(3)}(z) = \frac{1}{D \etam^2} \biggl[2 \int\limits_{z}^{\etam} dx \frac{e^x-x-1}{x^2} - \etam (1 - x_0) \biggr],
\end{equation}
so that the MFPT from a random location to any point on the boundary
becomes
\begin{equation}
T_{\pi}^{(3)}(\kappa=\infty) = \frac{R^2}{D \etam^5} \biggl[2 (e^{\etam} (\etam-1) + 1)  - \frac{3 \etam^2 + 8 \etam + 12}{12} \etam^2 \biggr].
\end{equation}
Consequently, the MFPT to the EW from some fixed location has the
form:
\begin{align}
t(r,\theta)  = t_0(r) + a_0 &+ R \, t'_0(R)  \sum\limits_{n=1}^\infty \frac{M\left(n,2n+2,\dfrac{\etam \, r}{R}\right) (r/R)^n}{n M(n+1,2n+2,\etam)}  \phi_n(\ve) P_n(\cos(\theta)), 
\end{align}
where $a_0$ is given by (\ref{eq:a0}) and, explicitly,
\begin{equation}
\label{eq:uprime_strong}
t'_0(R) = \frac{R}{D\etam^3} \bigl(\etam^2 + 2 \etam + 2 - 2 e^{\etam}\bigr) \,.
\end{equation}

Consider next the behaviour of the inverse logarithmic derivative of
the radial functions at the confining boundary in the limit $n \to
\infty$ for arbitrary $r_0$.  First, we find that the first factor in
(\ref{eq:gn_Rgnprime2}) obeys
\begin{align}
&\frac{1}{n} ~ \frac{M(n,2n+2,\etam)}{M(n+1,2n+2,\etam)}  = \frac{1}{n} - \frac{\etam}{2 n^2}  + \frac{4 \etam + \etam^2}{8 n^3} + O\left(\frac{1}{n^4}\right) \,.
\end{align}
Second, we analyse the large-$n$ behaviour of the second factor in
(\ref{eq:gn_Rgnprime2}).  Taking into account the definition of
$w_n(z)$ in (\ref{w}), we observe that the correction term to unity
has the form of a product of ratios of two Kummer's and Tricomi's
functions with different arguments.  The asymptotic large-$n$
behaviour of the ratio of two Kummer's functions follows
\begin{align}
& \frac{M(n+1, 2 n + 3, \etam_0)}{(n+1) M(n, 2 n+2, \etam)} = \exp\left(- \frac{\etam}{2} \left(1 - x_0\right)\right) \nonumber\\
& \times \Bigg[ \frac{1}{n} + \frac{\etam_0^2  - 4 \etam_0 - \left(4 - \etam\right)^2}{16 n^2} + O\left(\frac{1}{n^3}\right)\Bigg] \,,
\end{align}
i.e., is an expansion in the inverse powers of $n$.  The ratio of two
Tricomi's functions is given by
\begin{align}
\label{hypU}
 & \frac{U(n, 2 n + 2, \etam)}{U(n+1, 2 n + 3, \etam_0)} = n \, \etam_0 \,  x_0^n  
\frac{\sum_{s=0}^{n+1} \binom{n+1}{s} \Gamma(n+s)/\etam^s}{\sum_{s=0}^{n+1} \binom{n+1}{s} \Gamma(n+s+1)/\etam_0^s} \,.
\end{align}
Noticing that in the latter expression the major contribution to the
sums in the numerator and the denominator stems from the terms with
$s=n+1$, we infer that the leading behaviour of the ratio in
(\ref{hypU}) in the limit $n \to \infty$ obeys
\begin{align} 
& \frac{U(n,2 n + 2, \etam)}{U(n+1, 2 n + 3, \etam_0)} \sim \frac{\etam_0  \, n}{(2 n + 1)} x_0^{2 n+1} \,, 
\end{align} 
which means that the ratio of two Tricomi's functions vanishes
\textit{exponentially} fast with $n$ as $n \to \infty$ for $x_0< 1$
(i.e., $r_0 < R$).  This implies, in turn, that the correction term to
unity in the second factor in (\ref{eq:gn_Rgnprime2}) is exponentially
small as $n\to \infty$ and hence, can be safely neglected.
  
Next, we consider the behaviour of the third factor in
(\ref{eq:gn_Rgnprime2}) which is also a product of ratios of two
Kummer's and Tricomi's functions.  We have
\begin{align}
& \frac{M(n+1, 2 n + 3, \etam_0)}{M(n+1, 2 n + 2, \etam)} = \exp\left(- \frac{\etam}{2}\left(1 - x_0\right)\right)  \nonumber\\
& \times \Bigg[1 - \frac{4 \etam_0 + \etam^2 (1 - x_0^2)}{16 n} + O\left(\frac{1}{n^2}\right)  \Bigg] 
\end{align}
and
\begin{align}
& \frac{U(n+1,2 n+2, \etam)}{U(n+1,2 n + 3, \etam_0)} = x_0^{n+1}  
\frac{\sum_{s=0}^n \binom{n}{s} \Gamma(n+1+s)/\etam^s}{\sum_{s=0}^{n+1} \binom{n+1}{s} \Gamma(n+1+s)/\etam_0^s} \,.
\end{align}
Noticing that in the $n \to \infty$ limit the major contribution to
the sums in the numerator and the denominator in the latter expression
is provided by the terms with $s = n$, we find eventually that the
leading behaviour of the ratio of two Tricomi's functions is defined
by
\begin{align}
& \frac{U(n+1,2 n+2, \etam)}{U(n+1,2 n + 3, \etam_0)} \sim \frac{\etam_0}{(2 n + 1)} x_0^{2 n + 1} \,.
\end{align}
Therefore, due to the factor $x_0^{2n+1}$, which vanishes
exponentially fast as $n \to \infty$, the third factor in
(\ref{eq:gn_Rgnprime2}) appears to be exponentially close to $1$ and
can be safely neglected.

\begin{figure}
\begin{center}
\includegraphics[width=80mm]{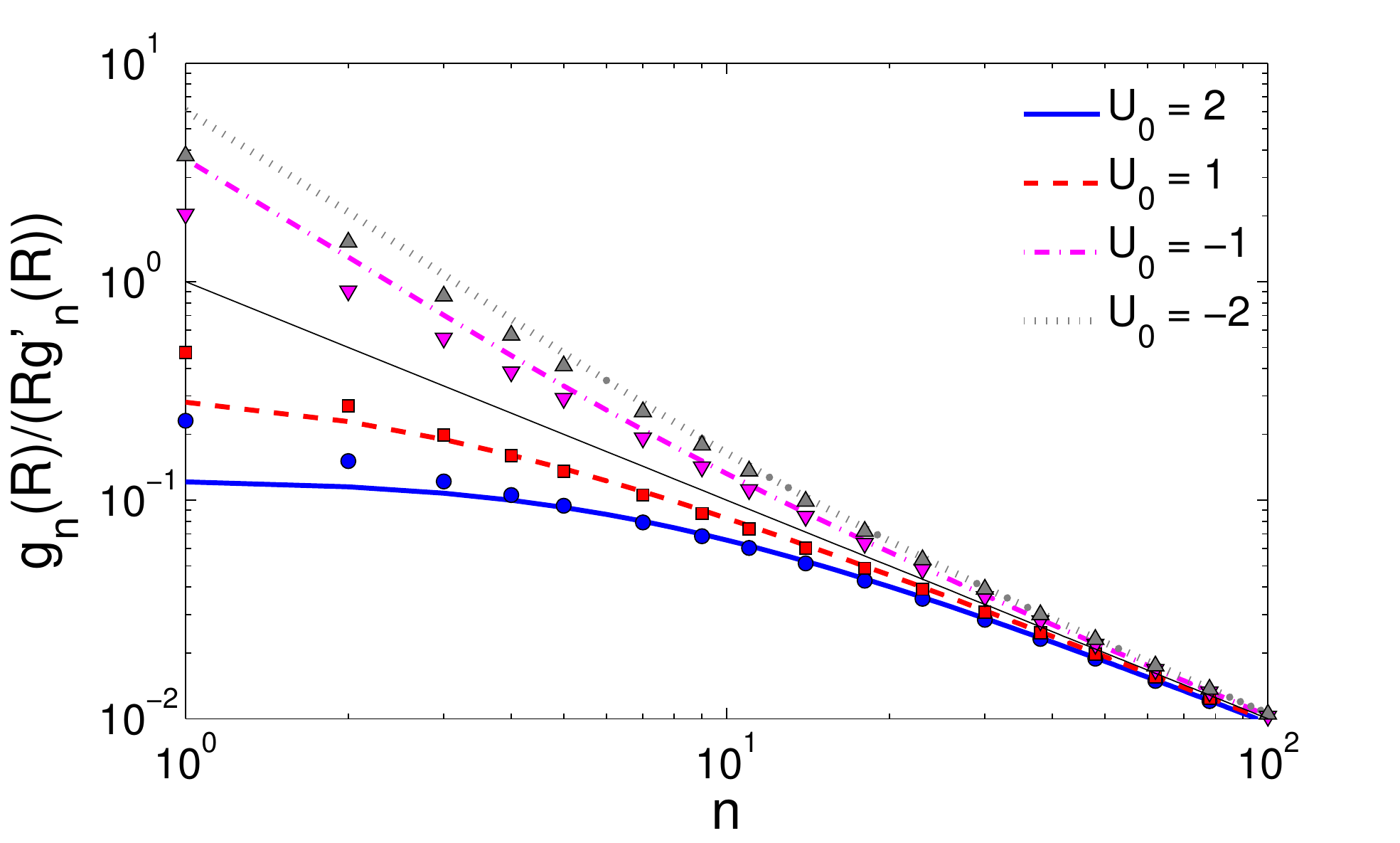} 
\end{center}
\caption{
The ratio $g_n(R)/(R g'_n(R))$ vs the order $n$ of the radial function
for several values of $U_0$ with $r_0 = 0.8$ and $R=1$.  Comparison of
the exact result in (\ref{eq:gn_Rgnprime2}) (symbols) and the
approximate expression in (\ref{apprx7}) (lines).  Thin solid line is
the $1/n$ asymptotics (solution of the problem with $U_0 \equiv 0$).}
\label{denis7}
\end{figure}

\begin{figure}
\begin{center}
\includegraphics[width=80mm]{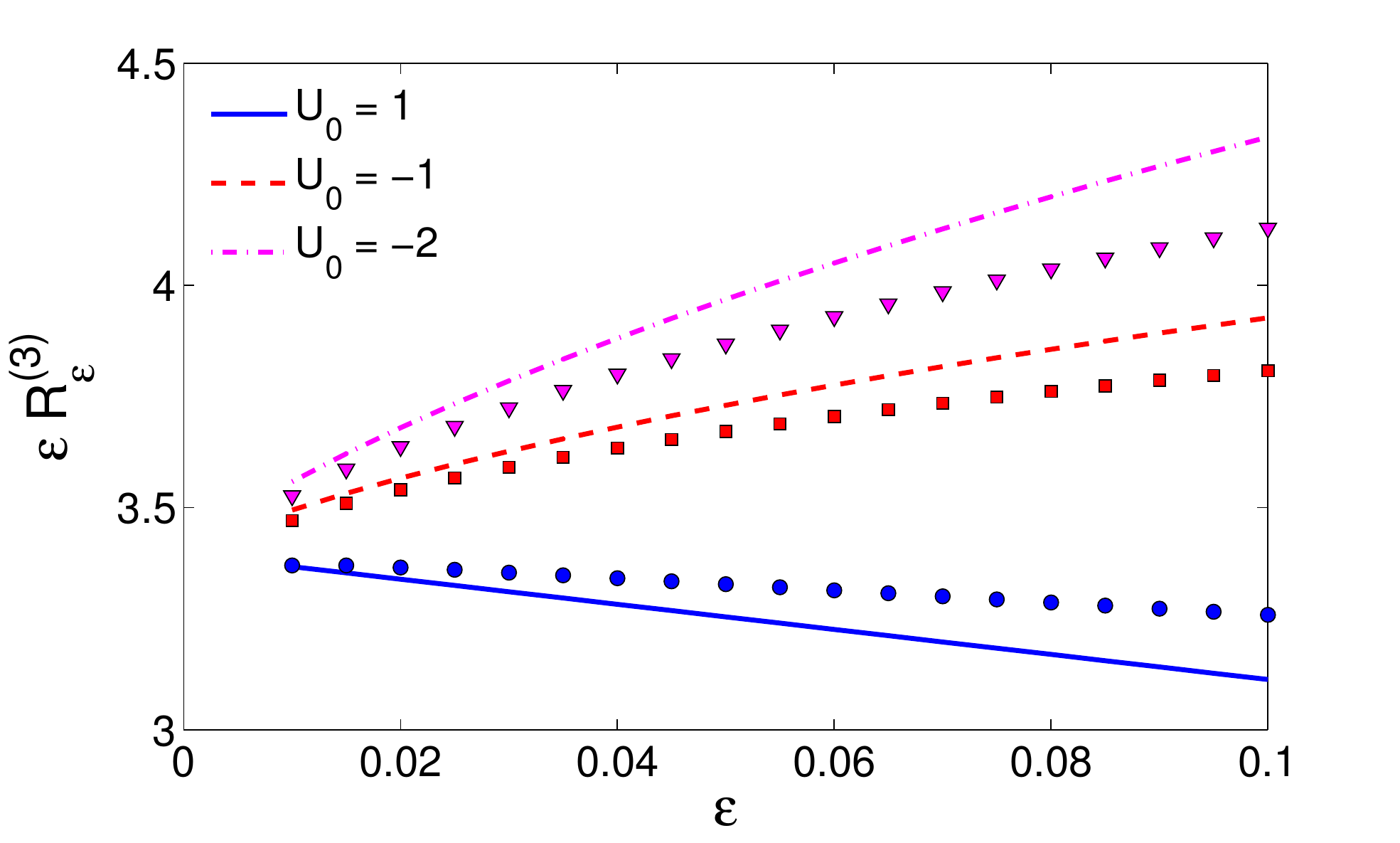}  
\end{center}
\caption{
Infinite series $\Rve^{(3)}$, multiplied by $\ve$, as a function of
the angular size $\ve$ of the EW for $U_0 = 1$, $U_0 = - 1$ and $U_0 =
- 2$ for $r_0 = 0$ and $R = 1$.  Symbols represent the exact result
obtained by numerical summation of (\ref{eq:Rd_3d}) involving the
expression in (\ref{eq:gn_Rgnprime2}), while curves show the
asymptotic relation (\ref{asympR3D}) without the last infinite sum.  }
\label{denis9}
\end{figure}

As a consequence, the leading asymptotic behaviour of $g_n(R)/(R
g'_n(R))$ in (\ref{eq:gn_Rgnprime2}) is entirely dominated by the
first factor and hence, we have
\begin{align}
\label{eq:gn_Rgnprime3}
&\frac{g_n(R)}{R g'_n(R)}  = \frac{1}{n} - \frac{\etam}{2 n^2} + \frac{4 \etam + \etam^2 }{8 n^3} + O\left(\frac{1}{n^4}\right) \,.
\end{align} 
The expansion in (\ref{eq:gn_Rgnprime3}) permits us to verify our
general argument on the asymptotic behaviour of the ratio
$g_n(R)/(Rg'_n(R))$ in the limit $n \to \infty$, presented in the
beginning of \ref{SM2} (see (\ref{cc})).  Recalling that $\etam = R
U'(R) = U_0/(1 - x_0)$ and that for such a potential $U''(R)=0$, we
observe a perfect coincidence of (\ref{cc}), based on an intuitive
(albeit quite plausible) argument, and the large-$n$ expansion of the
inverse of the logarithmic derivative, evaluated for an exactly
solvable case of a triangular-well potential $U(r)$.  Further, we note
that the large-$n$ behaviour of (\ref{eq:gn_Rgnprime3}) is dominated
by the first factor, which is the solution for a particular case $r_0
= 0$.  This implies, in turn, that in the large-$n$ limit the
dependence on $r_0$ is fully embodied in the dimensionless parameter
$\etam$.

Lastly, we compare the approximate expression (\ref{apprx7}) for
$g_n(R)/(R g'_n(R))$ and the exact result in (\ref{eq:gn_Rgnprime2})
obtained for the triangular-well potential, see Fig. \ref{denis7}.  We
observe a fairly good agreement between the approximate formula
(\ref{apprx7}) and the exact result already for quite modest values of
$n$, and notice that the agreement becomes even better for larger
values of $R |U'(R)|$.  Accordingly, our approximate small-$\ve$
expansion in (\ref{asympR3D}) (without the infinite sum in the last
line) and the exact result for $\Rve^{(3)}$ agree well with each
other, see Fig. \ref{denis9}.  The smaller $\ve$, the better agreement
is.

\section{Triangular-well potential in 2D case}
\label{SM5}

\subsection{Solution of the inhomogeneous problem (\ref{eq:u0_2d})}

Integrating Eq. (\ref{eq:u0_2d}), we get
\begin{equation}
\label{eq:t0_2d_A}
t_0(r) = \begin{cases} \displaystyle \frac{r_0^2 - r^2}{4D} + H^{(2)}(\etam_0) \qquad 0 \leq r \leq r_0 , \cr   
H^{(2)}\left(\dfrac{\etam \, r}{R}\right) \hskip 15mm r_0 < r \leq R, \end{cases}
\end{equation}
where
\begin{align}
\label{Hz}
H^{(2)}(z) &= \frac{R^2}{D\etam^2} \biggl[\bigl(\etam_0^2/2 + \etam_0 + 1 \bigr) e^{-\etam_0} \int\limits_z^{\etam} dx ~ \frac{e^x}{x} - \etam (1-x_0) + \ln x_0\biggr] ,
\end{align}
Integrating (\ref{eq:t0_2d_A}), one arrives at (\ref{eq:Tpi_2d}).  One
also gets
\begin{equation}
t'_0(R) = \frac{R}{D \etam^2} \biggl(\etam + 1 - \bigl(\etam_0^2/2 + \etam_0 + 1\bigr) e^{\etam - \etam_0}\biggr) \,.
\end{equation}

\subsection{Radial functions}

In two dimensions, the solutions of (\ref{eq:gn}) for a
triangular-well potential read
\begin{equation}
\begin{split}
u_n(z) & = z^n M(n,2n+1,z) , \\
v_n(z) & = z^{-n} U(-n,-2n+1,z) . \\
\end{split}
\end{equation}
Using the identities
\begin{equation}
\begin{split}
zu'_n(z) & = n z^n M(n+1,2n+1,z) , \\
zv'_n(z) & = -n^2 z^{-n} U(-n+1,-2n+1,z) , \\
\end{split}
\end{equation}
one gets
\begin{equation}
\begin{split}
zu'_n(z) -n u_n(z) & = \frac{n z^{n+1}}{2n+1} M(n+1,2n+2,z) , \\
zv'_n(z) -n v_n(z) & = -n z^{-n} U(-n,-2n,z) , \\
\end{split}
\end{equation}
so that $w_n(z)$ becomes
\begin{align}
\label{eq:w_2d}
& w_n(z)  = \frac{z u'_n(z) - n u_n(z)}{z v'_n(z) - n v_n(z)}  = - \frac{M(n+1,2n+2,z)}{(2n+1) U(n+1,2n+2,z)} .
\end{align}
Combining these equations we obtain the following explicit expression
for the inverse logarithmic derivative of the radial functions in the
2D case with the triangular-well potential:
\begin{align}
\label{eq:gn_Rgnprime2_2d}
\frac{g_n(R)}{R g'_n(R)} & = \frac{1}{n} ~ \frac{M(n,2n+1,\etam)}{M(n+1,2n+1,\etam)}  
\Bigg(1 - \frac{U(n,2n+1,\etam)}{M(n,2n+1,\etam)}~ w_n(\etam_0)\Bigg)  \nonumber\\
&\times  \Bigg(1 + \frac{n U(n+1,2n+1,\etam)}{ M(n+1,2n+1,\etam)}~ w_n(\etam_0)\Bigg)^{-1} \,.  
\end{align}
As earlier in the 3D case, another representation can be obtained
using (\ref{eq:bessels})
\begin{equation}
\frac{g_n(R)}{Rg'_n(R)} = \frac{1}{n} \, \frac{1-i_n(\etam) + (1+k_n(\etam)) \, j_n(\etam,\etam_0)}{1+i_n(\etam) + (1-k_n(\etam))\, j_n(\etam, \etam_0)} \,, 
\end{equation}
with
\begin{align}
i_n(z) &= \frac{I_{n+1/2}(z/2)}{I_{n-1/2}(z/2)} \,, \qquad   k_n(z) = \frac{K_{n+1/2}(z/2)}{K_{n-1/2}(z/2)} \,, 
\nonumber\\ j_n(z,z_0) &= \frac{K_{n-1/2}(z/2)}{K_{n+1/2}(z_0/2)} \, \frac{I_{n+1/2}(z_0/2)}{I_{n-1/2}(z/2)}  \,.
\end{align}

As in the 3D case, we consider first the solution in the particular
case when $r_0 = 0$.  One may readily observe that here $w_n(0) = 0$,
which implies that (\ref{eq:gn_Rgnprime2_2d}) attains a simpler form
\begin{equation}
\frac{g_n(R)}{R g'_n(R)} = \frac{1}{n} ~ \frac{M(n,2n+1,\etam)}{M(n+1,2n+1,\etam)} \,,
\end{equation}
which is again just the first factor in (\ref{eq:gn_Rgnprime2_2d}). 

%
%

\begin{figure}
\begin{center}
\includegraphics[width=80mm]{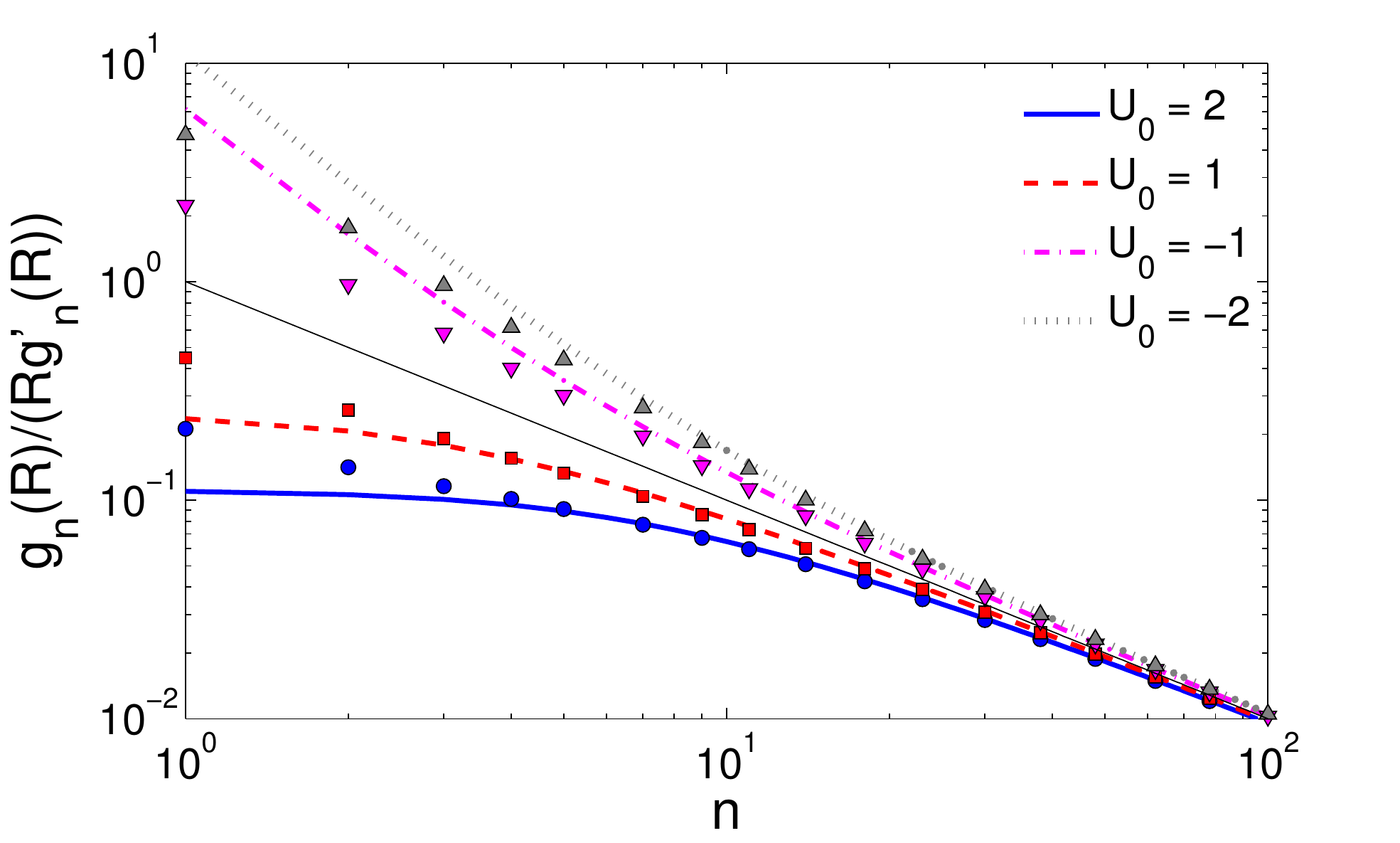} 
\end{center}
\caption{
The ratio $g_n(R)/(R g'_n(R))$ vs the order $n$ of the radial function
for several values of $U_0$, with $r_0 = 0.8$ and $R=1$.  Comparison
of the exact result in (\ref{eq:gn_Rgnprime2_2d}) (symbols) and the
approximate expression in (\ref{apprx8}) (lines).  Thin solid line is
the $1/n$ asymptotics (solution for $U_0 \equiv 0$). }
\label{denis8}
\end{figure}

Turning to the limit $n \to \infty$, we find that the first factor in
(\ref{eq:gn_Rgnprime2_2d}) obeys
\begin{align}
& \frac{1}{n} ~ \frac{M(n,2n+1,\etam)}{M(n+1,2n+1,\etam)} = \frac{1}{n} - \frac{\etam}{2 n^2} 
 + \frac{2 \etam + \etam^2}{8 n^3} + O\left(\frac{1}{n^4}\right) \,.
\end{align}
Further, considering the second and the third factors on the
right-hand-side of (\ref{eq:gn_Rgnprime2_2d}) we use a similar
analysis as in the 3D case to find that their deviation from unity is
exponentially small.  This yields the following result for the
behaviour of $g_n(R)/(R g'_n(R))$ in the limit $n \to \infty$:
\begin{align}
\label{2dex}
\frac{g_n(R)}{R g'_n(R)} & = \frac{1}{n} - \frac{\etam}{2 n^2}  
+ \frac{2 \etam + \etam^2}{8 n^3} + O\left(\frac{1}{n^4}\right) \,.
\end{align}
Recalling the definition of $\etam$ and noting that for the
triangular-well potential $U''(R)=0$, we again observe a perfect
agreement between our expansion in (\ref{cc2D}) and the exact result
in (\ref{2dex}).  We note as well that similarly to the 3D case, it
appears that the large-$n$ behaviour is dominated by the solution with
$r_0 = 0$, which implies that the dependence on this parameter of the
interaction potential is fully taken into account by the parameter
$\etam$.

Lastly, we compare the approximate expression (\ref{apprx8}) for
$g_n(R)/(R g'_n(R))$ and the exact result in
(\ref{eq:gn_Rgnprime2_2d}) obtained for the triangular-well potential.
We observe in Fig. \ref{denis8} a fairly good agreement between the
approximate formula (\ref{apprx8}) and the exact result already for
even smaller than in the 3D case values of $n$.  The agreement becomes
even better for larger values of $R |U'(R)|$.

\end{widetext}

\end{document}